\documentclass[3p,final]{elsarticle}

\usepackage[colorlinks]{hyperref}
\usepackage{graphicx,color}
\usepackage{amsmath}
\usepackage{amssymb}
\usepackage{latexsym}
\usepackage{subfigure}
\usepackage{caption}
\usepackage{multirow}
\usepackage{mdframed}
\usepackage[normalem]{ulem}
\DeclareMathOperator{\sgn}{sgn}

\usepackage{algorithm}
\usepackage{algorithmic}

\journal{J. Comp. Phys.}

\begin{document}

\begin{frontmatter}

 \title{A highly scalable massively parallel fast marching method for the Eikonal equation}

 \author{Jianming Yang\corref{cor}}
 \cortext[cor]{Corresponding author. Current affiliation: Fidesi Solutions LLC, PO Box 734, Iowa City, IA 52244, USA.} 
 \ead{jmyang@fidesisolutions.com} 
 \author{Frederick Stern}
 \address{IIHR -- Hydroscience and Engineering, University of Iowa, Iowa City, IA 52242, USA}

\begin{abstract}
The fast marching method is a widely used numerical method for solving the Eikonal equation arising from a variety of scientific and engineering fields. It is long deemed inherently sequential and an efficient parallel algorithm applicable to large-scale practical applications is not available in the literature. In this study, we present a highly scalable massively parallel implementation of the fast marching method using a domain decomposition approach. Central to this algorithm is a novel restarted narrow band approach that coordinates the frequency of communications and the amount of computations extra to a sequential run for achieving an unprecedented parallel performance. Within each restart, the narrow band fast marching method is executed; simple synchronous local exchanges and global reductions are adopted for communicating updated data in the overlapping regions between neighboring subdomains and getting the latest front status, respectively. The independence of front characteristics is exploited through special data structures and augmented status tags to extract the masked parallelism within the fast marching method. The efficiency, flexibility, and applicability of the parallel algorithm are demonstrated through several examples. These problems are extensively tested on six grids with up to 1 billion points using different numbers of processes ranging from 1 to 65536. 
Remarkable parallel speedups are achieved using tens of thousands of processes. Detailed pseudo-codes for both the sequential and parallel algorithms are provided to illustrate the simplicity of the parallel implementation and its similarity to the sequential narrow band fast marching algorithm.
\end{abstract}
\end{frontmatter}

\section{Introduction} \label{sec:intro}

The fast marching method is a widely used numerical method for solving the Eikonal (static Hamilton--Jacobi) equation, which is a first-order hyperbolic partial differential equation arising from a variety of applications, such as computational geometry, computational fluid dynamics, computer vision, materials science, optimal control, etc. \cite{Sethian99b} It is a non-iterative algorithm based on upwind difference schemes, which resembles Dijkstra's method \cite{Dijkstra59} for finding the shortest path on a network. 
Such a method was first proposed by Tsitsiklis \cite{Tsitsiklis95} using an optimal control approach. The same algorithm was independently derived in \cite{Sethian96} (where the name ``fast marching'' was introduced) and \cite{HelmsenPCD96} based on upwind difference schemes. 
The fast marching method has theoretically optimal complexity in its operation count by exploring the causality of the Eikonal equation and adapting a one-pass updating strategy. The Eikonal equation describes nonlinear boundary value problems in which the information from the boundary propagates away along characteristics. In the fast marching method, upwind difference schemes are used to discretize the Eikonal equation at a given grid point, such that the stencil contains only neighboring points with valid values (or, upwind points) and the causality of the equation is strictly followed. Moreover, a heap priority queue is used to march the solution in a rigorous increasing (decreasing for the negative solution) order. Therefore, the number of times that a point is visited is minimized and no iterations are involved in the whole process. Since the run-time complexity of reordering of a heap of length $n$ is $O(\log n)$, the fast marching method has a total operation count of $O(N \log N)$ for a case involving $N$ grid points.

The fast marching method has been applied to a wide range of scientific and engineering problems; also, numerous improvements and extensions have been developed since its introduction. For details, the reader is referred to \cite{Sethian99b}. Unfortunately, the lack of efficient parallel algorithms has severely limited its extensive usage in large-scale simulations usually found in geoscience and computational fluid dynamics applications among others. For example, Gillberg et al. \cite{Gillberg2014} solved the Eikonal equation in earth modeling problems using parallel iterative algorithms with grids up to $139$ million points; in \cite{Bhushan2011}, the wall distance function required in turbulence modeling was calculated by using the sequential fast marching method on grids up to $540$ million points. A non-iterative parallel Eikonal solver could greatly accelerate the simulations. Two major advantages of the fast marching method over other techniques that makes its parallelization particularly desirable are the monotonic increasing order of the solution and the narrow band formulation. The former, consistent with the propagation direction of the information from the boundary, is inevitable in some applications, e.g., field extension of information (e.g., velocity, scalars, etc.) from the boundary/interface to the surrounding domain. The latter can be taken advantage of to save the computational cost in many applications. For instance, in turbulence modeling, accurate wall distance function is critical within a certain distance (depending on the turbulence model used) away from the wall boundary; and the number of points requiring such a distance calculation usually is much smaller than that of the whole domain. Of course, it is possible to run the sequential fast marching algorithm within each subdomain with enough layers of ghost points to avoid the interdependence among neighboring subdomains. However, for turbulence computations, tens of layers of points are usually clustered near the wall and such a strategy is not feasible. 

The outstanding issue of parallelizing the fast marching method has received much less attention. One possible reason is that the fast marching method is long deemed inherently sequential and has no straightforward parallelism as found in an iterative method such as the fast sweeping method \cite{Zhao05,Zhao07}. Note there are several parallel implementations of iterative methods (e.g., 
\cite{JeongW08} and \cite{WeberDBBK08}, among others) for shared memory parallel architectures, especially, graphics processing units (GPU) most recently. On the other hand, Tsitsiklis \cite{Tsitsiklis95} developed two single-pass algorithms using an optimal control approach: an $O(N \log N)$ algorithm with a binary heap data structure (a Dijkstra-like method similar to the fast marching method) and an $O(N)$ algorithm using a bucket data structure, and provided a shared-memory parallel implementation for the latter. In general, these shared-memory based parallel algorithms are very difficult to be extended to distributed memory parallel architectures, on which the coarse-grain parallelization prototypes, usually based on domain decompositions, are prevalent. 

The first attempt to parallelize the fast marching method based on a domain decomposition technique was reported by Herrmann \cite{Herrmann03}. Essentially, the computational domain was decomposed into non-overlapping subdomains and ghost points (one layer for a first-order scheme) are used for communications between subdomains. 
For instance, if a just-accepted grid point is also in the ghost point zone of a neighboring block, then the information of this point will be sent to the target neighbor. In the mean time, each process repeatedly checks for ghost point updates from neighbors. A rollback mechanism was introduced to revoke the valid status of all pre-accepted points whenever a ghost point turned to valid status with a smaller value than these points. 
The asynchronous communications in this algorithm
were quite involved and difficult to implement. The rollback operations introduced significant communication and computation overheads and considerably limited the parallel performance. 
The test case of a spherical interface placed in the center of a unit cubic domain was studied with different domain decompositions in \cite{Herrmann03}. For optimal configurations without inter-dependence between subdomains, a nearly ideal linear speedup was obtained for up to $8$ processes. For non-optimal domain decomposition configurations, a parallel efficiency of $0.34$ was reported for $27$ processes. 

Breuss et al. \cite{BreussCGV2011} proposed a shared-memory domain decomposition parallelization of the fast marching method. The main idea was to split the boundaries/interfaces instead of the computational domain among the processes (parallel threads) at the beginning of the computation. An initialization procedure was given to split the boundaries/interfaces. Thread interaction rules were imposed to satisfy the causality principle. They tested the approach with several two-dimensional (2D) cases of unity speed up to $16$ threads and it was concluded that such an approach was useful for computers with two to four CPU cores. 

Tugurlan \cite{Tugurlan2008} developed a distributed-memory parallelization of the fast marching method based on a domain decomposition approach. In each iteration, the sequential fast marching algorithm is performed in on each subdomain, then ghost points are synchronized through MPI communications. The iterations continue until the convergence conditions are satisfied. An ordered overlap strategy with a sorted listed and a fast sweeping \cite{Zhao05} strategy were developed to update the points in the overlapped regions of neighboring subdomains. Some 2D cases were demonstrated with up to $36$ CPU cores. 

Chacon and Vladimirsky \cite{ChaconV2012} developed several hybrid two-scale methods for solving the Eikonal equation. In these methods, the domain is split into small grid blocks (cells) with approximately constant speed functions and the fast sweeping method is performed within cells. A fast marching like procedure with a min-heap data structure (heap) is used to determine the order of cells-to-be-processed and the sweeping directions within a cell. The author provided a shared-memory parallelization of their heap-cell method in \cite{ChaconV2015}. Each thread performs the sequential heap-cell method in a subdomain with its own local cell-heap. A reactivated cell will be added to the thread with less number of cells for better load balance. Several three-dimensional (3D) examples were demonstrated with grids up to $320^3$ ($32.8$ million) points using up to $32$ threads on a multi-core supercomputer. 

Gillberg et al. \cite{Gillberg2014} also developed two parallel algorithms based on a domain decomposition approach. Instead of a fast sweeping method, a 3D version of the parallel marching algorithm \cite{WeberDBBK08} is applied in each subdomain. Ghost point synchronization is used to exchange boundary conditions for subdomains. An updated subdomain will be locked for further iterations until its boundary conditions are changed by neighboring subdomains, and this subdomain will be added to the list of subdomains for further computations. In the list of active subdomains method, all active subdomains are placed in a list and updated in parallel. Therefore it is similar to the fast iterative method \cite{JeongW08} in some characteristics. In the semi-ordered list of active subdomains method, the lists are created to make the list of subdomains better follow the isosurfaces of the solution. Computations on grids up to $518^3$ points were performed on multicore CPUs with up to 16 cores and a 2496-core NVIDIA GPU with 5120 MBytes of memory.

In this study, a highly scalable massively parallel algorithm of the fast marching method based on a domain decomposition technique, which was first briefed in \cite{YangMBHWS10}, is discussed in detail. Developed with serious large-scale practical applications (e.g., \cite{Gillberg2014} and \cite{Bhushan2011}, among others) in mind, this massively parallel fast marching method can give a remarkable parallel performance on billion-point grids using tens of thousands of processes, whereas its implementation is surprisingly simple and straightforward. Actually, the sequential fast marching algorithm is directly incorporated into the parallel algorithm with only a few minor plain modifications. In particular, the procedure central to our parallel algorithm is a novel restarted narrow band approach, in which the fronts advance at a specified stride during each restart. Therefore, it is fully consistent with the narrow band idea in the fast marching level set method \cite{Sethian96}. Basically, for each restart of front advancing, a global bound is first determined according to the given stride size, the (essentially sequential) fast marching algorithm is then executed; updated points in the overlapping regions of neighboring subdomains are collected and exchanged; with the new data from neighboring subdomains the fast marching algorithm is carried out once more to bring the fronts everywhere up to the designated bound. Only simple synchronous communication modes are employed in the whole process. This algorithm exploits the independence of front characteristics to extract the parallelism deeply buried under the apparent sequentiality of the fast marching method. For example, special data structures are designed for two-sided interface problems, such that both the positive and negative fronts of the interface can be advanced concurrently. In addition, when a subdomain receives updated function values for grid points in the overlapping regions, none of the grid points with larger (absolute) values in this subdomain will be reset uniformly as if with a rollback mechanism. This is because many of these larger values might be computed following other characteristics that are independent of points with lower values received from neighboring processes. Augmented tags are introduced to define the point status precisely. Associated with a few slight modifications in the sequential algorithm, points that are influenced by the incoming data can be refreshed without being singled out for any special treatments.

The rest of this paper is organized as follows: In the next section the sequential fast marching method is given. Then the data structures for two-sided interface problems are introduced. In the parallel fast marching method part the parallel algorithm is described thoroughly, detailed pseudo-codes are provided for a side-by-side comparison and straightforward implementations of the sequential and parallel algorithms. 
Six test cases with different stride sizes on six Cartesian grids ranging from less than $40$ thousand to more than $1$ billion points using one to $65536$ CPU cores are performed to demonstrate the parallel speedups and efficiencies. 
Some concluding remarks are provided in the final section. 

\section{Sequential fast marching method}\label{sec:fmm}

\subsection{Eikonal equation and finite difference discretization}

The fast marching method solves the stationary boundary value problem defined by the Eikonal equation:
\begin{equation}
\begin{array}{cc}
 | \nabla \psi (\mathbf{x}) | F (\mathbf{x}) = 1, & \mathbf{x} \in \varOmega \, \backslash \, \varGamma, \\
\psi (\mathbf{x}) = 0, & \mathbf{x} \in \varGamma \subset \varOmega,
\end{array}
\label{eq:eikonal}
\end{equation}
where $\varOmega$ is a domain in $\mathcal{R}^n$, $\varGamma$ is the initial interface (boundary), and  $F (\mathbf{x})$  is a positive speed function, with which the interface information propagates in the domain. 

\begin{figure}[htbp!]
\begin{center}
 \includegraphics[angle=0,width=0.5\textwidth]{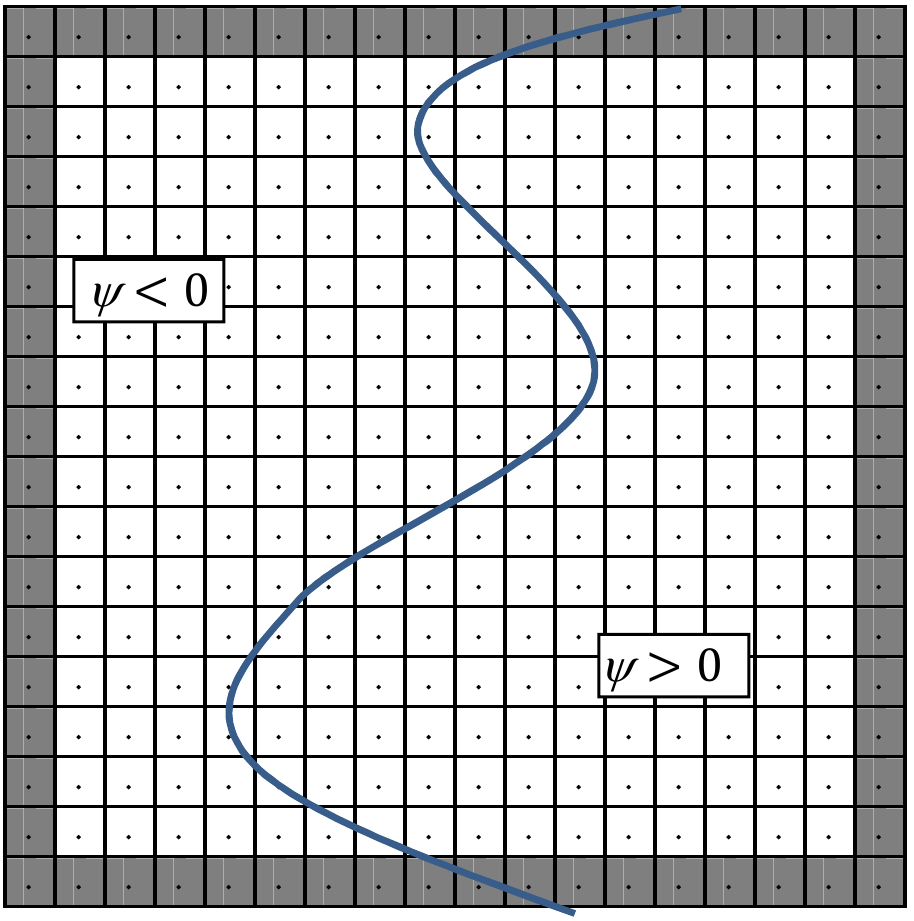}\\
\end{center}
 \caption{The computational domain $\varXi$ and the interface $\varGamma$. Blank and grey-shaded cells represent grid points inside and outside the physical domain, respectively.}
\label{fig:domain1}
\end{figure}

To solve Eq. (\ref{eq:eikonal}) numerically, domain $\varOmega$ has to be discretized first. Here a regular domain in  $\mathcal{R}^3$ defined by $\varOmega = [x_{\text{min}},x_{\text{max}}] \times[y_{\text{min}},y_{\text{max}} ]\times[z_{\text{min}},z_{\text{max}} ]$ is partitioned as $\varOmega = \cup  \Delta \varOmega_{i,j,k}$, where $1 \leq i \leq nx, 1 \leq j \leq ny, 1 \leq k \leq nz$, and $\Delta V_{i,j,k} = [x_{i-1/2},x_{i+1/2}] \times [y_{j-1/2},y_{j+1/2}] \times [z_{k-1/2},z_{k+1/2}]$. $x_{1/2} = x_{\text{min}}, x_{nx+1/2} = x_{\text{max}}, y_{1/2} = y_{\text{min}}, y_{ny+1/2} = y_{\text{max}}$, and $z_{1/2} = z_{\text{min}}, z_{nz+1/2} = z_{\text{max}}$ define $\partial \varOmega$, the boundary of $\varOmega$. Notice that the function $\psi$ is defined at the center of each computational cell (i.e., $\Delta \varOmega_{i,j,k}$). Therefore, no function is defined at the domain boundary $\partial \varOmega$, ghost points are used to facilitate the imposition of boundary conditions. Fig. \ref{fig:domain1} shows the interface $\varGamma$ and the computational domain, $\varXi = \varOmega + \varTheta $ (i.e., the combination of the discretized physical domain $\varOmega$ and ghost point zone $\varTheta$). To simplify the discussion, uniform grid distribution in each direction is considered in this study, i.e., $\Delta \varOmega_{i,j,k} = \Delta x \times \Delta y \times \Delta z$, although the methodology to be discussed is not limited to a uniform grid. 

The Godunov-type finite difference scheme given in \cite{RouyT92}, which satisfies the entropy condition in hyperbolic conservation laws, can be used to approximate Eq. (\ref{eq:eikonal}) on the computational domain as follows
\begin{equation}
\left[ 
 \begin{array}{cc}
  &\max (D^{-x}_{i,j,k} \psi, -D^{+x}_{i,j,k} \psi, 0) ^2\\
+ &\max (D^{-y}_{i,j,k} \psi, -D^{+y}_{i,j,k} \psi, 0) ^2\\
+ &\max (D^{-z}_{i,j,k} \psi, -D^{+z}_{i,j,k} \psi, 0) ^2
\end{array}
\right] ^{1/2} = \dfrac{1}{F_{i,j,k}},
\label{eq:fds}
\end{equation}
where the operators $D^{-x}_{i,j,k}$ and $D^{+x}_{i,j,k}$ define the backward and forward difference approximations to the spatial derivative $\partial \psi / \partial x$, respectively. In this study, first-order schemes are used:
\begin{equation}
D^{-x}_{i,j,k} \psi = \dfrac{\psi_{i,j,k} - \psi_{i-1,j,k}}{\Delta x}, \quad D^{+x}_{i,j,k} \psi = \dfrac{\psi_{i+1,j,k} - \psi_{i,j,k}}{\Delta x}.
\label{eq:fos}
\end{equation}
The operators for the $y$ and $z$ directions are defined similarly. Eq. (\ref{eq:fds}) gives a quadratic equation for $\psi_{i,j,k}$.

\subsection{Algorithm} \label{sec:sfmm}

\begin{figure}[htbp!]
\begin{center}
 \includegraphics[angle=0,width=0.5\textwidth]{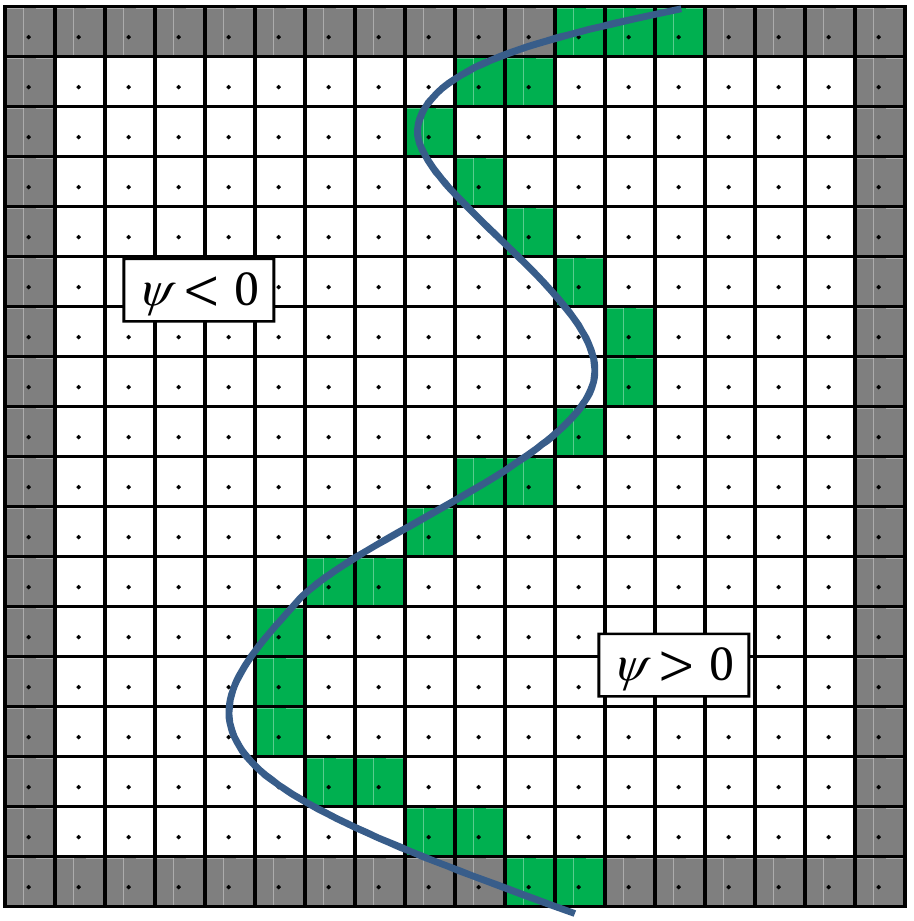}\\
\end{center}
 \caption{Interface initialization. Green-shaded cells represent $\texttt{KNOWN}$ points in the $\psi > 0$ region.}
\label{fig:domain2}
\end{figure}

Observe that Eq. (\ref{eq:fds}) has a very special upwind structure, i.e., $\psi_{i,j,k}$ only depends on the neighboring points of smaller value. The fast marching method takes advantage of this fact by solving Eq. (\ref{eq:fds}) using only the upwind points and building up the whole solution following a systematical manner from the point of the smallest value. To identify the upwind directions and establish the order of updating, each grid point $(i,j,k)$ is labeled by a status tag $G_{i,j,k}$: i) $G_{i,j,k} = \texttt{KNOWN}$, if this point contains a final function value, thus give the upwind direction; ii) $G_{i,j,k} = \texttt{BAND}$, if this point contains a function value updated by its neighboring $\texttt{KNOWN}$ point(s), but may be further updated by any new $\texttt{KNOWN}$ neighbors; and iii) $G_{i,j,k} =  \texttt{FAR}$, if this point is in the downwind side and does not have any $\texttt{KNOWN}$ neighbors yet. The point with the smallest value in the $\texttt{BAND}$ set is located and moved into the $\texttt{KNOWN}$ set, and then its neighboring $\texttt{BAND}$ and $\texttt{FAR}$ points can be updated and re-categorized. This step repeats until all points in the domain or within a pre-defined narrow band become $\texttt{KNOWN}$.

\begin{algorithm}
\algsetup{indent=1em}
\caption{Interface initialization:\newline $\textsc{Initialize\_Interface}$.}
\label{alg:init_inter}
\begin{algorithmic}[1]
  \STATE $\psi \leftarrow +\infty$
  \STATE $G  \leftarrow  \texttt{FAR}$
  \FORALL{$(i,j,k)\in \varXi $ such that $\psi_{i,j,k} \in \psi^0$ is adjacent to $\varGamma$} 
    \STATE $\psi_{i,j,k} \leftarrow  \psi^0$
    \STATE $G_{i,j,k}  \leftarrow  \texttt{KNOWN}$ 
  \ENDFOR
\end{algorithmic}
\end{algorithm}
\begin{figure}[htbp!]
\begin{center}
 \includegraphics[angle=0,width=0.5\textwidth]{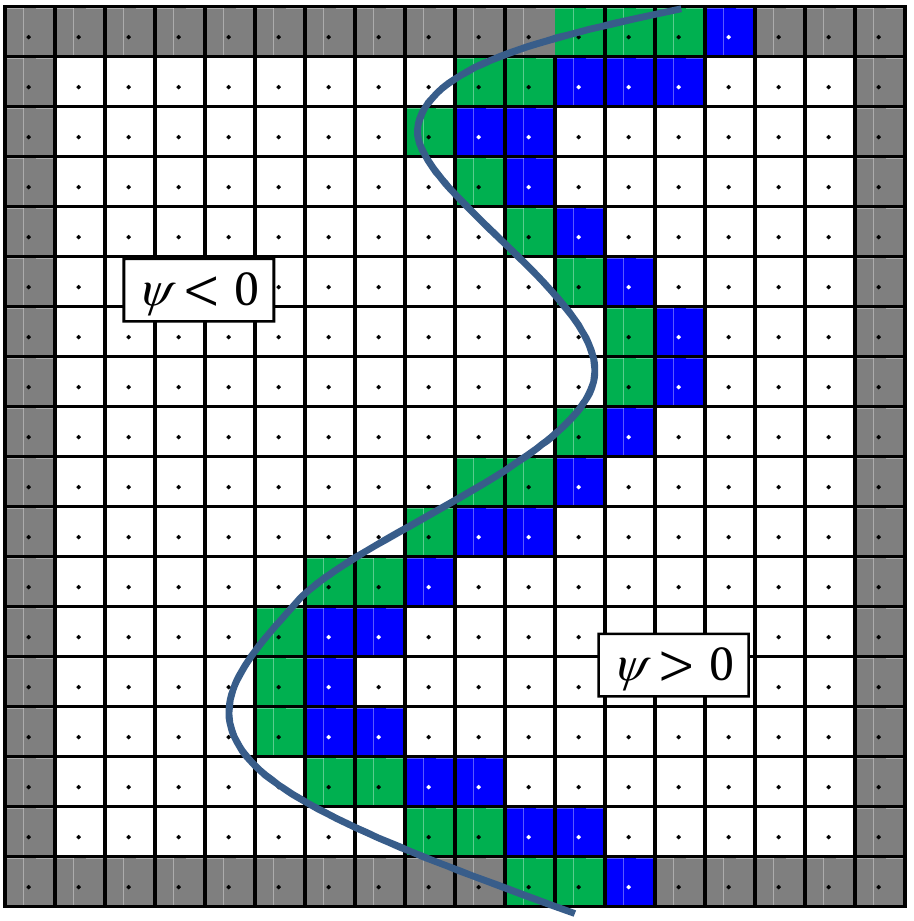}\\
\end{center}
 \caption{Heap initialization. Blue-shaded cells represent $\texttt{BAND}$ points in the $\psi > 0$ region.}
\label{fig:domain3}
\end{figure}

In the present study, detailed pseudo-codes are given to emphasize the similarities and differences between the sequential and parallel methods. In many applications, such as the first-arrival traveltime calculation for seismic wave propagation, wall distance calculation for turbulence modeling, etc., only the solution in the positive region $\Omega^+$ is required. Thus the negative values in $\psi^0$ are not needed for a first-order scheme. For brevity, in this section the algorithms are presented for one-sided boundary value problems. The algorithm for two-sided interface problems will be addressed in the next section. 

As a boundary value problem, the boundary or interface condition has to be specified for the discretized Eikonal equation in the solution procedure. In the actual implementation, this is fulfilled by the interface initialization procedure. Fig. \ref{fig:domain2} shows the interface initialization step, and the corresponding Algorithm \ref{alg:init_inter} details the operations for initializing the interface. Initially, $ + \infty$ and $\texttt{FAR}$ are assigned to each point $\psi_{i,j,k}$ as its function value and status tag, respectively. A $\texttt{FAR}$ point is identified by the blank and grey-shaded cells shown in Fig. \ref{fig:domain2} and other figures. Then, all grid points immediately adjacent to the interface are assigned values $\psi^0$ (analytical solution in the present study) and tagged as $\texttt{KNOWN}$ as shown in Fig. \ref{fig:domain2}.

\begin{algorithm}
\algsetup{indent=1em}
\caption{Heap initialization:\newline $\textsc{Initialize\_Heap}$.}
\label{alg:init_heap}
\begin{algorithmic}[1]
  \STATE $\texttt{size}_{\mathfrak{H}} \leftarrow 0$ 
    \FORALL{$(i,j,k)\in \varXi $ such that $G_{i,j,k} = \texttt{KNOWN}$} 
      \STATE $\textsc{Update\_Neighbors}(i,j,k)$
    \ENDFOR
\end{algorithmic}
\end{algorithm}

The fast marching method relies on a binary heap structure, which has to be initialized too. In the heap initialization step given in Algorithm \ref{alg:init_heap}, the initial heap size is set to zero; then, for each $\texttt{KNOWN}$ point in the computational domain, its neighboring points are updated by solving Eq. (\ref{eq:fds}) and tagged as $\texttt{BAND}$ points. It should be noted that all points in the domain including both physical and ghost regions are treated in the same manner as shown in Fig. \ref{fig:domain3}. 

The procedure for updating neighboring points of a $\texttt{KNOWN}$ point is described in Algorithm \ref{alg:update1}, which is the major operation involved in the fast marching method. In 3D, for a $\texttt{KNOWN}$ point $(i,j,k)$, its neighbors to be considered are $(i-1,j,k)$, $(i+1,j,k)$, $(i,j-1,k)$, $(i,j+1,k)$, $(i,j,k-1)$, and $(i,j,k+1)$. For each point among these neighbors, if it is inside the computational domain (ghost points are included) and it is not a $\texttt{KNOWN}$ point, then Eq. (\ref{eq:fds}) is solved at this point to obtain a new function value. If the new value is smaller than the present value at this point, then its function value will be updated with the new value. The final step is to check the status of this point, if it is a $\texttt{FAR}$ point, it will be added to the heap; otherwise, as an existing $\texttt{BAND}$ point its position in the heap will be updated. 

The heap data structure guarantees a strict order of increasing function values for solving the Eikonal equation in the fast marching method. In this study, a binary heap data structure similar to what described in \cite{Sethian99b} is used. The standard heap operations in an implementation of the indexed priority queue algorithm, i.e., $\texttt{Insert\_Heap}$, $\texttt{Locate\_Min}$, $\texttt{Remove\_Min}$, $\texttt{Up\_Heap}$ (also used in $\texttt{Insert\_Heap}$), and $\texttt{Down\_Heap}$ (used in $\texttt{Remove\_Min}$), are available in textbooks for algorithms (e.g., \cite{Sedgewick11}).

\begin{algorithm}
\algsetup{indent=1em}
\caption{Update the neighbors of a $\texttt{KNOWN}$ point:\newline $\textsc{Update\_Neighbors}(i,j,k)$.}
\label{alg:update1}
\begin{algorithmic}[1]
\FORALL{$(l,m,n)$ such that $\left( |l-i|+|m-j|+|n-k|\right) = 1$}
  \IF{$(l,m,n) \in \varXi $}
    \IF{$G_{l,m,n} \neq \texttt{KNOWN}$}
      \STATE $\psi_{\text{temp}} \leftarrow \textsc{Solve\_Quadratic}(l,m,n)$
      \IF{$\psi_{\text{temp}} < \psi_{l,m,n}$}
        \STATE $\psi_{l,m,n} \leftarrow  \psi_{\text{temp}}$
        \STATE $G_{l,m,n} \leftarrow \texttt{BAND}$
        \IF{$(l,m,n) \not\in \mathfrak{H}$} 
                  \STATE $\textsc{Insert\_Heap}(l,m,n)$ 
        \ELSE 
          \STATE $\textsc{Up\_Heap}(l,m,n)$ 
        \ENDIF
      \ENDIF
    \ENDIF
  \ENDIF
\ENDFOR
\end{algorithmic}
\end{algorithm}

\begin{figure}[htbp!]
\begin{center}
 \includegraphics[angle=0,width=0.45\textwidth]{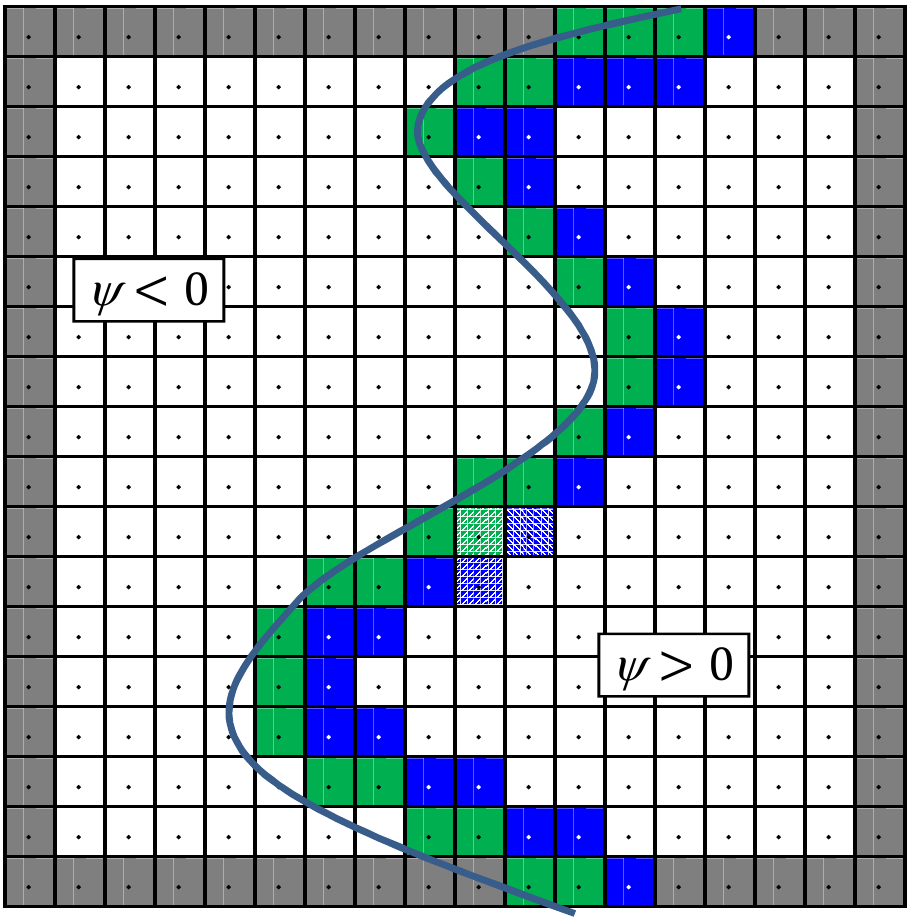} 
 \hspace{2em}
 \includegraphics[angle=0,width=0.45\textwidth]{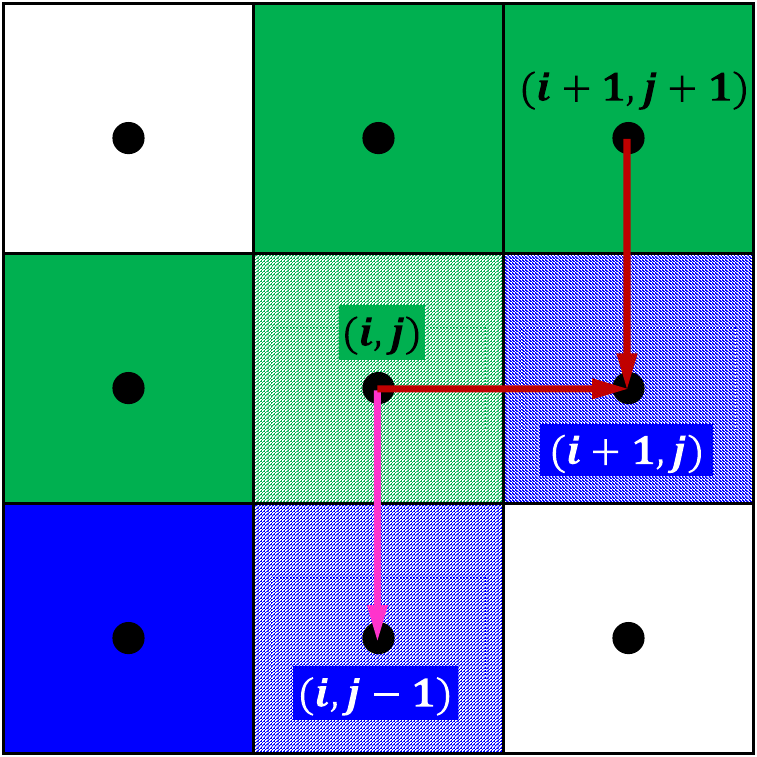} 
  \makebox[0.49\textwidth]{(a) Downwind neighbors of a new $\texttt{KNOWN}$ point}
  \makebox[0.49\textwidth]{(b) Close-up view}
\end{center}
 \caption{One step of the narrow band fast marching method. The green-shaded cell with a pattern represents a new $\texttt{KNOWN}$ point upgraded from the $\texttt{BAND}$ set. The two adjacent blue-shaded cells with patterns, to its right and bottom, represent a new $\texttt{KNOWN}$ point and a new $\texttt{BAND}$ point upgraded from the $\texttt{BAND}$ and the $\texttt{FAR}$ sets, respectively.}
\label{fig:neighbor1}
\end{figure}

The solution procedure of the quadratic equation, Eq. (\ref{eq:fds}), is described in Algorithm \ref{alg:alg3}. Only the operation for the $x$ direction is detailed as other two directions are very similar. For simplicity, assume that the only upwind point is $(l-1,m,n)$ in the quadratic equation, then Eq. (\ref{eq:fds}) will become the following form
\begin{equation}
 (D^{-x}_{l,m,n} \psi ) ^2  = \dfrac{1}{F ^2_{l,m,n}},
\label{eq:fds_x1}
\end{equation}
or, 
\begin{equation}
 (\dfrac{\psi_{l,m,n} - \psi_{l-1,m,n}}{\Delta x}) ^2  = \dfrac{1}{F ^2_{l,m,n}},
\label{eq:fds_x2}
\end{equation}
since $\psi_{l+1,m,n} = +\infty$ in Eq. (\ref{eq:fds}). Then a standard quadratic equation $a \psi ^2_{l,m,n} + b \psi _{l,m,n} + c = 0$ can be obtained with the following coefficients:
\begin{equation}
a = \dfrac{1}{\Delta x ^2}, \quad b = -\dfrac{2 \psi_{l-1,m,n}}{\Delta x ^2}, \quad c = \dfrac{\psi ^2_{l-1,m,n}}{\Delta x ^2} - \dfrac{1}{F ^2 _{l,m,n}}.
\label{eq:fds_x3}
\end{equation}
And only the solution 
\begin{equation}
    \psi_{\text{temp}} = \dfrac{-b+\sqrt{b^2-4ac}}{2a}
\label{eq:fds_x4}
\end{equation}
is acceptable if available, since $\psi_{\text{temp}} > \psi_{l-1,m,n} = -b/2a$. The above example shows the case in which the left neighbor in the $x$ direction is the only upwind point. In practice, all directions are checked for possible upwind points. And the solution from the quadratic equation will be checked against all source points to make sure that the causality is not violated. For example, assume the front is parallel to the $y-$axis and moves toward the right direction with a unity speed in Fig. \ref{fig:neighbor1}(b), then the equation will be solved at point $(i+1,j)$ after $(i+1,j+1)$ is moved to the set $\texttt{KNOWN}$. Here the solution at $(i+1,j)$ should be equal to that at $(i+1,j+1)$; and apparently point $(i+1,j+1)$ should be rejected as an upwind point to be used in the quadratic solver. With this check, therefore, 
 the whole algorithm only uses upwind points in set $\texttt{KNOWN}$ to advance the front, which guarantees that the final result is the correct viscosity solution to the Eikonal equation.

\begin{algorithm}
\algsetup{indent=1em}
\caption{Solve the quadratic equation:\newline $\textsc{Solve\_Quadratic}(l,m,n)$.}
\label{alg:alg3}
\begin{algorithmic}[1]
\STATE $\psi_{\text{temp}} \leftarrow + \infty$
\STATE {Check the $x$ direction to set $\psi_1$ and $h_1$:}
  \STATE $d \leftarrow 0$
  \IF {$(l-1,m,n) \in \varXi _p$}
    \IF {$G_{l-1,m,n} = \texttt{KNOWN}$}
      \STATE $d \leftarrow -1$
    \ENDIF
  \ENDIF
  \IF {$(l+1,m,n) \in \varXi _p$}
    \IF {$G_{l+1,m,n} = \texttt{KNOWN}$}
      \IF {$d = 0$}
        \STATE $d \leftarrow +1$
      \ELSIF {$\psi_{l+1,m,n} < \psi_{l-1,m,n}$}
        \STATE $d \leftarrow +1$
      \ENDIF
    \ENDIF
  \ENDIF
  \IF{$d \neq 0$}
    \STATE $\psi_1 \leftarrow \psi_{l+d,m,n}$
    \STATE $h_1    \leftarrow  \Delta x ^ {-1} $
  \ELSE
    \STATE $\psi_1 \leftarrow 0$
    \STATE $h_1    \leftarrow 0 $
  \ENDIF
\STATE {Check the $y$ direction to set $\psi_2$ and $h_2$}
\STATE {Check the $z$ direction to set $\psi_3$ and $h_3$}
  \STATE $a \leftarrow     \sum_i (h_i ^2)$
  \STATE $b \leftarrow - 2 \sum_i (h_i ^2 \psi_i)$
  \STATE $c \leftarrow     \sum_i (h_i ^2 \psi_i ^2 )- F_{l,m,n} ^{-2}$
  \IF {$(b^2-4ac) \geq 0$}
    \STATE $\psi_t \leftarrow \frac{-b+\sqrt{b^2-4ac}}{2a}$
    \IF {$\psi_1 < \psi_t$ \AND $\psi_2 < \psi_t$ \AND $\psi_3 < \psi_t$}
      \STATE $\psi_{\text{temp}} \leftarrow \psi _t$
    \ENDIF
  \ENDIF
\end{algorithmic}
\end{algorithm}

Algorithm \ref{alg:march} shows the loop for propagating the front with a narrow band defined by its width $\texttt{width}_{\text{band}}$. It is evident that the full field algorithm can be obtained by removing the bandwidth related termination condition in the above algorithm or simply setting the band width to $+\infty$. An empty heap is the other loop termination condition, i.e., in the given region, all $\texttt{BAND}$ points have been given $\texttt{KNOWN}$ status and there are no more $\texttt{FAR}$ points which can be added to the $\texttt{BAND}$ category. In each loop, this termination condition is checked first; if not satisfied then the $\texttt{BAND}$ point on the top of the heap is located and its value is checked against the band width termination condition; if still not satisfied then this point is tagged as $\texttt{KNOWN}$ and removed from the heap. For this newly added $\texttt{KNOWN}$ point, its downwind neighbors will be checked and updated if possible. 
Fig. \ref{fig:neighbor1} gives a 2D example of this procedure for a newly added $\texttt{KNOWN}$ point $(i,j)$.
As shown in the close-up view, $(i+1,j)$ is an existing $\texttt{BAND}$ point, but its function value was calculated using Eq. (\ref{eq:fds}) solely from $\texttt{KNOWN}$ point $(i+1,j+1)$. Now it has two $\texttt{KNOWN}$ neighbors, and Eq. (\ref{eq:fds}) is re-solved to possibly update its function value. For point $(i,j-1)$, it was a $\texttt{FAR}$ point, now with a $\texttt{KNOWN}$ neighbor, Eq. (\ref{eq:fds}) is solved to update its function value and its status is changed to $\texttt{BAND}$. 

Algorithm \ref{alg:sfmm} gives the overall solution procedure for the sequential fast marching method. It simply consists of two initialization steps and the narrow band marching procedure. For further details, the reader is referred to \cite{Sethian99b} and the references therein.

\begin{algorithm}
\algsetup{indent=1em}
\caption{Front propagation within the narrow band:\newline $\textsc{March\_Narrow\_Band}$.}
\label{alg:march}
\begin{algorithmic}[1]
    \LOOP 
      \IF {$\texttt{size}_{\mathfrak{H}} = 0$}
        \STATE \textbf{exit loop}
      \ENDIF
      \STATE $(i,j,k) \leftarrow  \textsc{Locate\_Min} $ 
      \IF{$\psi_{i,j,k} > \texttt{width}_{\text{band}}$}
        \STATE \textbf{exit loop}
      \ENDIF
      \STATE $G_{i,j,k}  \leftarrow  \texttt{KNOWN}$ 
      \STATE $\textsc{Remove\_Min}$ 
      \STATE $\textsc{Update\_Neighbors}(i,j,k)$
    \ENDLOOP
\end{algorithmic}
\end{algorithm}
\begin{algorithm}
\algsetup{indent=1em}
\caption{Sequential narrow band fast marching method:\newline $\textsc{Narrow\_Band\_Fast\_Marching}$.}
\label{alg:sfmm}
\begin{algorithmic}[1]
  \STATE $\textsc{Initialize\_Interface}$
  \STATE $\textsc{Initialize\_Heap}$
  \STATE $\textsc{March\_Narrow\_Band}$
\end{algorithmic}
\end{algorithm}

\section{Data structures for two-sided interface problems}

As discussed in the previous section, for interface problems with both positive and negative regions, the sequential fast marching method can be applied in a region-by-region manner; and the data structures required in the algorithm is exactly the same as those for single region problems. Here is a simple approach to apply the min-heap data structure introduced above to two-sided interface problems: first, the signs of the $\texttt{KNOWN}$ points from the interface initialization step are inverted, i.e., the $\texttt{KNOWN}$ points in the negative region become positive and vice versa for the positive region; then, the fast marching method with the min-heap data structure is applied to the negative region; after that, the signs of all points in the negative region and the $\texttt{KNOWN}$ points in the positive region (carry a negative sign from the operation in the first step) are inverted again to return to the correct signs; and finally, the fast marching method is applied to the positive region. It is evident in the above procedure for two-sided interface problems, a strict order of increasing (positive) function values for updating the solution is guaranteed. 

A simple parallelization can be implemented by following the same philosophy such that the parallel algorithm is applied to a single region at a time and the information propagation in this region has to be completed before moving to the region of an opposite sign. This approach essentially further decomposes the computational domain into positive and negative subdomains. A major problem of this approach is that all communications between neighboring subdomains and synchronizations among all subdomains have to be performed separately for both the positive and negative regions, which practically doubles the number of communication calls. On the other hand, it greatly deteriorates the load imbalance inherently rooted in solving the Eikonal equation on distributed memory parallel computers, as the information only propagates away from an interface.

In this study, novel data structures are designed to update both the positive and negative regions concurrently within our parallel algorithm. One particularly attractive property of the new data structure is that the fast marching algorithm given in the previous sections is barely changed. As shown in Algorithm \ref{alg:march_interface}, an outer \textbf{for} loop, in which the counter $s$ 
has a value of $-1$ for the negative region or $1$ for the positive region, is added to the $\textsc{March\_Narrow\_Band}$ Algorithm. Correspondingly, the scalar variables for the size of the binary heap and the width of the narrow band are changed into three-element arrays, i.e., $\texttt{size}_{\mathfrak{H}} (-1:1)$ and $\texttt{width}_{\text{band}}(-1:1)$, to be used inside the loop. However, the heap still keeps its one-dimensional array-backed structure through the incorporation of negative indices. That is, the heap for a two-sided interface problem will range from $\texttt{size}_{\mathfrak{H}} (-1)$ (a non-positive integer) to $\texttt{size}_{\mathfrak{H}} (1)$ (a non-negative integer). The side information $s$ is added to all the functions in the priority queue algorithm, i.e., $\texttt{Insert\_Heap}$, $\texttt{Locate\_Min}$, $\texttt{Remove\_Min}$, $\texttt{Up\_Heap}$, and $\texttt{Down\_Heap}$. For the increment operations to an index $p$ in the heap, instead of $p+1$ or $p-1$ in the original algorithm, now they are simply $p+s$ or $p-s$. For the comparison operations of indices and function values, $p$ will be replaced by $s \cdot p$ or $|p|$ and $\psi$ becomes $s \cdot \psi$ or $|\psi|$. With this treatment, the side information for a $\texttt{KNOWN}$ or $\texttt{BAND}$ point is always available from the sign of the function value at the given point. 

It is evident that the algorithm works for one-sided boundary value problems without any issues, as the heap size for the other side should be zero and the algorithm will not be executed for that side at all. On the other hand, the present approach is quite straightforward for two-sided interface problems. It gets rid of the positive-negative domain decomposition,
which has some significant impacts on the parallelization of the fast marching method. For example, the number of communications for data exchanges and reductions among processes is simply halved with the new data structures as two communication calls with one for each side of the interface can be combined into one now. In many applications, the load balance can be greatly improved as both sides are treated in one loop without involving any data communications within the loop. Because the processes spend more time in computations before data communications are required, this is very beneficial for reducing network congestion and improving parallel performance. 

\begin{algorithm}
\algsetup{indent=1em}
\caption{Front propagation within the narrow band for two-sided interface problems:\newline $\textsc{March\_Narrow\_Band\_Two\_Sided}$.}
\label{alg:march_interface}
\begin{algorithmic}[1]
  \FOR{$s \leftarrow -1$ \TO $1$ $\textbf{step}$ $2$ } 
    \LOOP 
      \IF {$\texttt{size}_{\mathfrak{H}} ( s ) = 0$}
        \STATE \textbf{exit loop}
      \ENDIF
      \STATE $(i,j,k) \leftarrow  \textsc{Locate\_Min} ( s ) $ 
      \IF{$|\psi_{i,j,k}| > \texttt{width}_{\text{band}} $}
        \STATE \textbf{exit loop}
      \ENDIF
      \STATE $G_{i,j,k}  \leftarrow  \texttt{KNOWN}$ 
      \STATE $\textsc{Remove\_Min}( s )$ 
      \STATE $\textsc{Update\_Neighbors}(i,j,k,s)$
    \ENDLOOP
  \ENDFOR
\end{algorithmic}
\end{algorithm}

\section{Parallel fast marching method}\label{sec:pfmm}

\subsection{Overlapping domain decompositions}

\begin{figure}[htbp!]
\begin{center}
 \includegraphics[angle=0,width=0.45\textwidth]{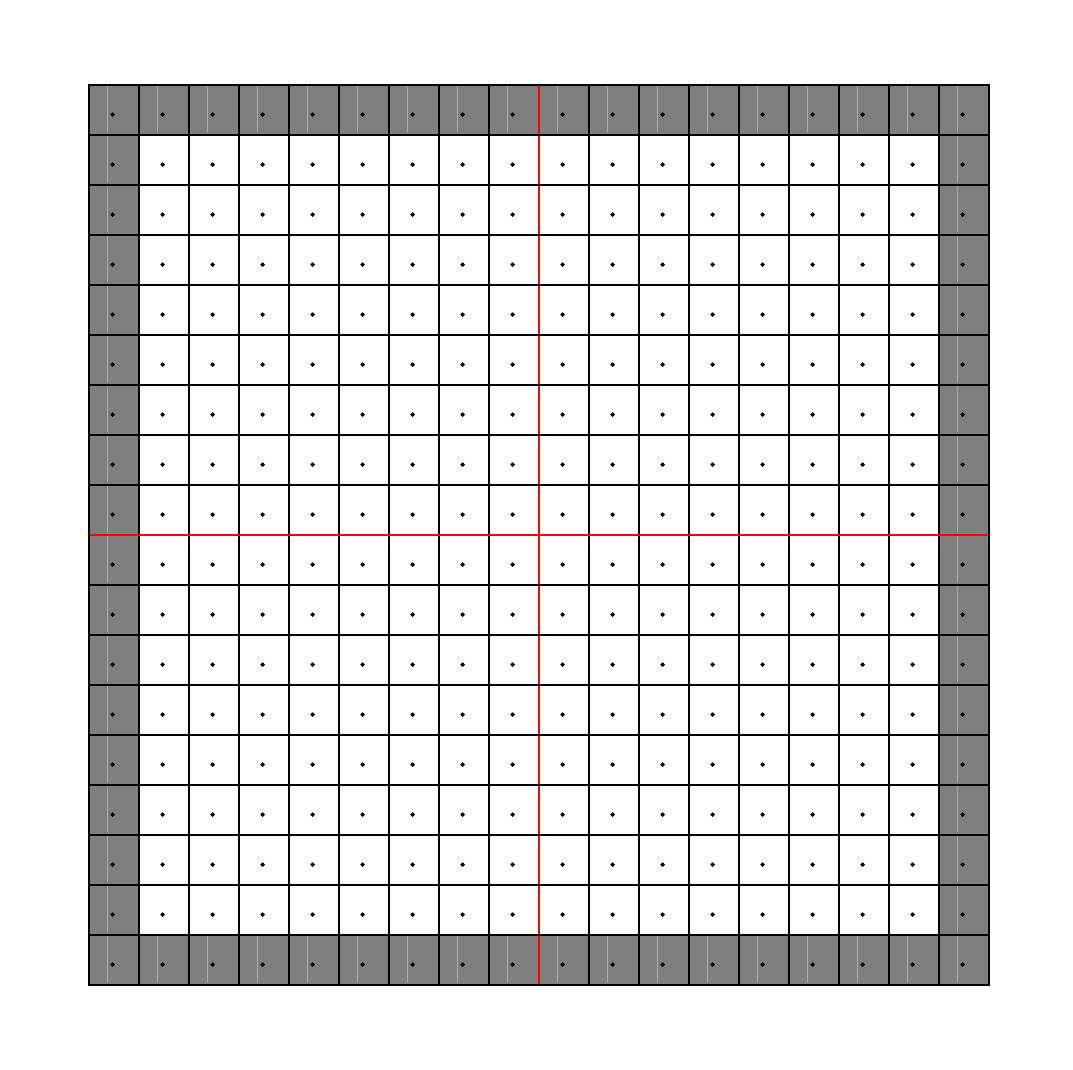}
 \includegraphics[angle=0,width=0.45\textwidth]{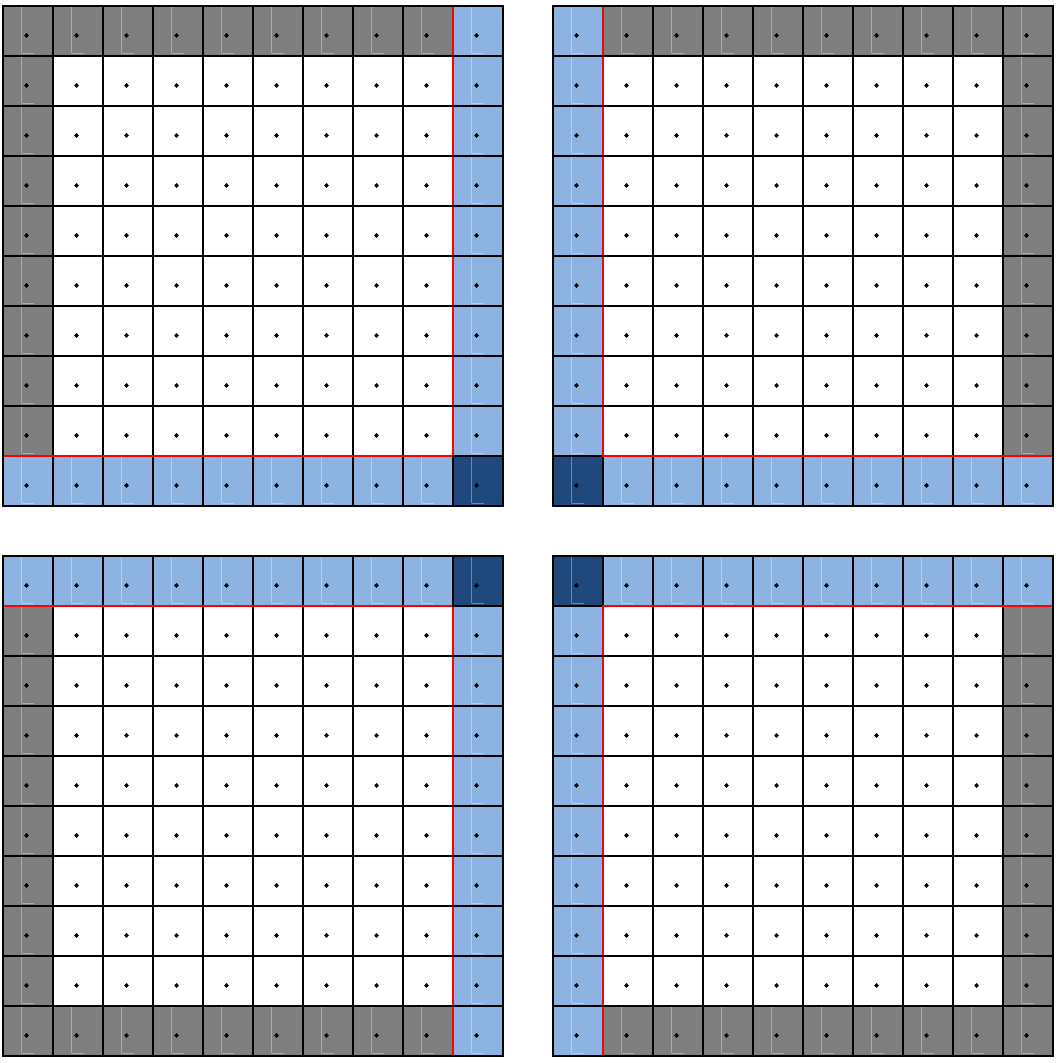}\\
\makebox[0.45\textwidth]{(a)}
\makebox[0.45\textwidth]{(b)} \\
\end{center}
 \caption{Domain decomposition in the parallel computation: the sequential (a) and the parallel (b) computational domains.}
\label{fig:domain5}
\end{figure}

The computational domain is divided into $p_i \times p_j \times p_k = np$ subdomains using a Cartesian process topology and mapped to $np$ processes. Each process $p = 0,\cdots, np-1$ works on a subdomain identified by its process coordinates $(ip,jp,kp)$ $(ip = 0, \cdots, p_i - 1; jp = 0, \cdots, p_j - 1$; and $kp = 0, \cdots, p_k - 1)$ in the Cartesian process grid. For simplicity, the domain is divided evenly in each direction. As shown in Fig. \ref{fig:domain5} for a 2D case, just like the case in a sequential computation, for each subdomain $\varOmega_p$ all subdomain boundaries, including those generated from the domain decomposition, are patched with one layer of ghost points to obtain a ghost point zone $\varTheta_p$ that encloses $\varOmega_p$. Similarly, with the parallel algorithm the basic scheme is executed in $\varXi _p$ without major changes. Each subdomain has its own heap structure and it is operated independently, regardless of the heaps residing in other subdomains. 

When a boundary value problem is solved using a discretization method, usually function values at ghost points are obtained from boundary conditions for physical domain boundaries or from neighboring processes for virtual boundaries generated in the domain decomposition procedure. For the Eikonal equation, however, it makes more sense to treat the ghost points in the same way as the internal points; because the interface, which contains the boundary values, is commonly embedded in the computational domain and the information propagates away from it does not depend on the domain boundary conditions at all. Even for the case when a domain boundary is the source of information (i.e., a Dirichlet boundary condition), it still can be considered as an interface embedded in the enlarged domain with the ghost points counted in. Therefore, in the present work, the Eikonal equation is solved everywhere without distinguishing the ghost points from the others. In both the sequential and parallel algorithms, it is only necessary to make sure that a discretization stencil does not involve an inaccessible point for the current process (i.e., to avoid array out-of-bounds errors) and boundary conditions are not implemented at all. It should be noted that an interface intersects with a domain boundary should be extended into the ghost points with appropriate values before the equation is being solved. 
\begin{figure}[htbp!]
\begin{center}
 \includegraphics[angle=0,width=0.5\textwidth]{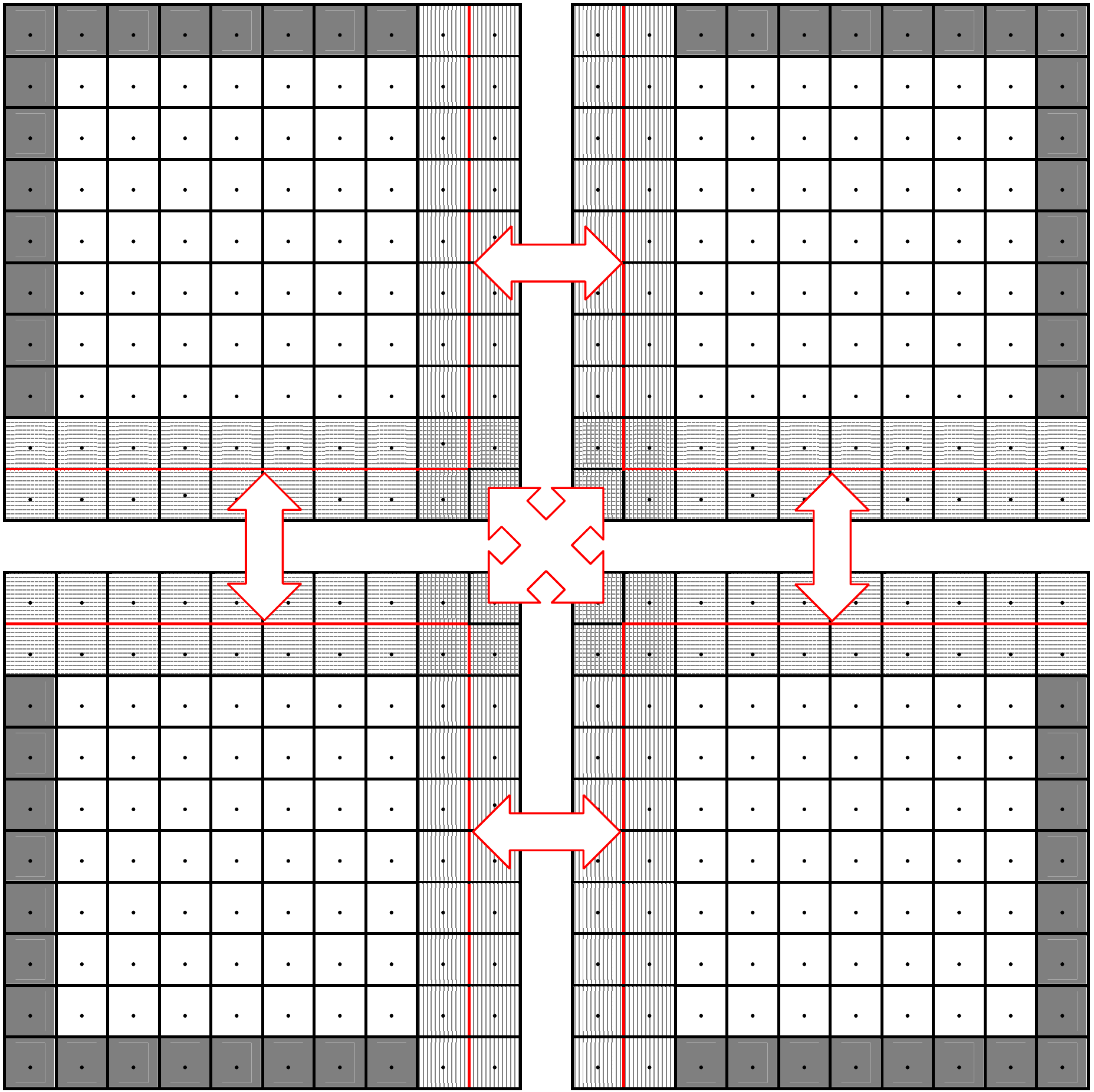}
\end{center}
 \caption{Data exchange between neighboring processes.}
\label{fig:domain6}
\end{figure}

It is obvious that the present strategy is an overlapping domain decomposition approach, since each process solves the equation everywhere including its ghost points. And the ghost points may contain better approximations to the solution than their corresponding physical domain points residing in a different process. Therefore, besides sending out values at physical domain points that are ghost points of neighboring processes, it is also necessary for a process to transfer its latest ghost point values to its neighboring processes that share these points with the specific process. As shown in Fig. \ref{fig:domain6}, a process has to exchange information with all neighboring processes that share subdomain faces, edges, and corners with it.  

At a first glance, it may seem there are unnecessary additional computations and extra communications by including the ghost points in the equation solution and data exchange procedures, respectively. Actually, in the fast marching method, only the points in the downwind direction will be updated, which means that very likely a grid point shared by two or more processes will be visited only in one of them and the updated value of this point will be sent from this process to others. Also, only the updated points, whose size is usually much less than that of the entire shared points, will be involved in the data exchange. Moreover, for the exchange of small size data such as ghost point updates, usually the number of message passing function calls determines the communication overhead instead of the actual data size of each exchange. On the other hand, the present overlapping domain decomposition strategy and the corresponding data exchange mechanism can greatly simplify the code structure by unifying the treatment of ghost points and physical domain points.

\subsection{Augmented status tags}

As mentioned above, only portion of the shared points are involved in the data exchanges in the parallel algorithm. This is realized by further distinguish the status tag of a grid point in the $\texttt{BAND}$ and $\texttt{KNOWN}$ categories. In the present parallel algorithm, the $\texttt{FAR}$ category remains unchanged from its definition in the sequential algorithm. The $\texttt{BAND}$ category is divided into two sub-categories: a) $\texttt{BAND\_NEW}$, which is the tag for a new $\texttt{BAND}$ point elevated from a $\texttt{FAR}$ status; and b) $\texttt{BAND\_OLD}$, which is the tag assigned to a $\texttt{BAND\_NEW}$ point in the shared regions after the position and value of this point are collected for data exchanges. The $\texttt{KNOWN}$ category is divided into three sub-categories: a) $\texttt{KNOWN\_FIX}$, which is the tag for those grid points that obtain their functions during the interface initialization procedure and their values are fixed during the solution process; b) $\texttt{KNOWN\_NEW}$, which is the tag assigned to the point at the top of the heap
with a $\texttt{BAND}$ tag
when it is to be removed from the heap; and c) $\texttt{KNOWN\_OLD}$, which is the tag assigned to a $\texttt{KNOWN\_NEW}$ point in the shared regions after the position and value of this point are collected for data exchange. 

\begin{algorithm}
\algsetup{indent=1em}
\caption{Interface initialization:\newline $\textsc{Initialize\_Interface\_Parallel}$.}
\label{alg:init_inter_parallel}
\begin{algorithmic}[1]
  \STATE $\psi \leftarrow +\infty$
  \STATE $G  \leftarrow  \texttt{FAR}$
  \FORALL{$(i,j,k)\in \varXi _p $ such that $\psi_{i,j,k} \in \psi^0$ is adjacent to $\Gamma$} 
    \STATE $\psi_{i,j,k} \leftarrow  \psi^0$
    \STATE $G_{i,j,k}  \leftarrow  \texttt{KNOWN\_FIX}$ 
  \ENDFOR
\end{algorithmic}
\end{algorithm}
\begin{algorithm}
\algsetup{indent=1em}
\caption{Heap initialization:\newline $\textsc{Initialize\_Heap\_Parallel}$.}
\label{alg:init_heap_parallel}
\begin{algorithmic}[1]
  \STATE $\texttt{size}_{\mathfrak{H}}(-1:1) \leftarrow 0$ 
    \FORALL{$(i,j,k)\in \varXi _p $ such that $G_{i,j,k} = \texttt{KNOWN\_FIX}$}
      \STATE $s \leftarrow \sgn(\psi_{i,j,k})$
      \STATE $\textsc{Update\_Neighbors\_Parallel}(i,j,k,s)$
    \ENDFOR
\end{algorithmic}
\end{algorithm}
\begin{algorithm}
\algsetup{indent=1em}
\caption{Update the values of neighbors of a point newly added to set $\texttt{KNOWN}$:\newline $\textsc{Update\_Neighbors\_Parallel}(i,j,k,s)$.}
\label{alg:update_parallel}
\begin{algorithmic}[1]
\FORALL{$(l,m,n)$ such that $\left( |l-i|+|m-j|+|n-k|\right) = 1$}
  \IF{$(l,m,n) \in \varXi _p$}
    \IF{$G_{l,m,n} \neq \texttt{KNOWN\_FIX}$ and $|\psi_{l,m,n}| > |\psi_{i,j,k}| $}
      \STATE $\psi_{\text{temp}} \leftarrow \textsc{Solve\_Quadratic}(l,m,n)$
        \IF{$\psi_{\text{temp}} < |\psi_{l,m,n}|$}
          \STATE $\psi_{l,m,n} \leftarrow s \cdot \psi_{\text{temp}}$
          \STATE $G_{l,m,n} \leftarrow \texttt{BAND\_NEW}$
          \IF{$(l,m,n) \not\in \mathfrak{H}(s)$} 
            \STATE $\textsc{Insert\_Heap}(l,m,n,s)$ 
          \ELSE   
            \STATE $\textsc{Up\_Heap}(l,m,n,s)$ 
        \ENDIF
      \ENDIF
    \ENDIF
  \ENDIF
\ENDFOR
\end{algorithmic}
\end{algorithm}

The introduction of these new status tags enables minimizing the point refreshing computations and the data exchanged between processes, but barely changes the main elements of the sequential algorithm. Algorithm \ref{alg:init_inter_parallel} shows the parallel version of the interface initialization procedure. Compared with the sequential version, the only differences are the replacements of $\varXi$ and $\texttt{KNOWN}$ with $\varXi _p$ and $\texttt{KNOWN\_FIX}$, respectively. Likewise, the parallel version of the heap initialization procedure, i.e., Algorithm \ref{alg:init_heap_parallel}, follows the same modifications. As discussed in the previous section, the sign of a point value is obtained to determine which side of the heap a $\texttt{BAND\_NEW}$ point should be inserted to in the $\textsc{Update\_Neighbors\_Parallel}$ procedure given in Algorithm \ref{alg:update_parallel}. In this part the condition $G_{l,m,n} \neq \texttt{KNOWN\_FIX}$ is of significance for the present parallel algorithm. It allows the function value at point $(l,m,n)$ with a tag $\texttt{KNOWN\_OLD}$ or $\texttt{KNOWN\_NEW}$ to be updated, just like a $\texttt{BAND}$ or $\texttt{FAR}$ point, as long as $|\psi_{l,m,n}| > |\psi_{i,j,k}| $. Also if its function value does be updated, then its tag will be reset to $\texttt{BAND\_NEW}$ no matter what tag it has previously. 

The strong resemblance is also seen in the sequential and parallel versions of the solution procedure of the quadratic equation. As shown in Algorithm \ref{alg:quadratic_parallel}, the $\texttt{KNOWN}$ tag in the parallel version is used to include a neighboring point with a $\texttt{KNOWN\_FIX}$, $\texttt{KNOWN\_OLD}$, or $\texttt{KNOWN\_NEW}$ tag in the stencil. However, it is still necessary to make sure that the specific neighboring point is an upwind point by a comparison of its value with that of point $(l,m,n)$. This is because that, as discussed above, point $(l,m,n)$ could be a $\texttt{KNOWN\_OLD}$ or $\texttt{KNOWN\_NEW}$ point and might carry a function value lower than that of its neighbor with a $\texttt{KNOWN\_OLD}$ or $\texttt{KNOWN\_NEW}$ tag. In addition, different solutions of the quadratic equation will be calculated as follows: a) if available, a solution using all three source points from the three coordinate directions (3D); b) if available, a solution using two source points (with the smaller function values for a 3D case) (2D); and c) a solution using the source point with the minimum function value (one-dimensional, 1D). The minimum of these solutions, which must also satisfy the causality principle, is chosen as the final solution. Unlike the sequential version, in which the points elevated to the $\texttt{KNOWN}$ set follows a strict order of increasing function values, here it is necessary to compare solutions from different configurations of source points because the $\texttt{KNOWN\_OLD}$ or $\texttt{KNOWN\_NEW}$ tag does not mean a source point contains the final or updated characteristic in the present parallel algorithm.

The parallel procedure for the front propagation within a narrow band is given in Algorithm \ref{alg:march_new}. Here, a $\texttt{BAND}$ point will be elevated to a $\texttt{KNOWN\_NEW}$ status because of the augmented tag sets, whereas the a $\texttt{KNOWN\_OLD}$ point will retain its status because, as to be explained in the next part, such a point was updated by a data exchange with a neighboring subdomain instead of from solving the quadratic equation within this subdomain. Also $\texttt{width}_{\text{band}}$ is replaced by $\texttt{bound}_{\text{band}}$, which is not a constant specified beforehand any more and will be discussed later. Other than these small differences, this procedure is almost the same as the sequential version given in Algorithm \ref{alg:march_interface}.

\begin{algorithm}
\algsetup{indent=1em}
\caption{Solve the quadratic equation:\newline $\textsc{Solve\_Quadratic\_Parallel}(l,m,n)$.}
\label{alg:quadratic_parallel}
\begin{algorithmic}[1]
\STATE $\psi_{\text{temp}} \leftarrow + \infty$
\STATE {Check the $x$ direction to set $\psi_1$ and $h_1$:}
  \STATE $d \leftarrow 0$
  \IF {$(l-1,m,n) \in \varXi _p$}
    \IF {$G_{l-1,m,n} \in \texttt{KNOWN}$ \AND $|\psi_{l-1,m,n}| < |\psi_{l,m,n}|$}
      \STATE $d \leftarrow -1$
    \ENDIF
  \ENDIF
  \IF {$(l+1,m,n) \in \varXi _p$}
    \IF {$G_{l+1,m,n} \in \texttt{KNOWN}$ \AND $|\psi_{l+1,m,n}| < |\psi_{l,m,n}|$}
      \IF {$d = 0$}
        \STATE $d \leftarrow 1$
      \ELSIF {$|\psi_{l+1,m,n}| < |\psi_{l-1,m,n}|$} 
        \STATE $d \leftarrow 1$
      \ENDIF
    \ENDIF
  \ENDIF
  \IF{$d \neq 0$}
    \STATE $\psi_1 \leftarrow |\psi_{l+d,m,n}|$
    \STATE $h_1    \leftarrow \Delta x ^{-1} $
  \ELSE
    \STATE $\psi_1 \leftarrow 0$
    \STATE $h_1    \leftarrow 0$
  \ENDIF
\STATE {Check the $y$ direction to set $\psi_2$ and $h_2$}
\STATE {Check the $z$ direction to set $\psi_3$ and $h_3$}
\STATE $nd \leftarrow$ number of nonzero $h _i, \; (i=1,2,3)$
\WHILE {$nd \neq 0$}
  \STATE $a \leftarrow     \sum_i (h_i ^2)$
  \STATE $b \leftarrow - 2 \sum_i (h_i ^2 \psi_i)$
  \STATE $c \leftarrow     \sum_i (h_i ^2 \psi_i ^2 )- F_{l,m,n} ^{-2}$
  \IF {$(b^2-4ac) \geq 0$}
    \STATE $\psi_t \leftarrow \frac{-b+\sqrt{b^2-4ac}}{2a}$
    \IF {$\psi_1 < \psi_t$ \AND $\psi_2 < \psi_t$ \AND $\psi_3 < \psi_t$}
      \STATE $\psi_{\text{temp}} \leftarrow \min (\psi_{\text{temp}}, \psi _t)$
    \ENDIF
  \ENDIF
  \STATE $j \leftarrow$ index of maximum $\psi_i, \; (i=1,2,3)$
  \STATE $\psi_j \leftarrow 0$
  \STATE $h_j \leftarrow 0$
  \STATE $nd \leftarrow (nd - 1) $
\ENDWHILE
\end{algorithmic}
\end{algorithm}
\begin{algorithm}
\algsetup{indent=1em}
\caption{Front propagation within the narrow band:\newline $\textsc{March\_Narrow\_Band\_Parallel}$.}
\label{alg:march_new}
\begin{algorithmic}[1]
  \FOR{$s \leftarrow -1$ \TO $1$ $\textbf{step}$ $2$ } 
    \LOOP 
      \IF {$\texttt{size}_{\mathfrak{H}} ( s ) = 0$}
        \STATE \textbf{exit loop}
      \ENDIF
      \STATE $(i,j,k) \leftarrow  \textsc{Locate\_Min} ( s ) $ 
      \IF{$|\psi_{i,j,k}| > \texttt{bound}_{\text{band}} ( s ) $}
        \STATE \textbf{exit loop}
      \ENDIF
      \IF{$G_{i,j,k} \neq \texttt{KNOWN\_OLD}$}
        \STATE  $G_{i,j,k} \leftarrow \texttt{KNOWN\_NEW}$
      \ENDIF
      \STATE $\textsc{Remove\_Min} (s) $  
      \STATE $\textsc{Update\_Neighbors\_Parallel}(i,j,k,s)$
    \ENDLOOP
  \ENDFOR
\end{algorithmic}
\end{algorithm}

Up to this point, it should be evident that the major components of the sequential fast marching method are barely modified in the present parallel method. Actually, in a single-process setting, the augmented tag sets work exactly in the same way as the original $\texttt{BAND}$ and $\texttt{KNOWN}$ tags; and the additional upwind direction checks for two neighboring points are not something unexpected (they are implied by the one-way conversion of $\texttt{BAND}$ to $\texttt{KNOWN}$ status) in the sequential algorithm. This is essential for keeping all the desirable properties of the fast marching method as well as achieving a straightforward parallel implementation based on a sequential algorithm. 

\subsection{Synchronized data exchanges}

\begin{algorithm}
\algsetup{indent=1em}
\caption{Collect data in the overlapping region:\newline $\textsc{Collect\_Overlapping\_Data}$.}
\label{alg:collect}
\begin{algorithmic}[1]
  \STATE $\texttt{count}_{\text{new}} \leftarrow 0$   
  \FORALL{$(i,j,k) \in \sum _{q \in \mathfrak{N} _p} \varXi _p \cap \varXi _q$} 
    \IF{$G_{i,j,k} = \texttt{BAND\_NEW}$ or $G_{i,j,k} = \texttt{KNOWN\_NEW}$}
      \STATE $\texttt{count}_{\text{new}} \leftarrow \texttt{count}_{\text{new}} + 1$   
      \IF{$G_{i,j,k} = \texttt{BAND\_NEW}$}
        \STATE $G_{i,j,k}  \leftarrow \texttt{BAND\_OLD}$
      \ELSE 
        \STATE $G_{i,j,k}  \leftarrow \texttt{KNOWN\_OLD}$
      \ENDIF
      \FORALL{process $q \in \mathfrak{N} _p$}
        \IF{$(i,j,k)  \in \varXi _q$}
          \STATE Add $(i,j,k)$ to outgoing buffer: $\mathfrak{S} _p ^q (i,j,k) \leftarrow \psi_{i,j,k}$
        \ENDIF
      \ENDFOR
    \ENDIF
  \ENDFOR
\end{algorithmic}
\end{algorithm}
\begin{algorithm}
\algsetup{indent=1em}
\caption{Exchange data in the overlapping region:\newline $\textsc{Exchange\_Overlapping\_Data}$.}
\label{alg:exchange}
\begin{algorithmic}[1]
  \FORALL{process $q \in \mathfrak{N} _p$} 
    \STATE Send outgoing buffer $\mathfrak{S}_p ^q$ to $q$
  \ENDFOR
  \FORALL{process $q \in \mathfrak{N} _p$} 
    \STATE Receive incoming buffer $\mathfrak{R}_p ^q$ from $q$
  \ENDFOR
\end{algorithmic}
\end{algorithm}

The data exchanges at the boundaries of subdomains play a central role in a domain decomposition parallelization. Usually the ghost points for one subdomain are filled with solutions computed at the corresponding physical domain points from a neighboring subdomain. But in a fast marching algorithm, it is very likely that function values are only updated at a portion of physical domain points that coincide the ghost points of a neighboring subdomain. Apparently just this portion of points with updated values is to be conveyed to the neighboring subdomain. For a process $p$, it may have at most $26$ neighbors in a 3D case. The set of processes that are neighbors of $p$ is labeled as $\mathfrak{N} _p$. For the overlapping domain decomposition approach adopted in the present algorithm, the shared region between process $p$ and its neighbor $q$ is $\varXi _p \cap \varXi _q$. As shown in Algorithm \ref{alg:collect}, the status of every point in the shared regions, i.e., $\sum _{q \in \mathfrak{N} _p} \varXi _p \cap \varXi _q$, is checked to single out points with updated function values. It should be noted that this step includes both $\texttt{KNOWN\_NEW}$ and $\texttt{BAND\_NEW}$ points. The inclusion of the latter serves the purpose of propagating the latest information away from the upwind direction in a timely manner. The importance of this cannot be over-emphasized for the parallelization of the fast marching method as a sequential algorithm in nature. The size of the communicated data may be slightly increased because of it; but the associated penalty in message passing communication overhead should be negligible as explained earlier. After such a point is identified, its tag should be changed from a $\texttt{NEW}$ suffix to a $\texttt{OLD}$ one. This can avoid the inclusion of the same point in the next round of communication, unless its value is renewed again. 
(Also the reasoning for a $\texttt{KNOWN\_OLD}$ point retaining its status in Algorithm \ref{alg:march_new} should be apparent at this point.) 
An updated point may be shared by more than one neighboring processes. Therefore, it is necessary to check against all neighbors and add it to the corresponding outgoing data buffers. 
Here a counter $\texttt{count}_{\text{renew}}$ is used to determine the number of outgoing points that were elevated to $\texttt{NEW}$ status during the updating computations. If this counter is not zero, obviously the fast marching algorithm has to be carried on to check if these $\texttt{NEW}$ points can be used to refresh solutions in the neighboring subdomains. Therefore, it is necessary to achieve a zero count here for considering the termination of the whole parallel algorithm later.

\begin{algorithm}
\algsetup{indent=1em}
\caption{Integrate data received from neighboring processes:\newline $\textsc{Integrate\_Overlapping\_Data}$.}
\label{alg:integrate}
\begin{algorithmic}[1]
\FORALL{process $q \in \mathfrak{N} _p$}
    \FORALL{$(l,m,n) \in \mathfrak{R} _p ^q$}
      \STATE $\psi_{\text{new}} \leftarrow \mathfrak{R}_p ^q (l,m,n)$
      \IF{$|\psi_{\text{new}}| < |\psi_{l,m,n}|$}
        \STATE $s \leftarrow \sgn(\psi_{\text{new}})$
        \STATE $\psi_{l,m,n} \leftarrow \psi_{\text{new}}$
        \IF{$|\psi_{l,m,n}| > \texttt{bound}_{\text{band}} (s) $}
          \STATE $G_{l,m,n}  \leftarrow \texttt{BAND\_OLD}$
        \ELSE 
          \STATE $G_{l,m,n}  \leftarrow \texttt{KNOWN\_OLD}$
        \ENDIF
        \IF{$(l,m,n) \not\in \mathfrak{H}(s)$} 
            \STATE $\textsc{Insert\_Heap}(l,m,n,s)$ 
        \ELSE 
            \STATE $\textsc{Up\_Heap}(l,m,n,s)$ 
        \ENDIF
      \ENDIF
    \ENDFOR
\ENDFOR
\end{algorithmic}
\end{algorithm}

With a synchronized data exchange procedure as shown in Algorithm \ref{alg:exchange} following the data collection procedure, the present parallel algorithm retains the main elements of the sequential fast marching method and, additionally, has the benefit of a simple and straightforward implementation. The inclusion of $\texttt{BAND\_NEW}$ points in the data exchange also further makes the synchronized communications consistent with the present parallelization.

As shown in Algorithm \ref{alg:integrate}, the updated information from the shared regions is incorporated into the current solution field after the incoming data buffers from all neighboring processes are received. This procedure resembles Algorithm \ref{alg:update_parallel} for updating the neighboring points of a newly added $\texttt{KNOWN}$ point in several aspects except that the updated value is obtained from a neighboring process instead of from solving the quadratic equation. If the incoming value has a smaller magnitude than that of the local one, the function value at the local point will be replaced by the incoming one. Its tag should be updated consistently according to the tag of incoming point in its residing process as determined in Algorithms \ref{alg:update_parallel}, \ref{alg:march_new}, and \ref{alg:collect}. It is worth noting here that an updated local point with a newly assigned $\texttt{KNOWN\_OLD}$ tag will be inserted into the heap (or moved up in the heap if its previous status was $\texttt{BAND}$) for further treatment. In the spirit of the sequential fast marching method, Algorithm \ref{alg:update_parallel} could be directly used instead of the current heap operations. However, updating neighbors here might result in redundant computations as the neighboring points to be computed could also be updated by the incoming data. In addition, it could not utilize the full set of updated upwind points that is only available after the data integration step is completed. In the next part it will become clear that the treatment in the present algorithm only slightly delays the neighbor updating operations. And the seemingly unnecessary heap operations on these $\texttt{KNOWN\_OLD}$ points can actually reduce the amount of neighbor updating operations by keeping a strict order of these points in heap, since all $\texttt{KNOWN\_OLD}$ and $\texttt{KNOWN\_NEW}$ as well as $\texttt{BAND}$ points are to be considered in Algorithm \ref{alg:update_parallel}. 

\subsection{Restarted narrow band approach} 

Algorithm \ref{alg:pfmm} shows the main procedure for the present parallel fast marching method with a novel restarted narrow band approach proposed in this work. Compared with the sequential version given in Algorithm \ref{alg:sfmm}, the interface and heap initialization procedures remain the same, but the narrow band marching procedure (the first instance corresponds to that in the sequential version) is placed in a loop for a restarted scheme. Within this loop, the termination criterion is determined first. Two global minimum values (one for each side of the interface) of the $\texttt{BAND}$ points at the tops of the local heaps from all processes are obtained from a synchronized global reduction. Just like the case in the sequential version, these two values have to reach or surpass $\texttt{width}_{\text{band}}$ before the fast marching procedure can be stopped. In addition, the global maximum count of the $\texttt{NEW}$ points during the data collection procedure is also obtained from the same global reduction. A zero value of $\texttt{count}_{\text{global}}$, together with the former condition, means all $\texttt{KNOWN}$ points in one precess won't be updated by its neighboring processes through data exchanges and their function values can be considered as final.

\begin{algorithm}
\algsetup{indent=1em}
\caption{Parallel narrow band fast marching method:\newline $\textsc{Parallel\_Narrow\_Band\_Fast\_Marching}$.}
\label{alg:pfmm}
\begin{algorithmic}[1]
  \STATE $\textsc{Initialize\_Interface\_Parallel}$
  \STATE $\textsc{Initialize\_Heap\_Parallel}$
    \LOOP 
      \FOR{$s \leftarrow -1$ \TO $1$ $\textbf{step}$ $2$ }
        \IF {$\texttt{size}_{\mathfrak{H}} ( s ) \neq 0$}
          \STATE $(i,j,k) \leftarrow  \textsc{Locate\_Min} ( s ) $ 
          \STATE $\texttt{minval}_{\text{local}} ( s ) \leftarrow |\psi _{i,j,k}|$ 
        \ELSE 
          \STATE $\texttt{minval}_{\text{local}} ( s ) \leftarrow \texttt{width}_{\text{band}}$
         \ENDIF
      \ENDFOR
      \STATE  $\texttt{minval}_{\text{global}} \leftarrow \textsc{AllReduce}_{\textsc{min}} \left( \texttt{minval}_{\text{local}} \right)  $ 
      \STATE  $\texttt{count}_{\text{global}} \leftarrow \textsc{AllReduce}_{\textsc{max}} \left( \texttt{count}_{\text{new}} \right)  $ 
      \IF {$\texttt{minval}_{\text{global}} \left( \pm 1 \right) \geq \texttt{width}_{\text{band}} $ and $\texttt{count}_{\text{global}} = 0$}
        \STATE \textbf{exit loop}
      \ENDIF
        \STATE $\texttt{bound}_{\text{band}} \left( \pm 1 \right) \leftarrow  \min \left( \texttt{minval}_{\text{global}} \left(\pm 1 \right) + \texttt{stride}, \texttt{width}_{\text{band}} \right)$
      \STATE $\textsc{March\_Narrow\_Band\_Parallel}$
      \STATE $\textsc{Collect\_Overlapping\_Data}$
      \STATE $\textsc{Exchange\_Overlapping\_Data}$ 
      \STATE $\textsc{Integrate\_Overlapping\_Data}$
      \STATE $\textsc{March\_Narrow\_Band\_Parallel}$
    \ENDLOOP
\end{algorithmic}
\end{algorithm}

In this restarted narrow band approach, the restart frequency is determined by the parameter $\texttt{stride}$: one run-through can advance the front by the size of $\texttt{stride}$, or $\delta s$. Actually $\texttt{stride}$ is the only free parameter required in the present parallel algorithm. If it takes a zero value, the current parallel algorithm will be running in almost the same sequence as the sequential algorithm. The only exception will be the parallelism that could exist in those processes whose heap top $\texttt{BAND}$ points share the same global minimum function values. On the other hand, with $\delta s = \infty$ each process will be running the sequential fast marching algorithm until the heaps are empty in each restart (assume $\texttt{width}_{\text{band}} = \infty$ for a whole field computation). And the second instance of the parallel narrow band marching algorithm inside the loop can be omitted. 
However, in a restarted narrow band framework with a proper $\texttt{stride}$, the second marching step brings the local fronts residing in different processes to the same $\texttt{bound}_{\text{band}}$. This is essential for a restarted scheme with a synchronized global upper bound for each run-through. Without this step, the neighbor updating procedure will be required for each point that is updated by a neighboring process or an updated neighboring point and consequently receives a $\texttt{KNOWN}$ tag in Algorithm \ref{alg:integrate}. As discussed in the previous part, the resulting algorithm will be less efficient and far more complicated. It is also worth noting that the augmented tags proposed in this work facilitates the restarted narrow band approach with minimized data communications and makes the overall algorithm quite compact. 

\section{Results}\label{sec:benchmark}

\subsection{Test cases}

\begin{figure}[htbp!]
\begin{center}
 \includegraphics[angle=0,width=0.325\textwidth]{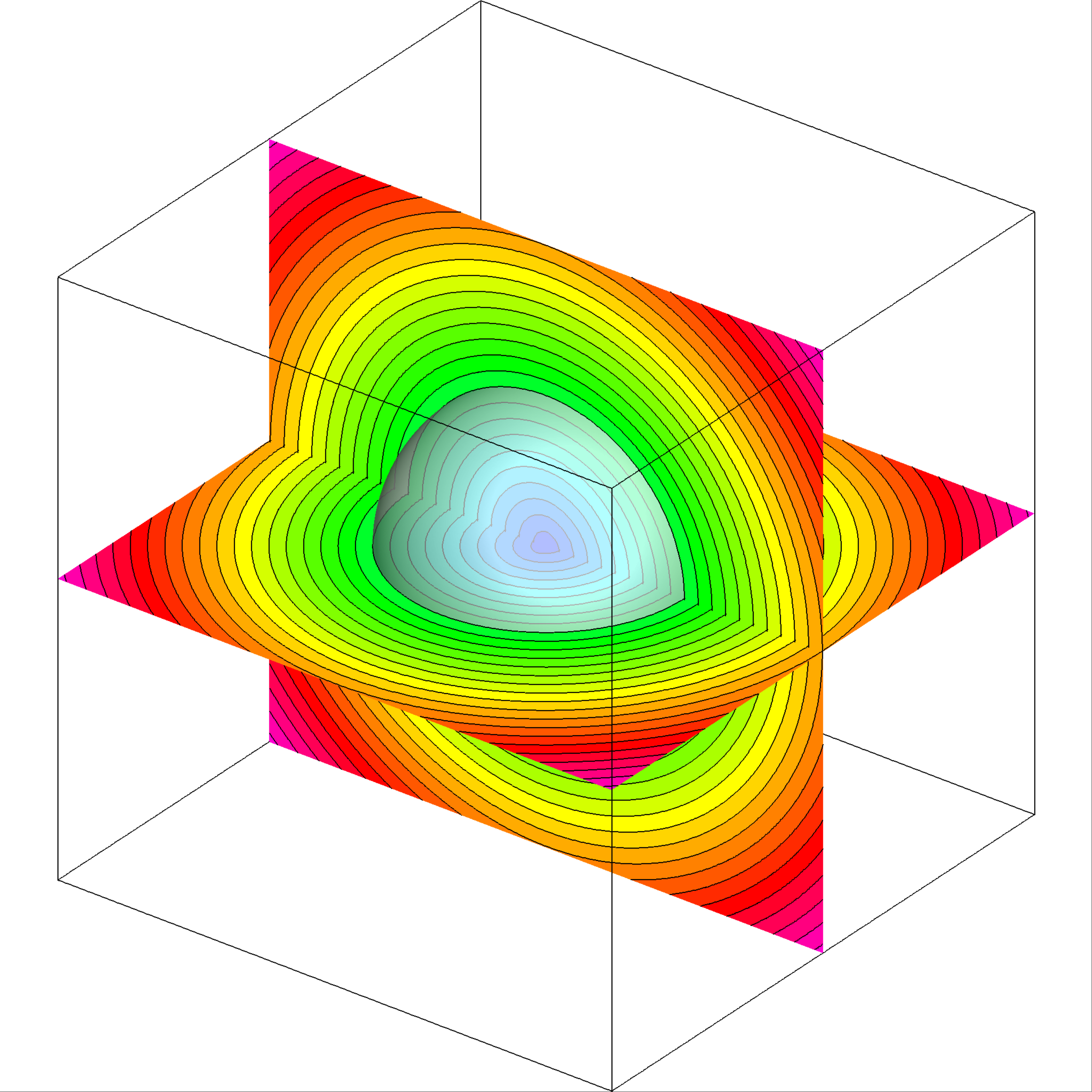}
 \includegraphics[angle=0,width=0.325\textwidth]{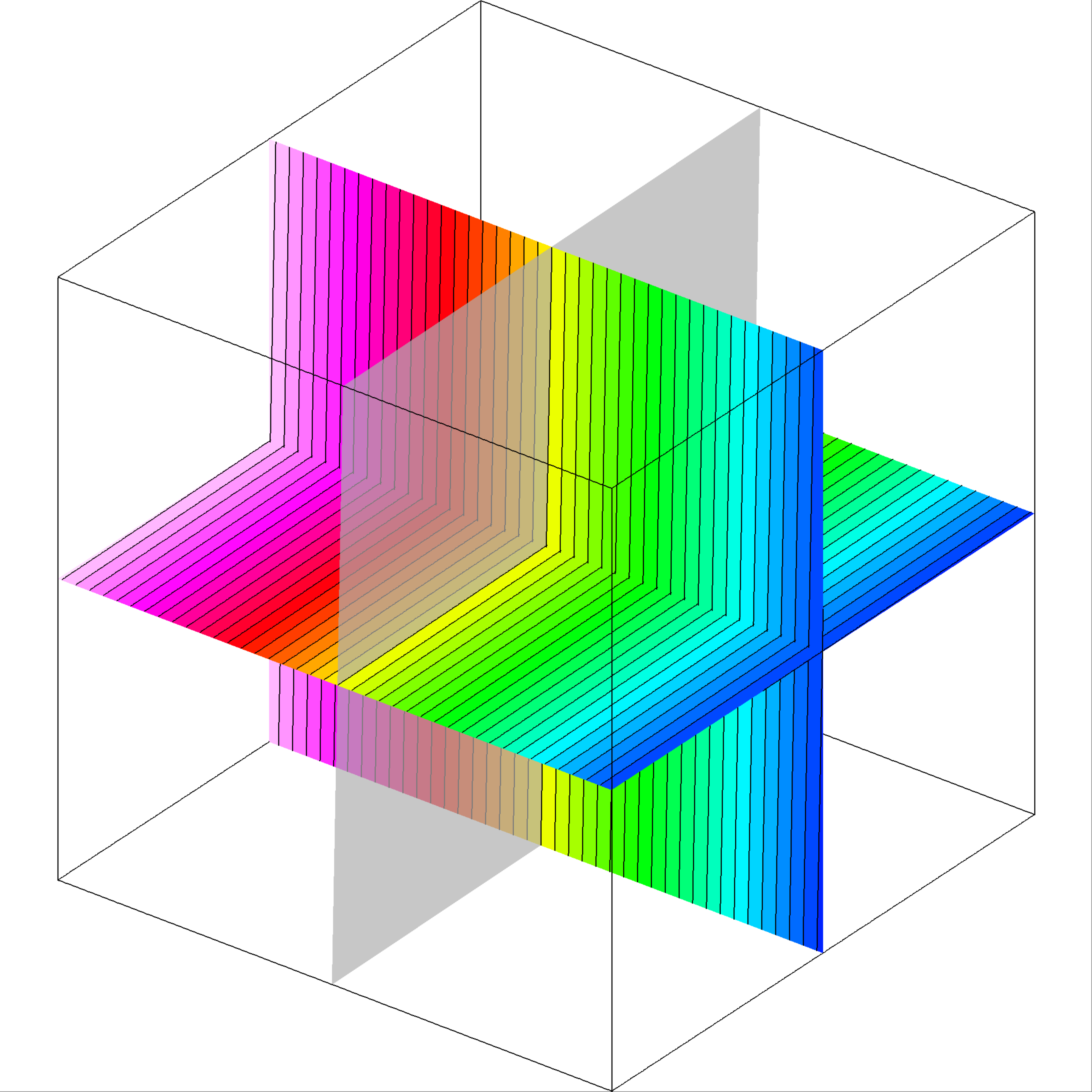}
 \includegraphics[angle=0,width=0.325\textwidth]{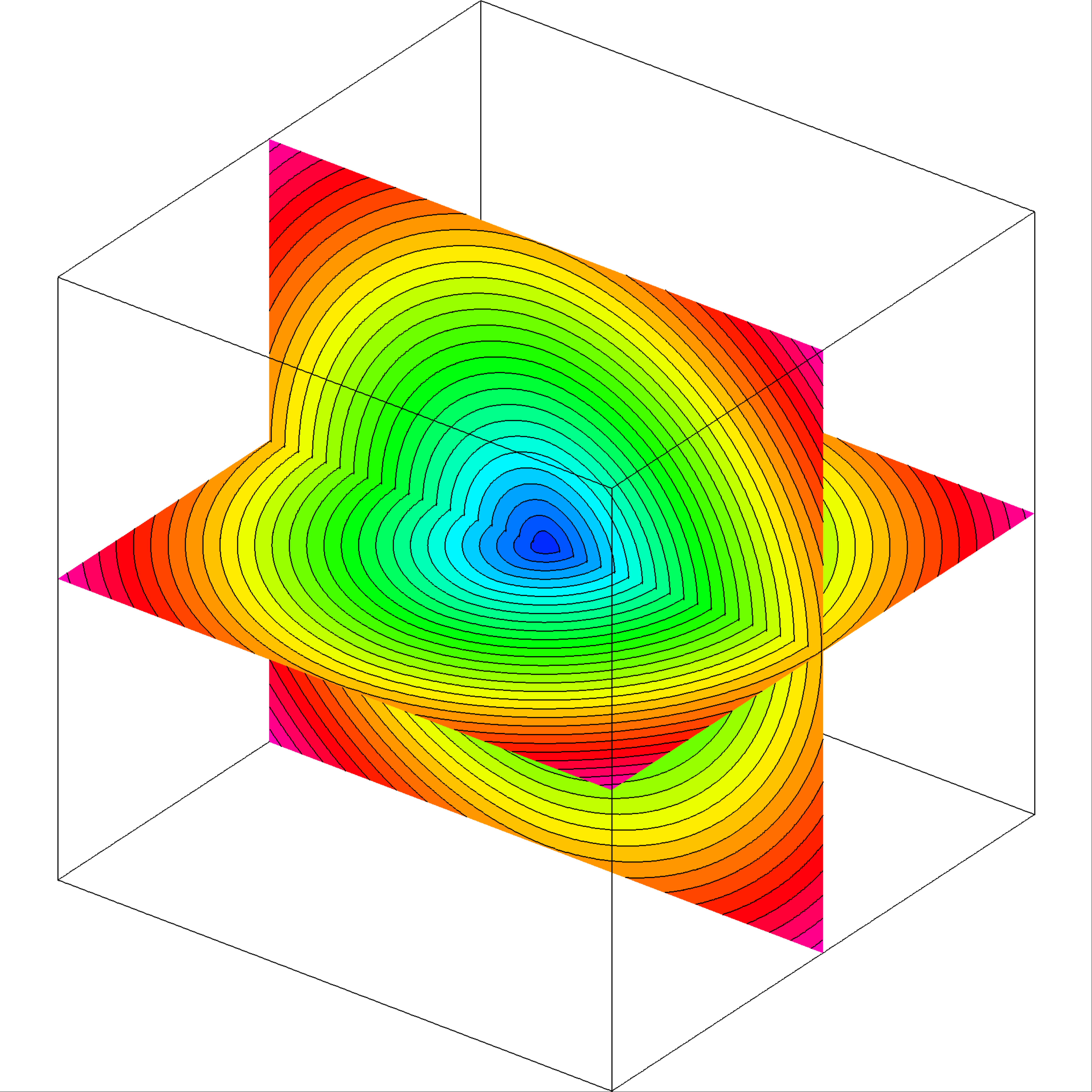}\\
  \makebox[0.325\textwidth]{(a) Case 1}
  \makebox[0.325\textwidth]{(b) Case 2}
  \makebox[0.325\textwidth]{(c) Case 3}
 \includegraphics[angle=0,width=0.325\textwidth]{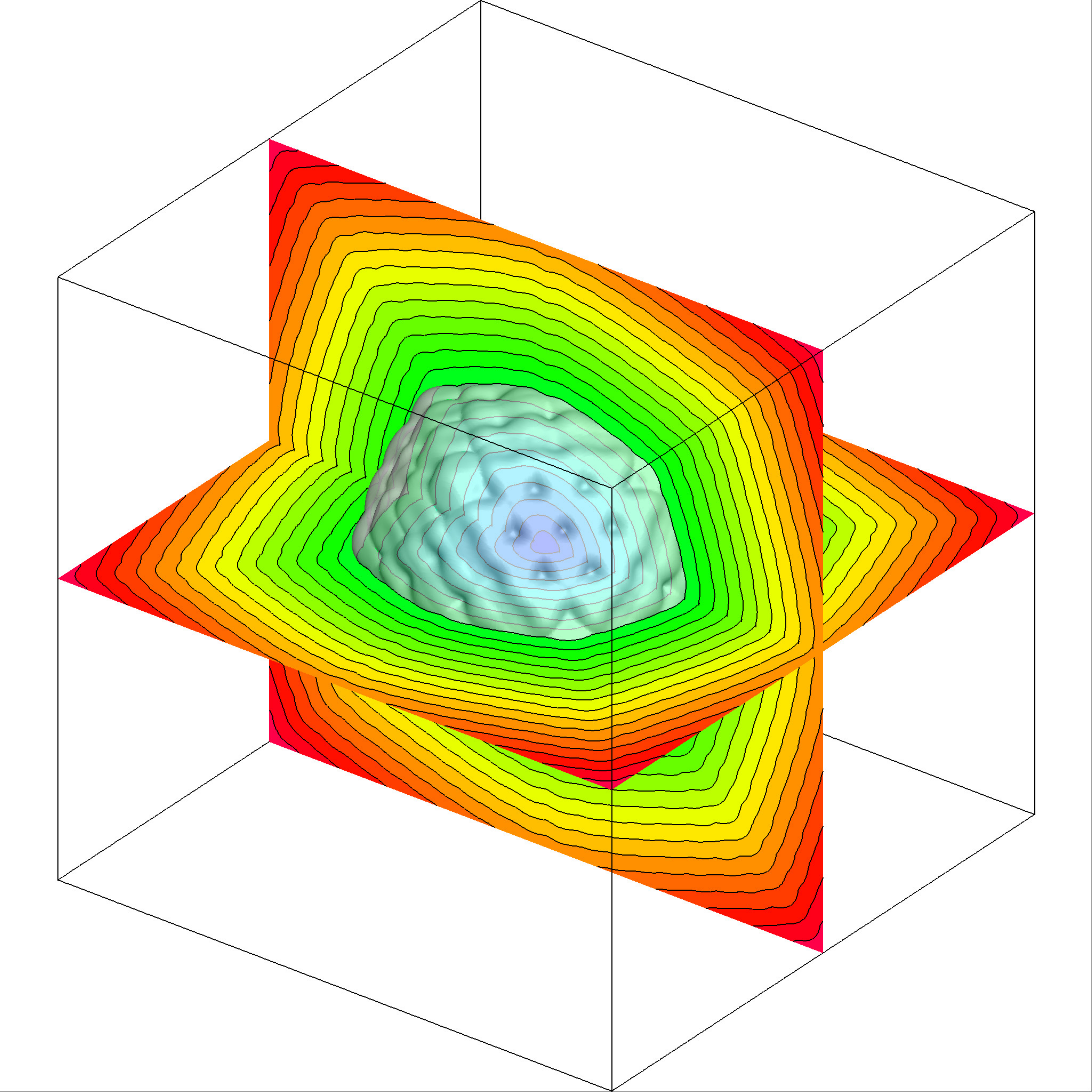}
 \includegraphics[angle=0,width=0.325\textwidth]{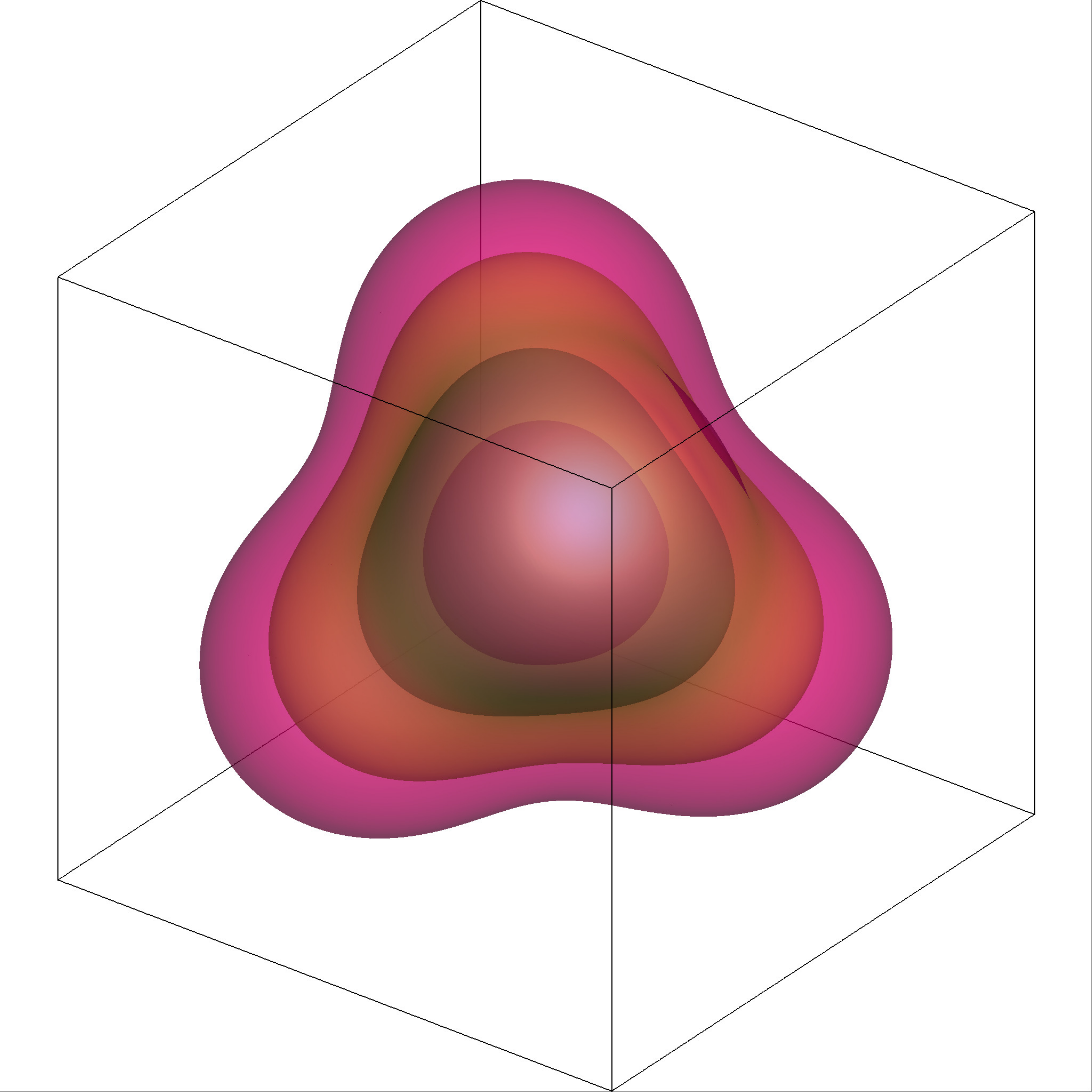}
 \includegraphics[angle=0,width=0.325\textwidth]{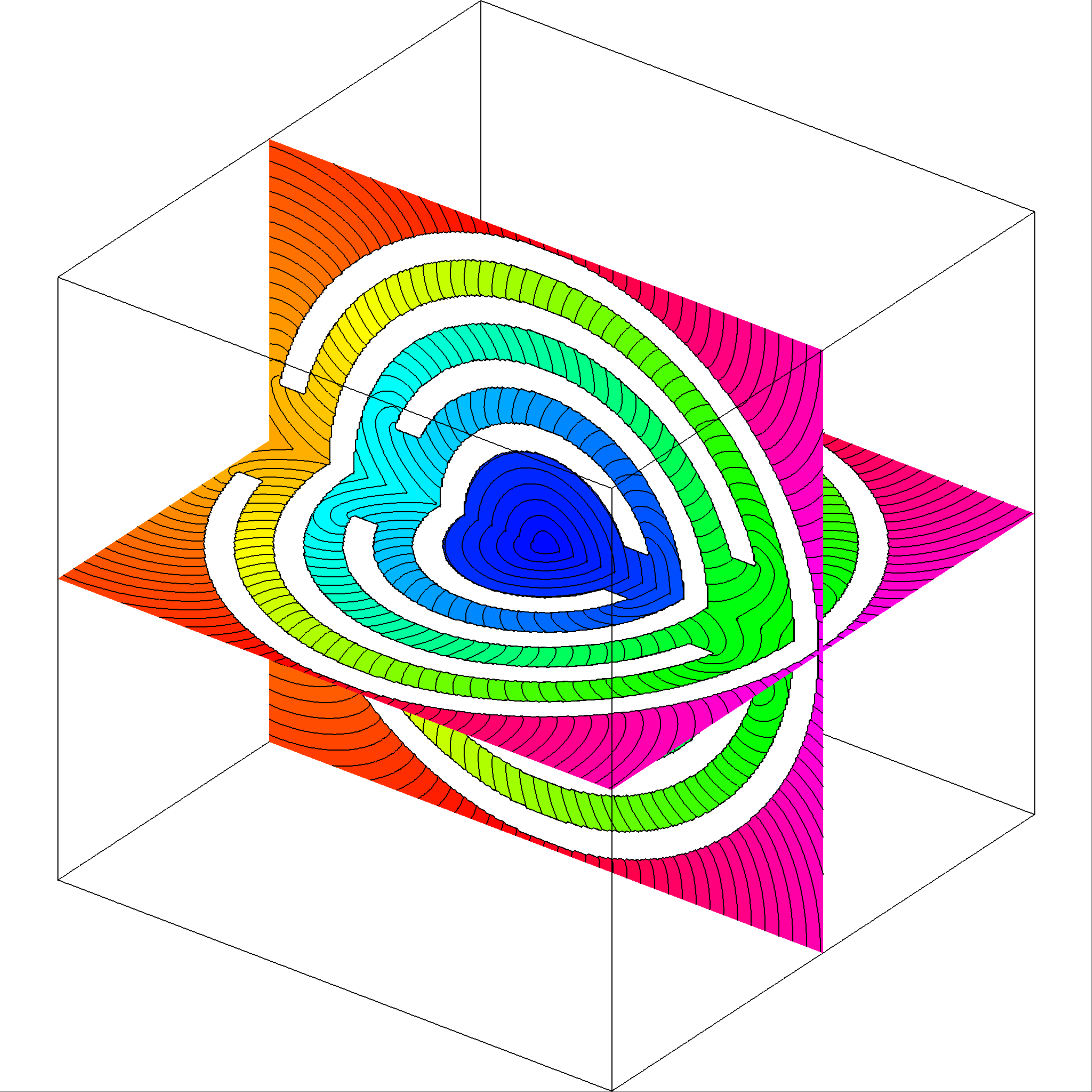}
  \makebox[0.325\textwidth]{(d) Case 4}
  \makebox[0.325\textwidth]{(e) Case 5}
  \makebox[0.325\textwidth]{(f) Case 6}
\end{center}
 \caption{Test cases: a) spherical interface; 
 b) planar interface; c) point source with $F$ defined by Eq. (\ref{eq:f1}); d) point source with $F$ defined by Eq. (\ref{eq:f2}); e) point source with $F$ defined by Eq. (\ref{eq:f3}); f) point source with four concentric spherical obstacles defined by Eq. (\ref{eq:barriers}).}
\label{fig:cases}
\end{figure}

An extensive systematical study, which involves widely different test cases, stride sizes, grid sizes, and numbers of processes, was carried out to illustrate the parallel performance of the present parallel fast marching algorithm. 

As shown in Fig. \ref{fig:cases}, six test cases were performed. For simplicity, a unit cube $[-0.5,0.5] \times [-0.5,0.5] \times [-0.5,0.5]$ was used as the computational domain. The first two cases are interface problems with unity speed functions. In the first case, a spherical interface of radius $0.25$ is centered in the domain and its interior is defined as the negative region. This case was used by Herrmann in his study \cite{Herrmann03}. 
The second case is defined by a plane past through the domain center: $100x+y+2z = 0$ and the positive region is given by $100x+y+2z > 0$. In the numerical implementation, the interfaces were initialized by assigning analytical function values to the two sets of grid points immediately adjacent to the interface from the positive and negative sides, respectively. This initialization was also applied to the ghost point zone to obtain strictly separated positive and negative regions in the computational domain. It should be noted that the characteristics from the planar interface directed inward the domain entail boundary conditions for obtaining the ideal global solution at every grid point, i.e., a signed distance to the interface itself instead of a point or a line on the interface. In this study, a ghost point was treated the same as an internal one without considering these characteristics. Such a simplification can be clearly justified, as the parallel performance would hardly be affected even if they were incorporated into the solutions through appropriate boundary conditions.
This case is particularly interesting for a parallel fast marching algorithm. It is evident that, each point depends on three upwind points of the same pattern for all points on one side of the interface, except for ghost cells with only one or two upwind points in this pattern. For each grid plane in the $x$ direction, information has to propagate from one corner to the diagonal corner in a point-by-point manner following a strict order of increasing magnitudes of the function values. 
The other four cases are point source problems with a point source at the domain center. 
Three cases were derived from \cite{ChaconV2015} with speed functions defined by 
\begin{eqnarray}
F(x,y,z) &=& 1, \label{eq:f1}\\  
F(x,y,z) &=& 1 + 0.50 \sin(20 \pi x)  \sin(20\pi y)  \sin(20\pi z), \label{eq:f2}\\  
F(x,y,z) &=& 1 - 0.99 \sin(2  \pi x)  \sin(2 \pi y)  \sin(2 \pi z), \label{eq:f3}
\end{eqnarray} 
respectively. The last case was derived from \cite{DetrixheGM2013}. Let $w = \tfrac{1}{24}$, $R = \sqrt{ x^2+y^2+z^2 }$, and $r= \sqrt{ x^2+y^2 }$, the four concentric spherical obstacles are defined by 
\begin{equation}
\begin{array}{ccc}
(0.15 < R < 0.15+w) &\setminus & ((r<0.05) \cap (z < 0));\\
(0.25 < R < 0.25+w) &\setminus & ((r<0.10) \cap (z > 0));\\
(0.35 < R < 0.35+w) &\setminus & ((r<0.10) \cap (z < 0));\\
(0.45 < R < 0.45+w) &\setminus & ((r<0.10) \cap (z > 0)),
\end{array}
\label{eq:barriers}
\end{equation}
where $\setminus$ represents the set difference operation. Within the obstacles, the speed function is $F = 0$; otherwise it is $F = 1$. 

The stride size in the restarted narrow band approach is the only essential free parameter in the present parallel fast marching algorithm. As shown in the previous section, it functions in the parallel algorithm just like the narrow band width in the sequential fast marching method. Apparently, the optimal stride size for achieving an ideal parallel performance is affected by many factors including the interface (source) properties, speed functions, domain decomposition configurations, grid sizes, and computing platforms. Nevertheless, the efficiency of a versatile algorithm should not be dramatically affected by a sensible choice of a free parameter. A parametric study was conducted to illustrate the effect of different stride sizes on the performance of the present parallel algorithm. First, a base choice of $\delta s = 2 \Delta h$ was determined simply according to the thickness of the overlapping regions. Then a series of stride sizes from $\delta s = 0.5 \Delta h$ to $3.5 \Delta h$ with an increment of $\Delta h$ were selected. A special case was also investigated, in which $\delta s$ was set to $\infty$ (a huge positive value in the actual implementation) such that only the termination condition $\texttt{size}_{\mathfrak{H}} ( s ) = 0$ in Algorithm \ref{alg:march_new} was to be met. In this case, therefore, all points in a subdomain will be computed once the computation is triggered. This seems to be a resemblance to many iterative algorithms for the Eikonal equation. But a salient feature distinguishes this case from those iterative algorithm: the completion of computation does not rely on a convergence check at all.
The other extreme case is $\delta s = 0$, in which the parallel algorithm runs exactly like the sequential version without activating the second marching step at all. This scenario was not tried in the present study due to the astronomic number of restarts required ($\sim N$) for fine grids. Instead, $\delta s = 0.5 \Delta h$ was tested to give a hint on this extreme scenario. 

Six uniform grids with $\Delta x = \Delta y = \Delta z =\Delta h$ were used in the computations and the total numbers of grid points (excluding the ghost points) were $ N = nh^3 = 32 ^ 3$, $64 ^ 3$, $128 ^ 3$, $256 ^ 3$, $512 ^ 3$, and $1024 ^ 3$, respectively. The finest grid has over $1$ billion points. 

The total number of domain decomposition configurations is $17$, i.e., $ p_i \times p _j \times p _k = 1 \times 1 \times 1$ ($np = 1$), $ 1 \times 1 \times 2$ ($np = 2$), $ 1 \times 2 \times 2$ ($np = 4$), $ 2 \times 2 \times 2$ ($np = 8$), $ 2 \times 2 \times 4$ ($np = 16$), $ 2 \times 4 \times 4$ ($np = 32$), $ 4 \times 4 \times 4$ ($np = 64$), $ 4 \times 4 \times 8$ ($np = 128$), $ 4 \times 8 \times 8$ ($np = 256$), $ 8 \times 8 \times 8$ ($np = 512$), $ 8 \times 8 \times 16$ ($np = 1024$), $ 8 \times 16 \times 16$ ($np = 2048$), $ 16 \times 16 \times 16$ ($np = 4096$), $ 16 \times 16 \times 32$ ($np = 8192$), $ 16 \times 32 \times 32$ ($np = 16384$), $ 32 \times 32 \times 32$ ($np = 32768$), and $ 32 \times 32 \times 64$ ($np = 65536$). Note that the coarsest grid $nh = 32$ cannot be decomposed with the last configuration due to $p_k = 64$.

All computations were performed on Garnet, a Cray XE6 supercomputer located at the U.S. Army Engineer Research and Development Center (ERDC) in Vicksburg, Mississippi, one of the five U.S. Department of Defense (DoD) Supercomputing Resource Centers (DSRCs) that are operated by the U.S. DoD High Performance Computing Modernization Program (HPCMP). Garnet has 150912 compute cores (4716 compute nodes each with 32 cores) and is rated at 1.5 peak PFLOPS. The compute nodes are populated by 2.5 GHz AMD Interlagos Opteron (6200 series) processors  with two processors per node, each with sixteen cores. Each node contains 64 GBytes of DDR3 memory shared by the 32 cores. Computer nodes are connected by the Cray Gemini Interconnect network. Garnet supports different parallel programming models. In this study, the algorithm was implemented in Fortran 2003 with the Message Passing Interface (MPI) for communications among processes. The MPI library on Garnet derives from the Argonne National Laboratory MPICH, which implements the MPI-3.0 standard. The code was compiled in double precision using the Intel Fortran Compiler XE version 14.0.2.144 with the $\texttt{-fast}$ optimization level. The code performance is affected by many different factors, especially the network throughput, since the system is shared by many users. In the present study, all computations were repeated five times and the CPU times reported here are the averaged values.  

\subsection{Accuracy of the parallel algorithm}

\begin{figure}[htbp!]
\begin{center}
 \includegraphics[angle=0,width=0.5\textwidth]{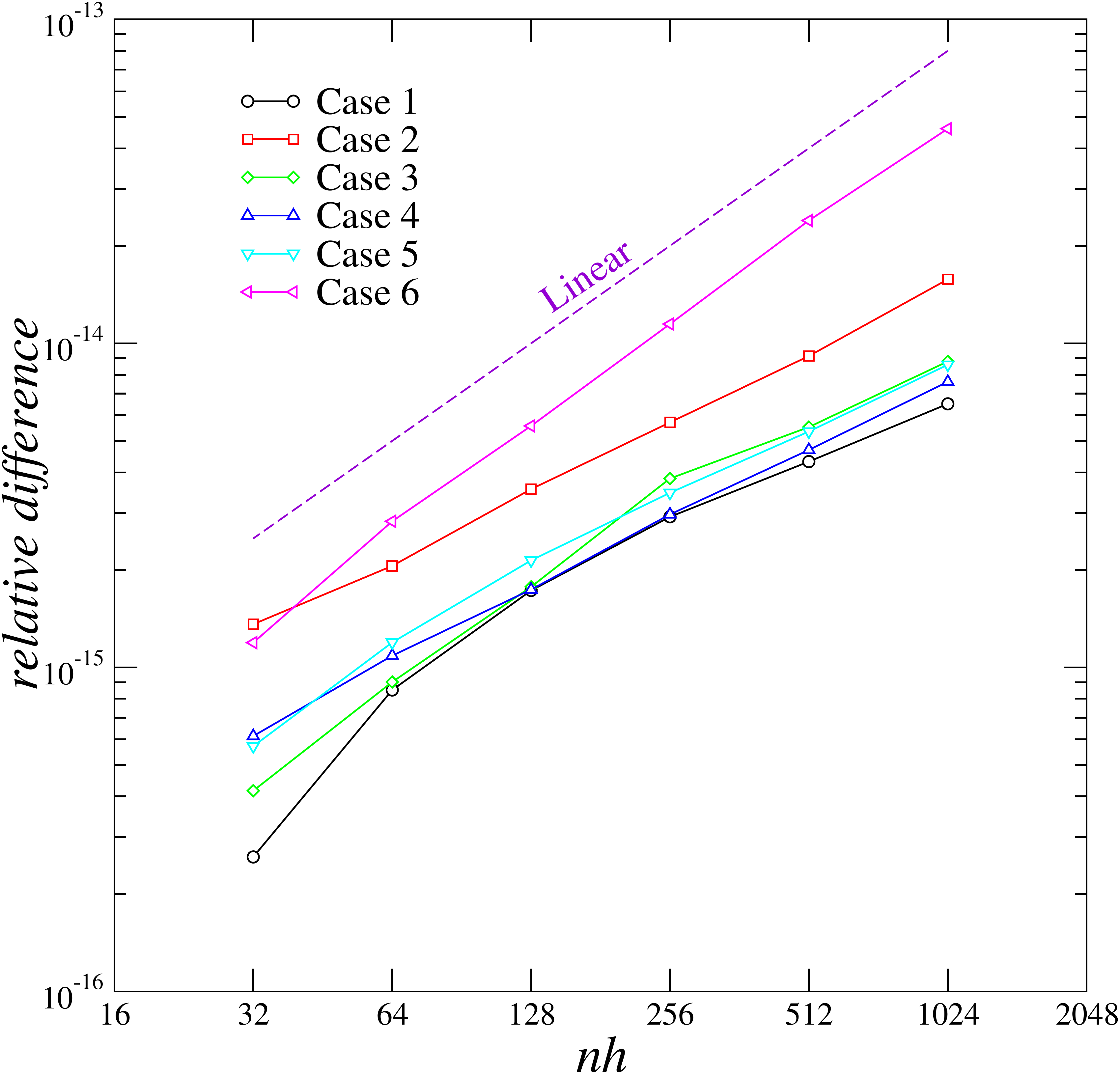}
\end{center}
 \caption{The maximum relative differences between the parallel and sequential solutions.}
\label{fig:differences}
\end{figure}

To verify that the parallel algorithm gives the correct solutions, a comparative study was conducted first. As defined above, the six test cases were solved on the six uniform Cartesian grids using both the sequential and parallel algorithms. In order to simplify the study, here the sequential algorithm was obtained by sequentializing the parallel algorithm.  That is, only $\textsc{Initialize\_Interface\_Parallel}$ (Algorithm \ref{alg:init_inter_parallel}), $\textsc{Initialize\_Heap\_Parallel}$ (Algorithm \ref{alg:init_heap_parallel}), and the first instance of $\textsc{March\_Narrow\_Band\_Parallel}$ (Algorithm \ref{alg:march_new}) in Algorithm \ref{alg:pfmm} were kept. Such a sequential algorithm is essentially the same as the original version given in Algorithm \ref{alg:sfmm}. A different set of tests were performed to demonstrate the expected first-order accuracy and the $O(N \log N)$ algorithm complexity of this sequential algorithm in the Appendix. 

For each case, a solution can be obtained on each grid by using the sequential algorithm. Such a sequential solution corresponds to $6 \times 17$ parallel solutions ($6 \times 16$ for the coarsest grid, $nh = 32$) from the parallel algorithm because of the $6$ different stride sizes and $17$ different domain decomposition configurations. The maximum relative difference between the sequential solution and a parallel solution among them is defined as $ \psi^{rd} = \max_{i,j,k} |\psi _{i,j,k} ^S - \psi _{i,j,k} ^P | / | \psi _{i,j,k} ^S | $ with $\psi _{i,j,k} ^S$ and $ \psi _{i,j,k} ^P$ represent the sequential solution and a parallel solution, respectively. Then the maximum value of $ \psi^{rd}$ from all parallel solutions on the same grid is obtained and shown in Fig. \ref{fig:differences}. In general, the maximum relative differences for all cases are within the order of machine accuracy for double precision floating point calculations. It is evident that the present parallel algorithm gives correct solutions to the Eikonal equation. As to the small differences, it is well-known that the numerical results may vary slightly depending on the order of floating point operations due to the limited precision in the floating point representation of real numbers. Apparently, the order of updating points in each subdomain for the parallel algorithm is quite different than that in the sequential algorithm. In addition, even with the same subroutines called in both algorithms and the same optimization options during compilation, the resulting object codes for these subroutines might be different due to the different calling sequences in the sequential and parallel algorithms. Nonetheless, such small differences do not alter the desirable non-iterative property of the fast marching method; and they are totally acceptable for a parallel algorithm, considering the usually much larger convergence criteria used in many iterative algorithms for the Eikonal equation.

Figure \ref{fig:differences} also shows that for each case the difference grows with a linear or even sub-linear rate as the grid is refined. This is consistent with the facts that the Eikonal equation is a first-order hyperbolic partial differential equation and it is solved with first-order upwind finite difference schemes here. Imaging that the first difference in the sequential and parallel solutions is generated at a certain grid point in the computational domain, this difference will propagate with the wave-front and slowly grow along with the solution process. When the number of grid points is doubled in each direction, for a first-order scheme the growth will be roughly doubled too. 
On the other hand, the first solution differences on different grids do not necessarily happen at similar spatial locations in the computational domain. If the first solution difference on a finer grid happens further away from the source, a sub-linear growth rate may be observed.

\subsection{Parallelization overheads}

\begin{figure}[htbp!]
\begin{center}
 \includegraphics[angle=-90,width=0.8\textwidth]{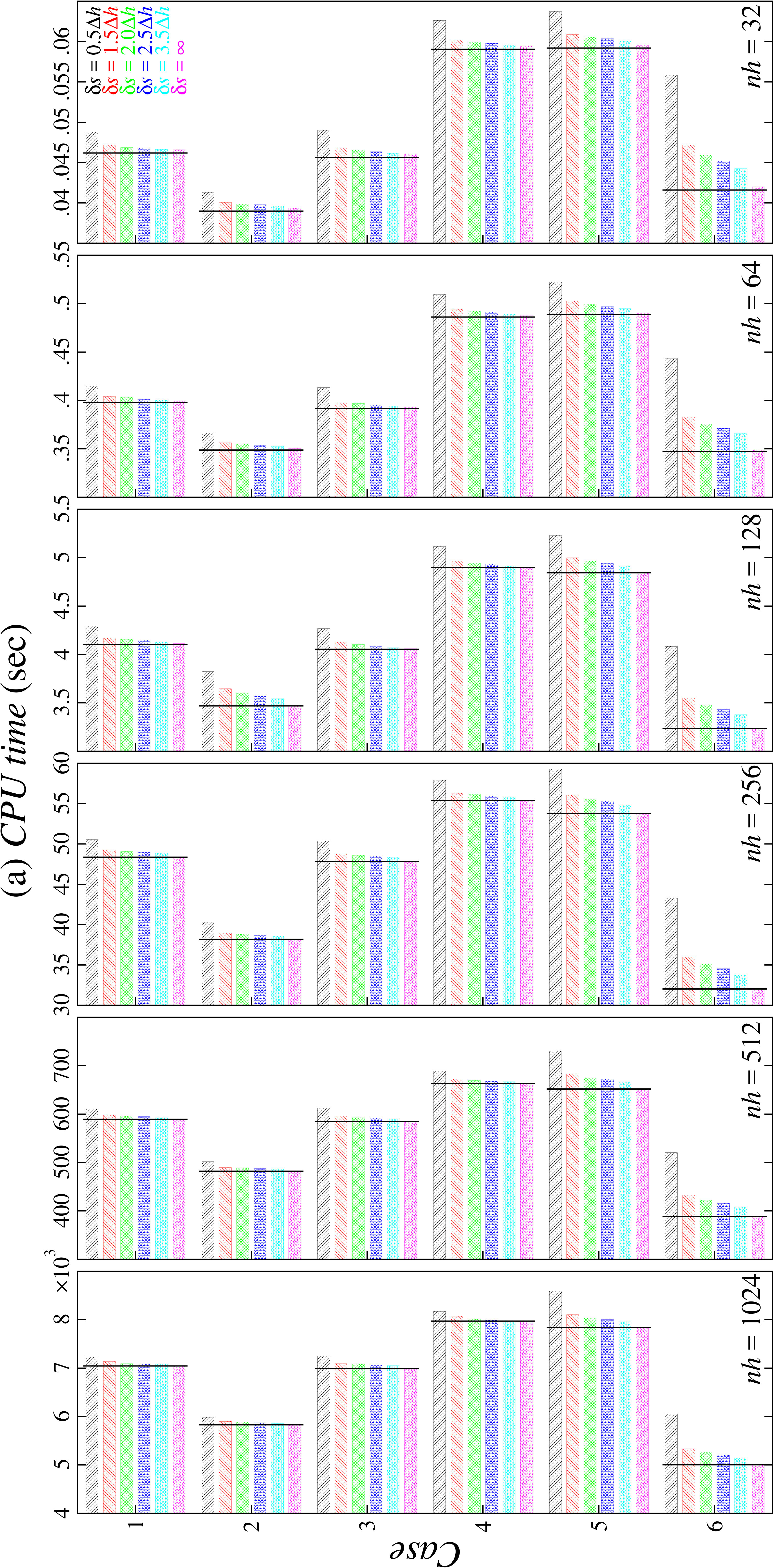}
 \includegraphics[angle=-90,width=0.8\textwidth]{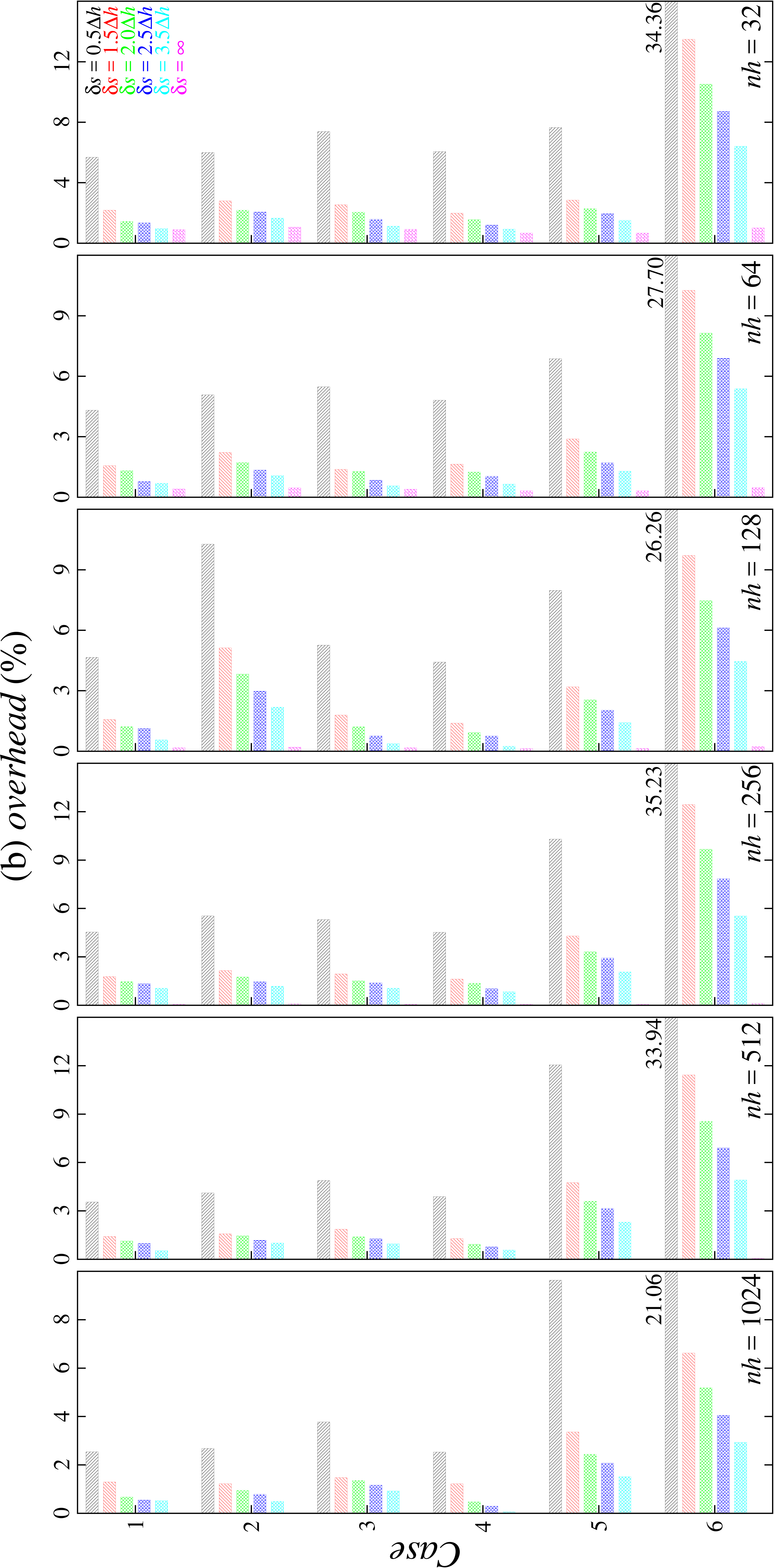}
 \includegraphics[angle=-90,width=0.8\textwidth]{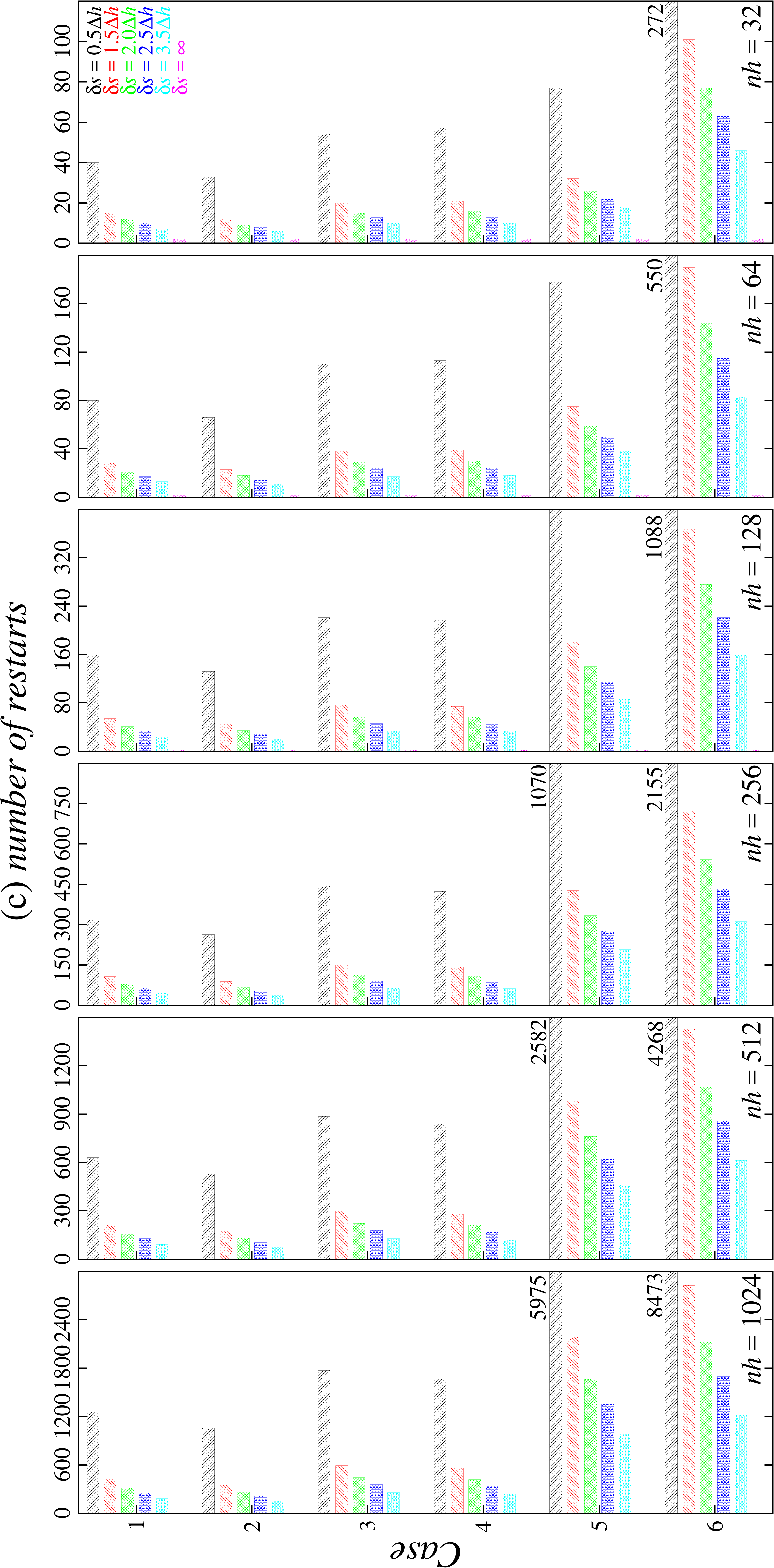}\\
 \end{center}
 \caption{Single-process runs of the parallel fast marching algorithm: the CPU time $T_1$ (a), the parallelization overhead (b), and the number of restarts $nr$ (c) as functions of the stride size $\delta s$ and the grid size $nh$. The vertical black lines in (a) represent CPU times $T_S$ from the sequential algorithm.}
\label{fig:singleprocess}
\end{figure}

The parallelization of a sequential algorithm comes with some overheads. A comparison of Algorithms \ref{alg:sfmm} and \ref{alg:pfmm} clearly shows the extra work required in the parallel algorithm and it is collectively called the parallelization overheads. Apparently, the introduction of the restarting loop has the major impact on the overheads. Besides a second marching step, this loop includes additional modules for data communication: global data reduction, and collection, local exchange, and integration of the data from the overlapping regions. Since there is no data exchange in a single-process run, the data integration and the second marching step do not involve any real operations on grid points. Therefore, the performance penalty of the parallel algorithm is mostly from overheads of setting up loops, function and subroutine calls, and $\texttt{if}$ statements. 

It should be noted that the sequential algorithm used in the results part is slightly different than that given in Sec. \ref{sec:sfmm} as discussed above. In theory, any operations added on top of the original sequential versions of all subroutines should be considered as part of the overheads. For instance, the data structures for two-sided interface problems, augmented status tags, additional function value comparisons, and extra solves of the quadratic equation are introduced in the present study to parallelize the sequential algorithm. However, these modifications do not substantially change the pure sequential algorithm, as evident by the discussions in Sec. \ref{sec:pfmm}. One major benefit of the present arrangement is that one set of subroutines can be used in both algorithms and the code maintenance is much simpler.

Here the CPU times required for solving a case with the sequential algorithm and the parallel algorithm running on a single process ($np = 1$, i.e., the domain decomposition configuration is $p_i \times p _j \times p _k = 1 \times 1 \times 1$) are defined as $T_S$ and $T_1$, respectively. For a specific case, its parallelization overhead is the relative difference between $T_S$ and $T_1$, i.e., $\frac{T_1 - T_S}{T_S} \times 100 \%$. It measures the amount of extra work required to parallelize a sequential algorithm. Fig. \ref{fig:singleprocess} shows the results of single-process runs of the parallel algorithm, in which (a) illustrates the CPU time $T_1$ as a function of the grid size $nh$ and the stride size $\delta s$ with the sequential CPU time $T_S$ for each case on each grid clearly marked, (b) gives the parallelization overheads for all cases with different stride sizes on different grids, and (c) shows the numbers of restarts $nr$ in all single-process runs.

In terms of the sequential CPU time $T_S$, cases 4 and 5 are the most expensive cases for the sequential algorithm on all six grids. It takes slightly more time to solve case 4 than case 5 on grids of $nh \geq 128$ because the speed function of case 4 varies in space at a much higher frequency than that of case 5. But the high frequency variation in the speed function cannot be captured on very coarse grids, thus case 5 is marginally more expensive than case 4 for grids $nh = 32$ and $nh = 64$. All the other cases have unity speed functions except that the speed function in the barriers is zero for case 6. For these cases, apparently the total numbers of grid points involved in the computations and the sizes of heaps determine the computational cost. Case 1 is slightly more expensive than case 3. It is evident that the former has two heaps, one for the positive region and the other for the negative region, whereas the latter has only one heap. Case 2 also has two heaps, but the heap sizes do not vary much during the solution process and are much smaller than those of cases 1 and 3 when the fronts are away from the sources. For case 6, the CPU times are generally lower than those of the other cases because the grid points within the barriers were not solved from the quadratic equation. However, on the coarsest grid ($nh = 32$) the numbers of grid points covered by the barriers is much reduced and case 6 gives a CPU time a little higher than that of case 2. 

For the single-process runs of the parallel algorithm, there is a direct correlation between the number of restarts and the parallelization overheads. For each case on each grid, as the stride size increases, the number of restarts decreases, thus the CPU time and the parallelization overhead decreases. With the present restarted narrow band formulation $\delta s = 0.5\Delta h$ incurs a significant number of restarts and gives much higher parallelization overheads in each case. Whereas with $\delta s = \infty$ the computations finish in two restarts for all cases. 
Apparently all operations on grid points were done in the first round; the second restart does not involve any essential operations and is required only because of one of the termination criteria for the parallel algorithm. 
This is also reflected in the CPU times of the $\delta s = \infty$ runs, which are only slightly higher than those from the sequential algorithm. The results for $\delta s = \infty$ in the single-process runs are not surprising, but they warrant further studies in multi-process runs. 

On the other hand, as the grid size $nh$ increases, the parallelization overheads for each test case with a specific stride size decrease due to the decreasing ratio of the cost from the additional modules and the cost of the first marching step. Except for case 6 and the $\delta s = 0.5\Delta h$ runs, the parallelization overheads are generally insignificant and well below $5\%$ for all grids. The barriers in case 6 make the travel-time for the wave-front to reach the domain boundary much larger than those in the other cases. This greatly increases the number of restarts for a given stride size as shown in Fig. \ref{fig:singleprocess}(c). Case 5 also gives larger numbers of restarts than the first four cases, because its speed function varies in a fairly wide range and may reach a minimum value of $F = 0.01$ in the domain, which again results in a larger travel-time and more restarts with a specific stride size than those from unity or smoother speed functions. 

\subsection{Parallel restarts}

\begin{figure}[htbp!]
\begin{center}
 \includegraphics[angle=0,width=0.3\textwidth]{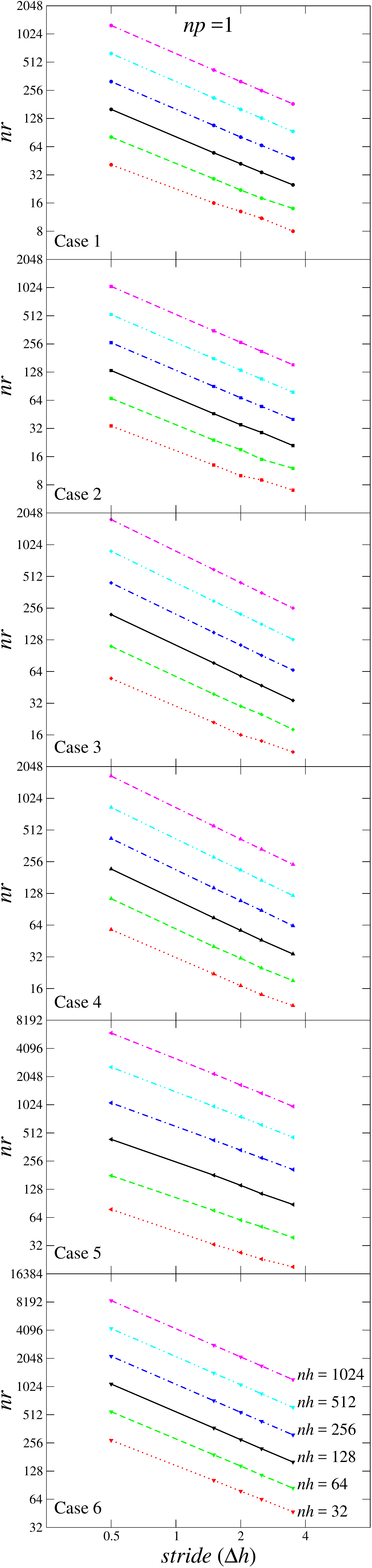}
 \includegraphics[angle=0,width=0.3\textwidth]{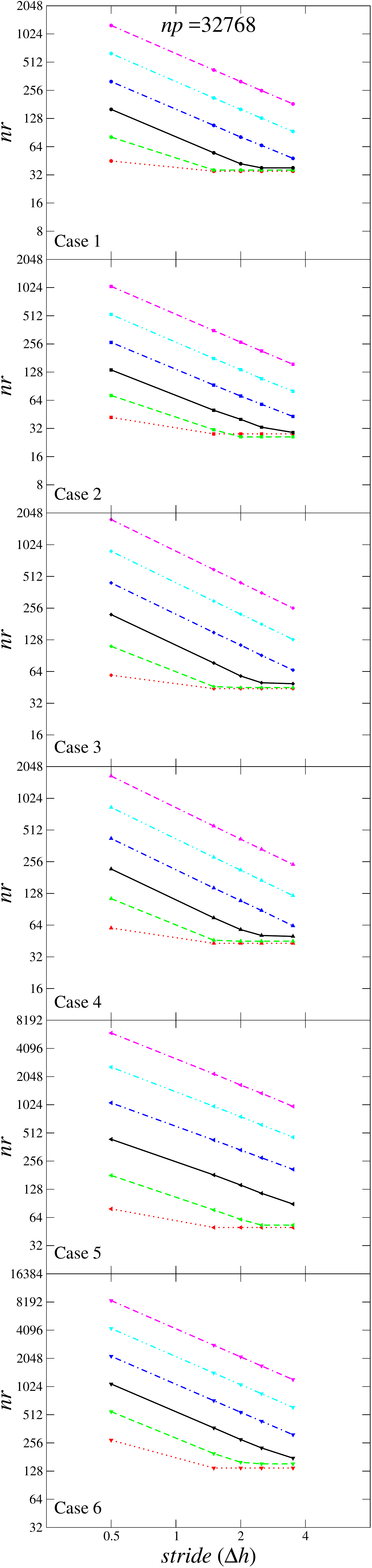}
 \includegraphics[angle=0,width=0.3\textwidth]{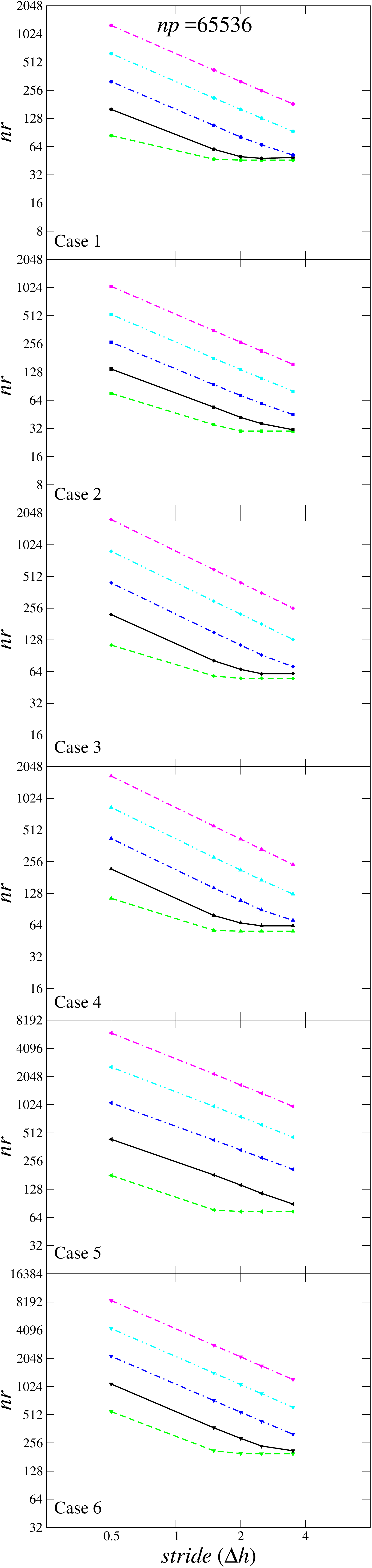}
\end{center}
 \caption{Parallel fast marching algorithm: the number of restarts $nr$ as a function of the grid size $nh$, the stride size $\delta s$ (from $\delta s = 0.5\Delta h$ to $\delta s = 3.5\Delta h$), and the number of processes $np$.}
\label{fig:stride}
\end{figure}

The results from the single-process runs show that the number of restarts in the present restarted narrow band approach determines the parallelization overheads of the parallel algorithm. These results are recast in Fig. \ref{fig:stride} together with those from the parallel runs with $np = 32768$ and $np = 65536$. Results from other domain decomposition configurations with $np$ between $1$ and $32768$ are not shown because they follow the same trend with $nr$ between those of $np = 1$ and $np = 32768$ given here. Also, the results from runs with $\delta s = \infty$ will be given separately below. As shown in Fig. \ref{fig:stride}, for $np = 1$, the variations of $nr$ with the stride size $\delta s$ for different grid sizes $nh$ are essentially parallel straight lines of slope $-1$ separated by approximately uniform gaps in the log-log plots. This clearly illustrates that $nr$ is inversely proportional to $\delta s$ and directly proportional to $nh$, respectively. As $np$ increases, the lowest side of the $nr$ distribution rises gradually, starting from the coarsest grid ($nh = 32$) and the largest stride size presented in this figure ($\delta s = 3.5 \Delta h$). However, the increase of $nr$ is very limited and the number of restarts from $\delta s = \infty$ defines the upper bound of the flattened distribution. Here $nr$ is only affected by the increase of $np$ in a small portion of the parametric space of the grid size $nh$ and the stride size $\delta s$. With the current domain decomposition configurations, apparently, the grid block sizes within a subdomain in these runs become really small, and for larger stride sizes the wave-fronts cannot reach the specified $\texttt{bound}_{\text{band}}$ without incurring extra restarts. Therefore, the increase of $nr$ in such circumstances is not a limitation of the present restarted narrow band approach, as it only indicates the grid block sizes of the subdomains are too small for the domain decomposition configuration employed in the parallel computation. 

\begin{figure}[htbp!]
\begin{center}
  \includegraphics[angle=0,width=0.95\textwidth]{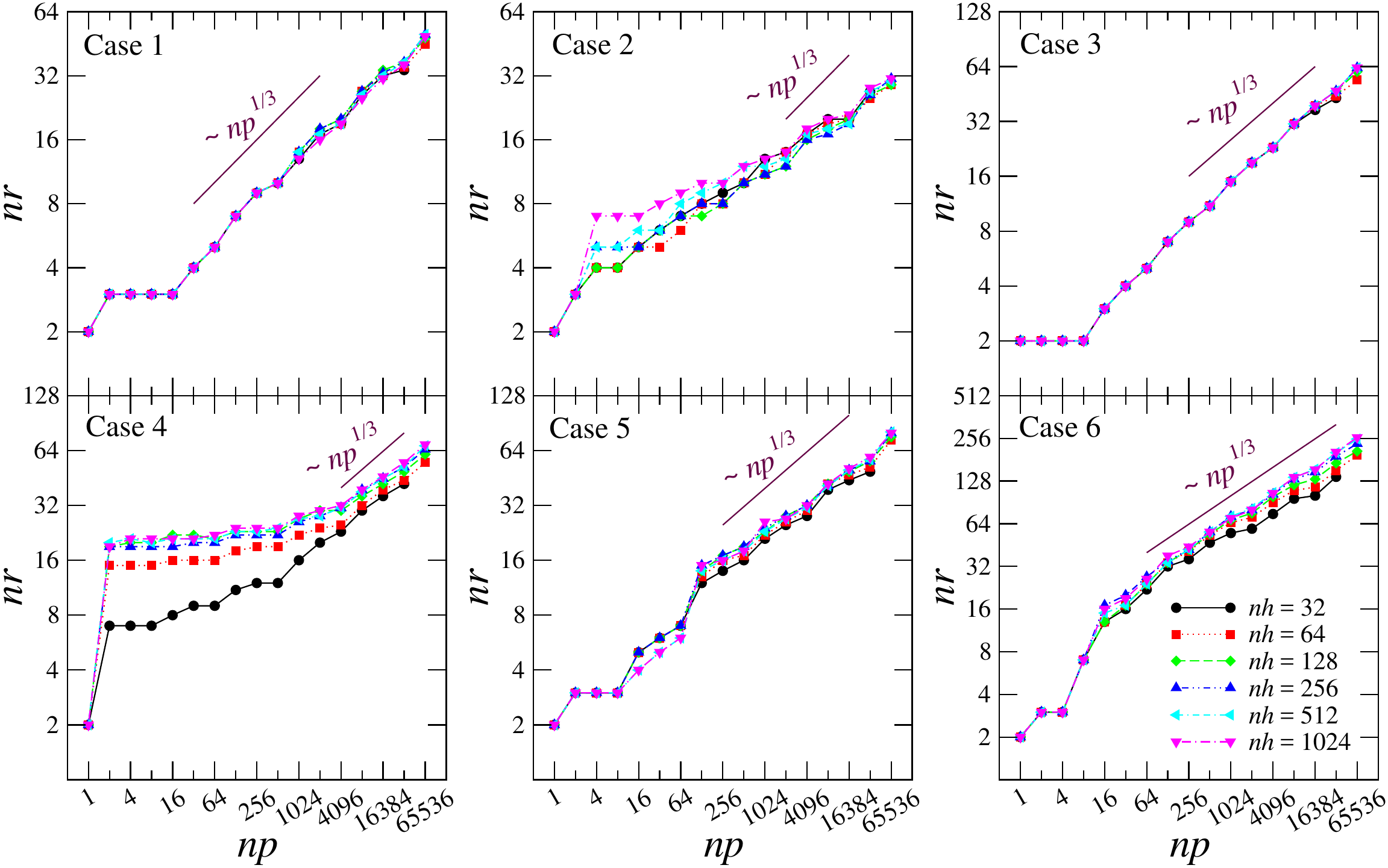}
\end{center}
 \caption{Parallel fast marching algorithm: the number of restarts $nr$ as a function of the grid size $nh$ and the number of processes $np$ for $\delta s = \infty$.}
\label{fig:stride2}
\end{figure}

Figure \ref{fig:stride2} presents the number of restarts $nr$ as a function of the grid size $nh$ and the number of processes $np$ for $\delta s = \infty$. It is evident that $nr$ remains finite for all cases on all grids with different $np$. This is totally different from many iterative algorithms for the Eikonal equation that the number of iterations is usually determined by some arbitrary user-specified convergence criteria. With $\delta s = \infty$, of course, whenever the computation is started or restarted in a subdomain, all points in this subdomain will be updated. Such a phenomenon in the parallel solution procedure does bear a resemblance to an iterative procedure. With the present parallel algorithm, however, the number of restarts asymptotically approaches $np^{1/3}$ (the number of subdomains in one direction) without using any iterations termination conditions. This verifies that the number of restarts is totally different from the number of iterations in iterative algorithms and the present restarted narrow band approach retains the highly desirable single-pass, non-iterative property of the sequential fast marching method.

On the other hand, the variations of $nr$ along with the increase of $np$ and the different behaviors among the six test cases illustrate the properties of different sources and speed functions. For cases 1 and 3, both with a unity speed function, $nr$ shows very small changes between different grids as $np$ increases from $1$ to $65536$. For case 2, its special source makes $nr$ very sensitive to the domain decomposition configurations. For the other cases with non-unity speed functions, refined grids resolve more variations in the speed functions, which results in a few more restarts. 

\subsection{Parallel Performance}
\begin{figure}[htbp!]
\begin{center}
 \includegraphics[angle=0,width=0.4\textwidth]{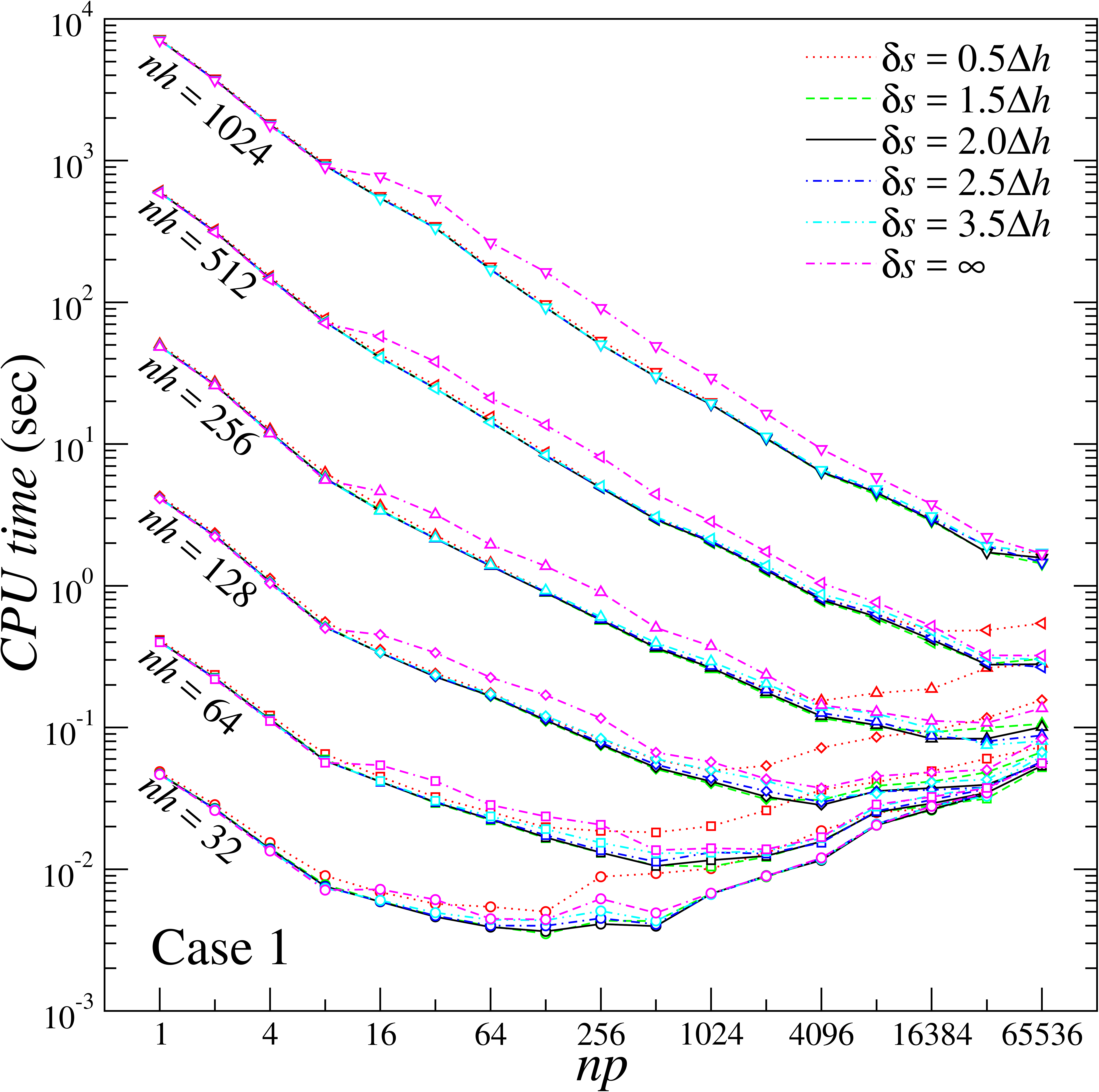}
 \includegraphics[angle=0,width=0.4\textwidth]{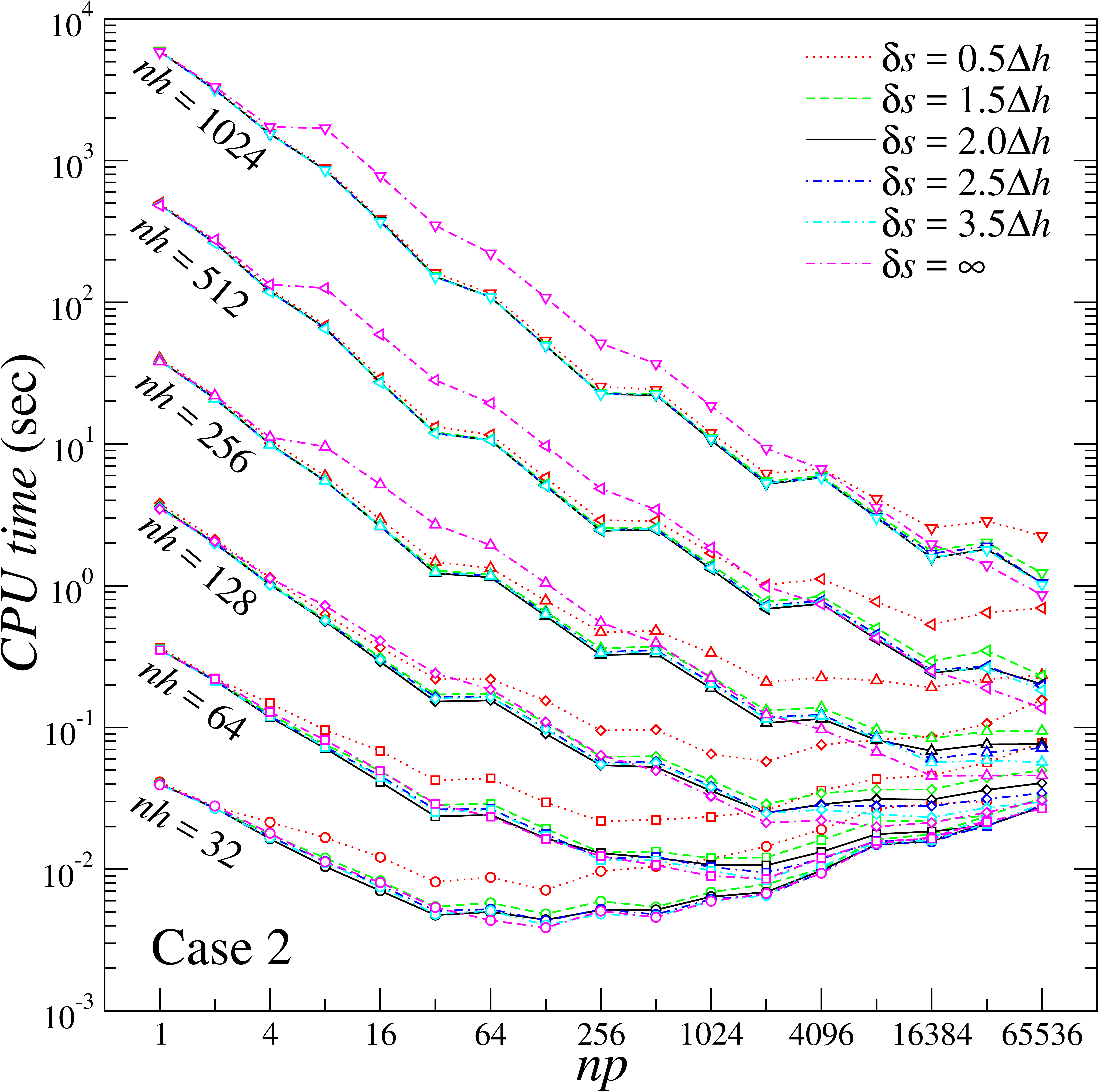}\\
 \vspace{1.5ex}
 \includegraphics[angle=0,width=0.4\textwidth]{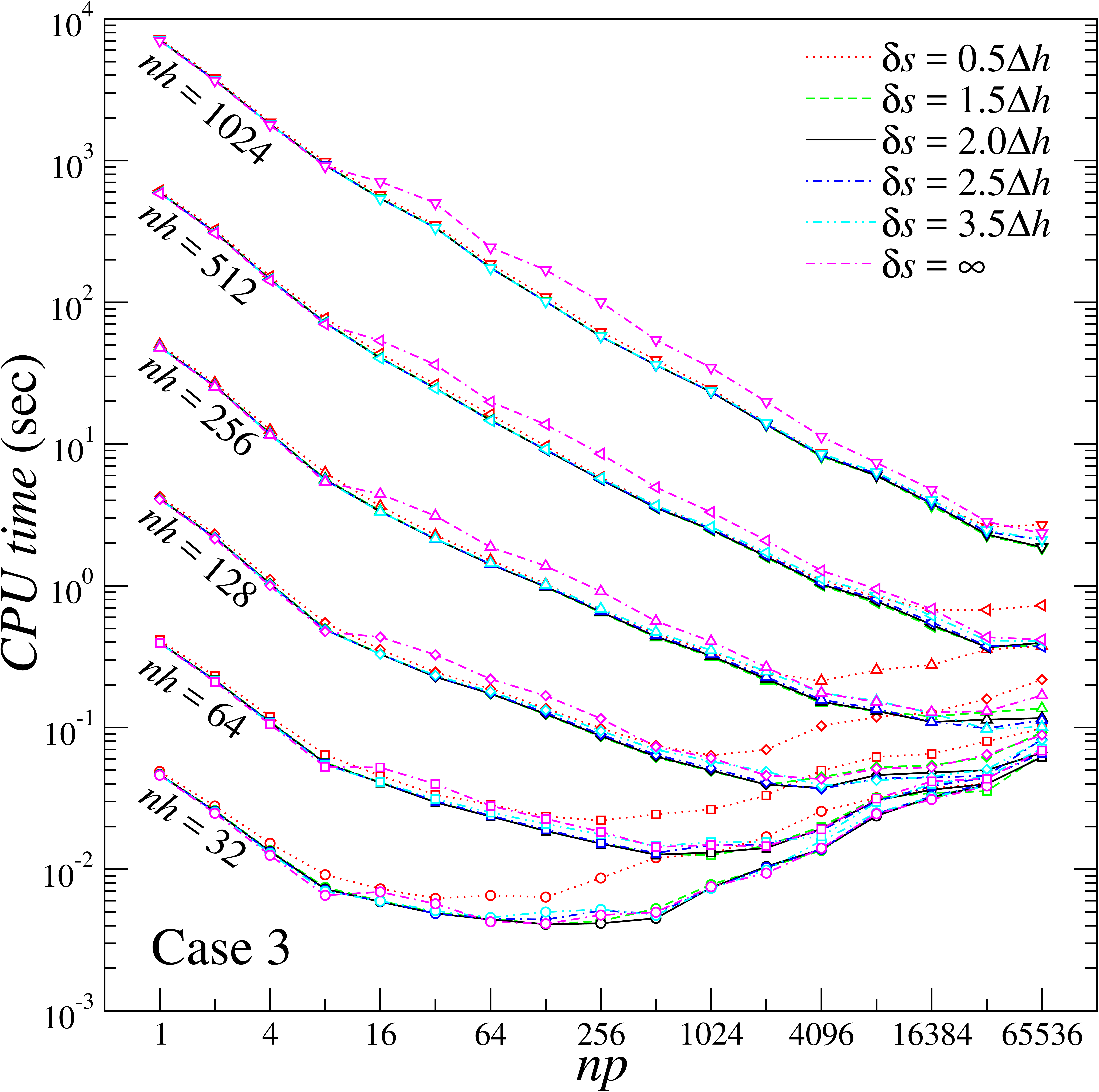}
 \includegraphics[angle=0,width=0.4\textwidth]{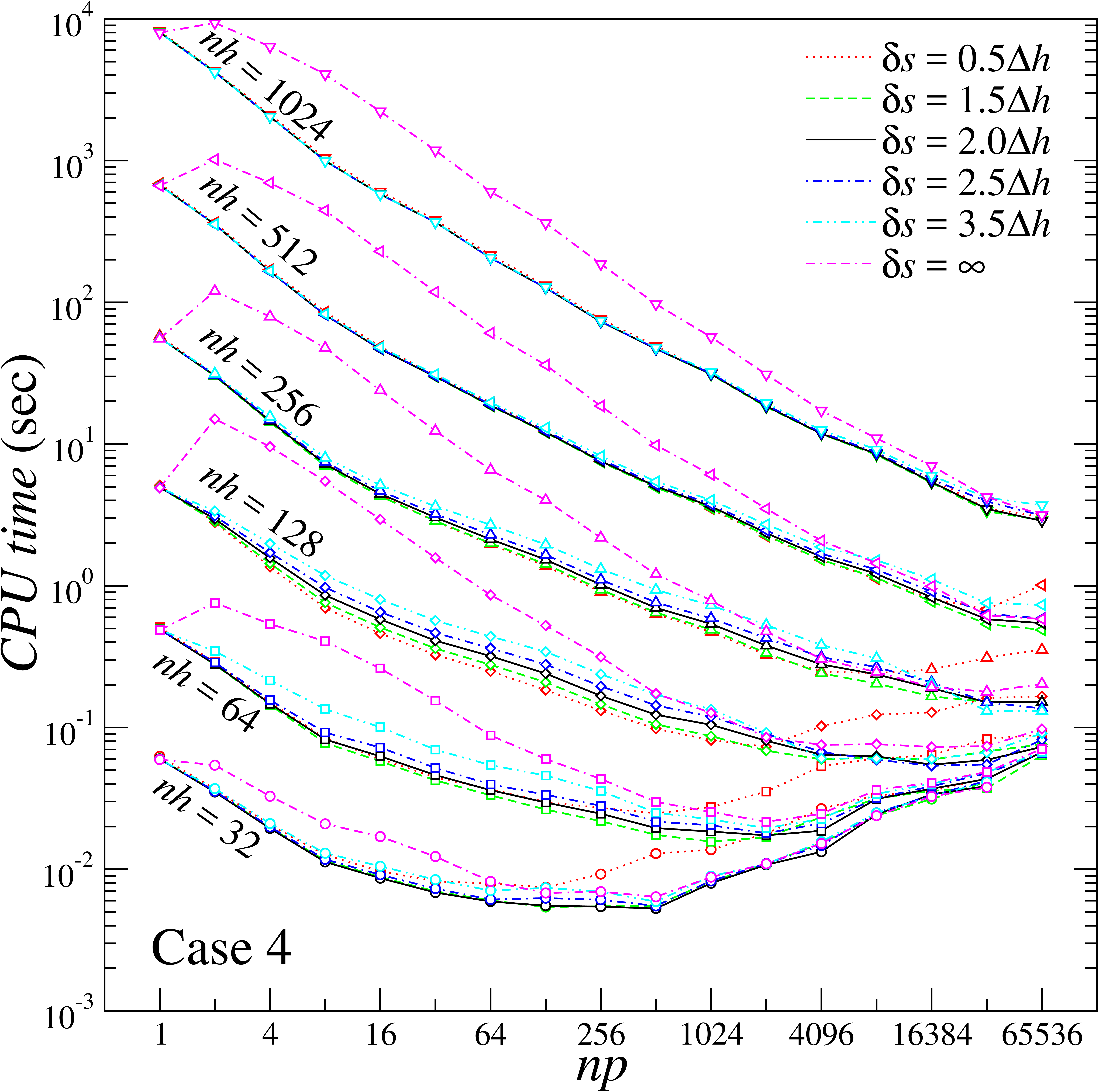}\\
 \vspace{1.5ex}
 \includegraphics[angle=0,width=0.4\textwidth]{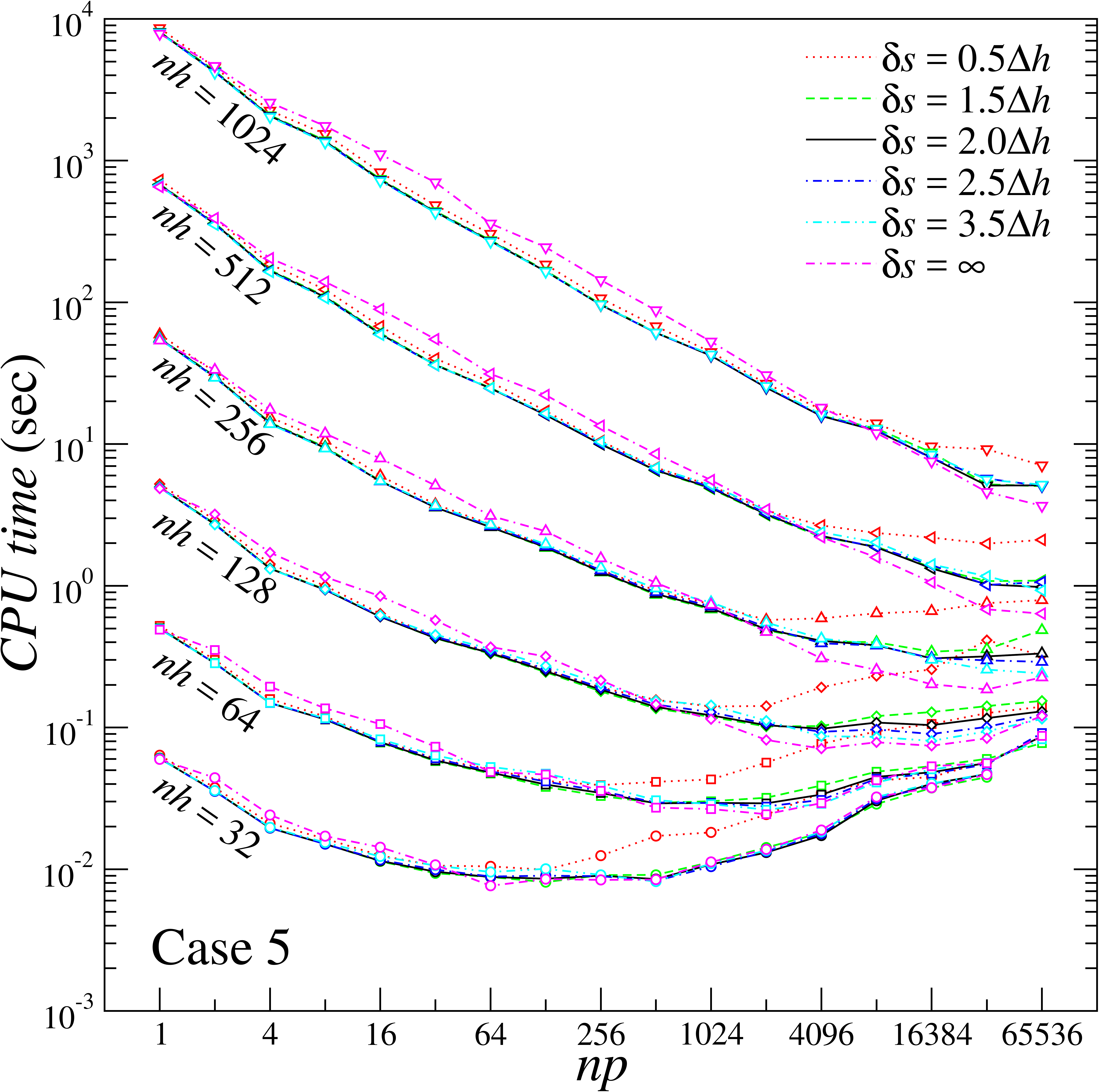}
 \includegraphics[angle=0,width=0.4\textwidth]{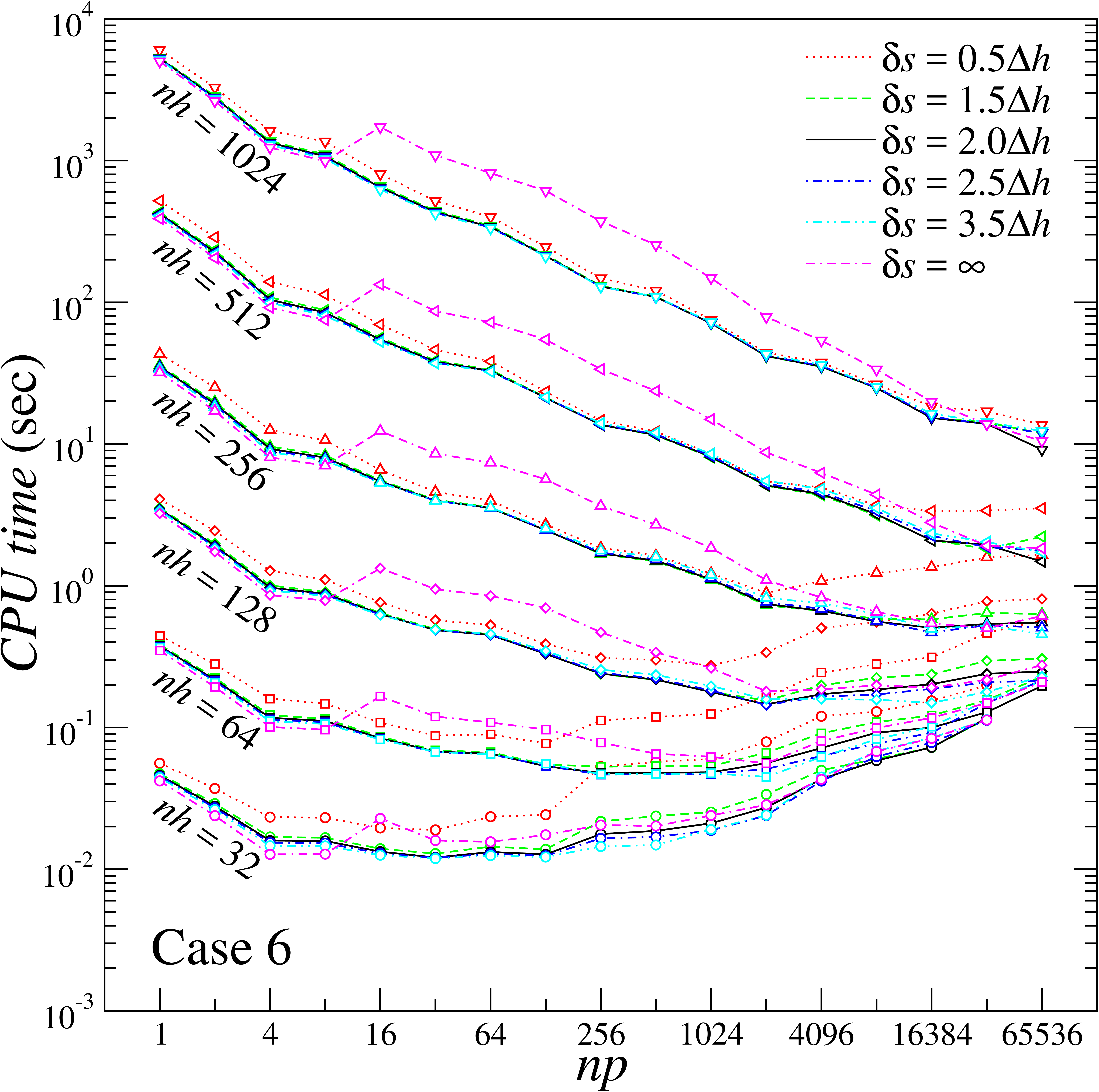}
\end{center}
 \caption{Parallel fast marching algorithm: the CPU time as a function of the stride size $\delta s$, the grid size $nh$, and the number of processes $np$.}
\label{fig:cputimes}
\end{figure}

Figure \ref{fig:cputimes} presents the CPU time $T_{np}$ as a function of the number of processes $np$ at different stride sizes $\delta s$ for different grid sizes $nh$. 
A remarkable parallel scalability can be observed in all cases for different stride sizes on all grids. A parallel algorithm is called scalable if $T_{np}$ decreases as $np$ increases. But this does not mean the CPU time will keep decreasing for any computational load per process in any test problems. 
On the coarsest grid ($nh = 32$), as $np$ increases, the overheads due to load imbalance and communications take over soon after the grid size of the subdomain block is below $8^3$. A further increase of $np$ results in a growing number of restarts and the CPU time begins to increase with a similar pattern as that of $nr$ discussed in the previous part. As soon as the the grid is refined, a much improved scalability is observed. For example, on grid $nh = 64$, the CPU time keeps decreasing until $np = 1024$ or $2048$. For grids of $nh \geq 128$, a grid size around $16 \times 16 \times 16$ for the subdomain block appears to the threshold for achieving a positive gain from the increased $np$. A sustained trend of decreasing CPU time for the full range of $np$ tested here is observed on finer grids of $nh \geq 256$. 

In terms of the effects of stride size on the parallel performance, apparently a smaller stride size will result in an increased number of restarts, thus the communication overheads will increase; whereas a larger stride size may deteriorate load imbalance in each restart. With the six stride sizes tested in this study, it is difficult to select an optimal one that performs better than the others on all grids for all $np$ in all cases.
Overall $\delta s = 2 \Delta h$ seems to a well-balanced choice between the trade-offs of load imbalance and communication overheads. Nonetheless, this doesn't mean that the performance of the present parallel algorithm becomes less satisfactory for other values of the stride size. As shown in the figure, stride sizes close to $2 \Delta h$, such as $1.5 \Delta h$ and $2.5 \Delta h$, give results very comparable with or even better than those at $\delta s = 2 \Delta h$. The results in this part verify that the present parallel algorithm is not exceedingly sensitive to the choice of the stride size. In general, on supercomputers with very fast interconnect networks, a stride size reasonably larger than $2 \Delta h$ can give a good performance for a large $np$ at which the subdomain size becomes rather small. Of course, it is always a worthwhile practice to experiment several different stride sizes to decide a good choice for new applications and/or new computing platforms. 

It is interesting to note that generally $\delta s = \infty$ gives a slightly better performance for some decompositions with optimally balanced loads, e.g., $np \leq 8$ for cases 1 and 3 and $np \leq 4$ for case 6. Evidently in these runs each subdomain can be solved independently, which is ideal for $\delta s = \infty$. For cases 4 and 5, although the source is symmetrical for $np \leq 8$, but in both cases the speed functions are nonsymmetrical with regard to these domain decompositions. On the other hand, as $np$ increases, the grid block size per process becomes smaller and eventually the computational cost with $\delta s = \infty$ decreases to the same order of magnitude as those from a stride size around $2 \Delta h$. This is quite reasonable for a large $np$ because of the relative scale variations between the computation and communication costs. For instance, at $np = 32768$ the subdomain grid size is $32 ^3$ for grid $nh = 1024$. With such a small computational load the communication cost becomes a substantial part; and with $\delta s = \infty$ the computation takes less restarts although each restart covers more grid points. 

Even for $\delta s = 0.5 \Delta h$, surprisingly, it performed fairly well in many runs. Especially for case 4, it gives the lowest CPU times for several grids in different ranges of $np$. Such a small stride size can greatly reduce the CPU time spent on the second marching step, which is a major source of the load imbalance overheads. On the other hand, after the subdomain size becomes very small, the CPU time saved from a swift second marching step cannot offset the increased communication overheads any more.  

\begin{figure}[htbp!]
\begin{center}
 \includegraphics[angle=0,width=0.3\textwidth]{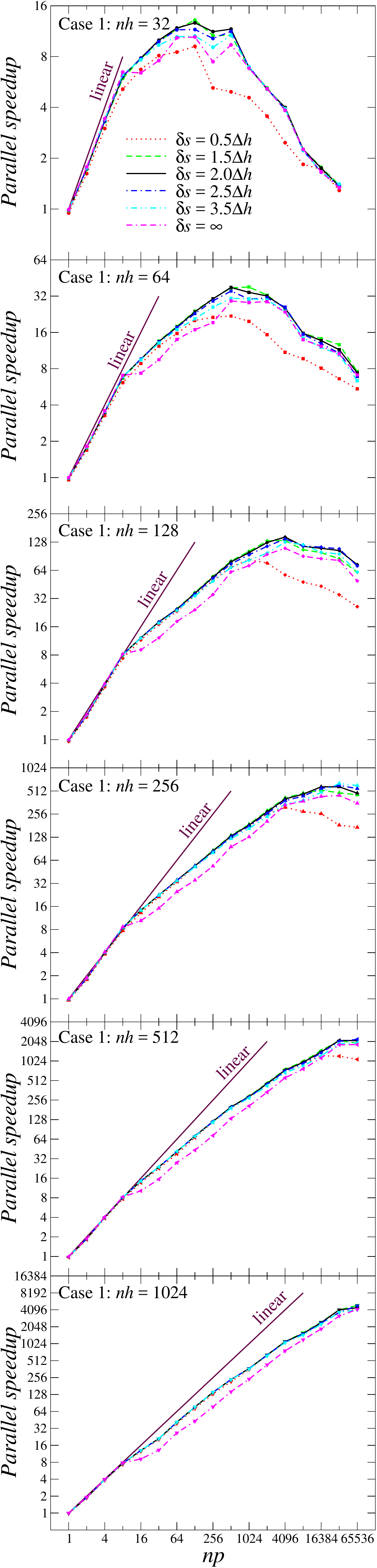}
 \includegraphics[angle=0,width=0.3\textwidth]{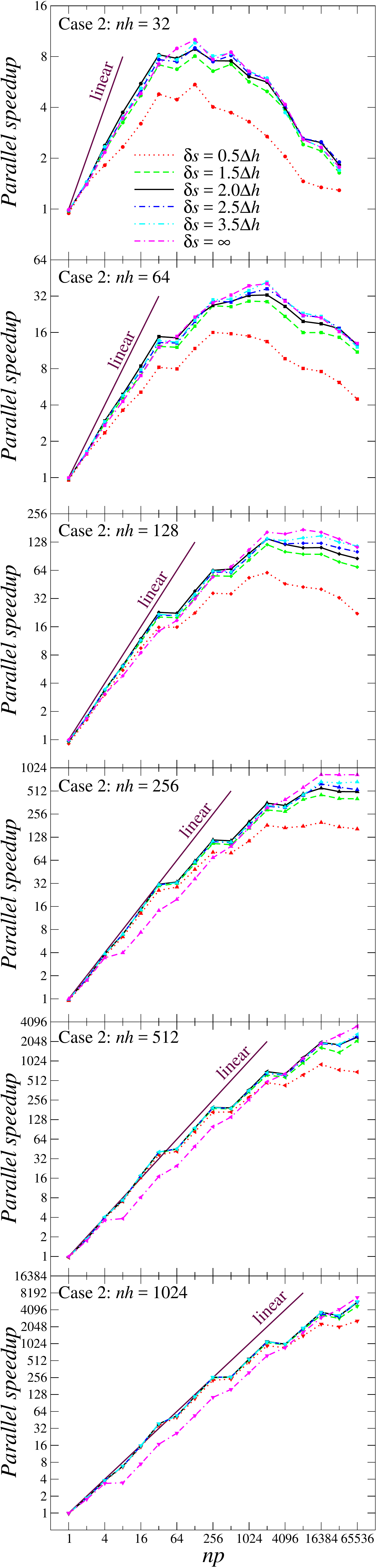}
 \includegraphics[angle=0,width=0.3\textwidth]{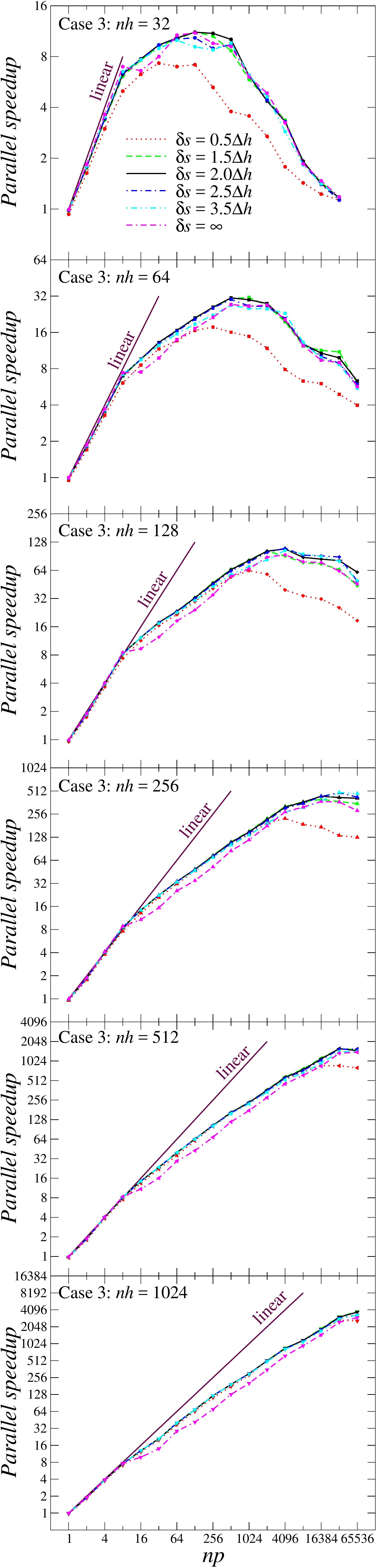}
\end{center}
 \caption{Parallel fast marching algorithm: the parallel speedup as functions of the stride size $\delta s$, the grid size $nh$, and the number of processes $np$.}
\label{fig:speedups}
\end{figure}
\begin{figure}[htbp!]
\ContinuedFloat
\begin{center}
 \includegraphics[angle=0,width=0.3\textwidth]{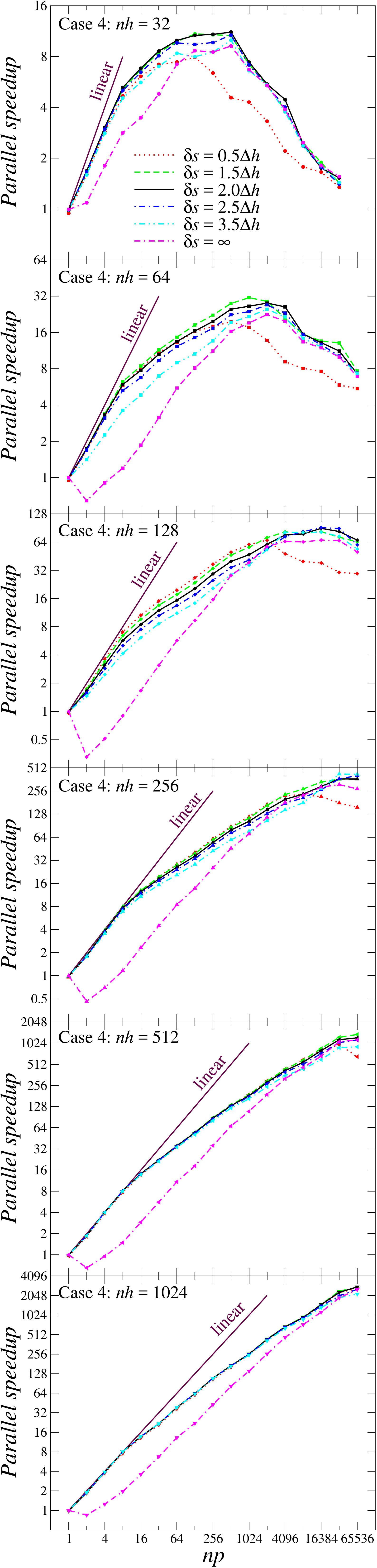}
 \includegraphics[angle=0,width=0.3\textwidth]{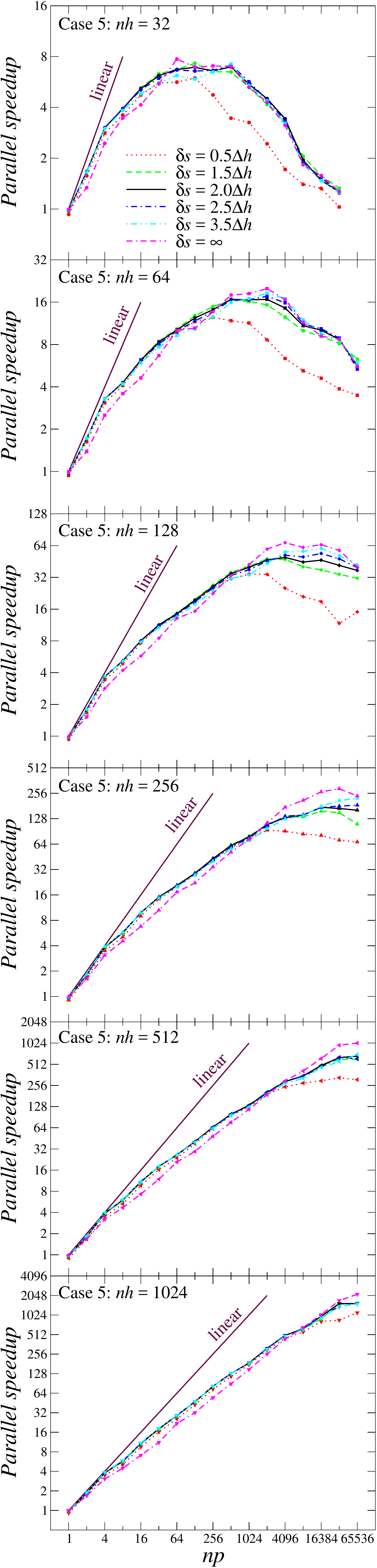}
 \includegraphics[angle=0,width=0.3\textwidth]{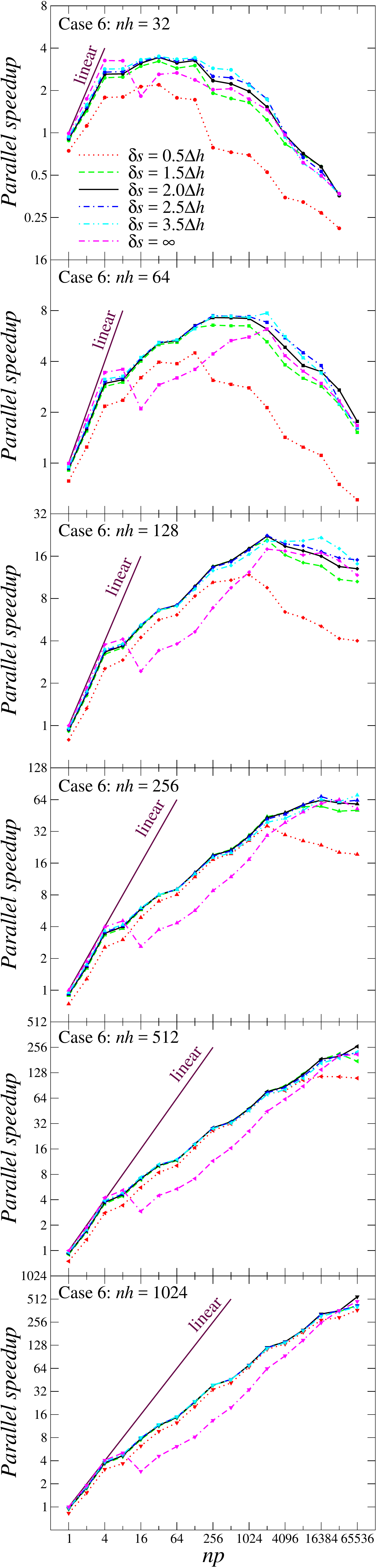}
\end{center}
 \caption{Continued. Parallel fast marching algorithm: the parallel speedup as functions of the stride size $\delta s$, the grid size $nh$, and the number of processes $np$. }
\label{fig:speedups}
\end{figure}

Figure \ref{fig:speedups} shows the parallel speedups of the present parallel fast marching algorithm for the six test cases. Here the absolute speedup defined as $S = \frac{T _S}{T _{np}}$ is used. The relative speedup defined using $T _1$ instead of $T _S$ will be slightly higher than the absolute one. As the grid refines, the speedup tends to approach the ideal linear slope. The parallel scalability of the present parallel algorithm becomes more evident in this figure. 

In general, cases 1 and 2 demonstrate higher speedups because both of them are interface problems and the load imbalance issues are less pronounced. In addition, as the range of variations in the speed function increases, the speedup decreases as shown in cases 3, 4, and 5. In case 6, the wave-front has to propagate through narrow passages between multiple barriers. For a domain decomposition parallelization, such a configuration greatly worsens the load imbalance issue in a point source problem. However, the present parallel fast marching method still performs very well and demonstrates impressive speedups in wide ranges of $np$ on different grids. 

\begin{figure}[htbp!]
\begin{center}
 \includegraphics[angle=0,width=0.3\textwidth]{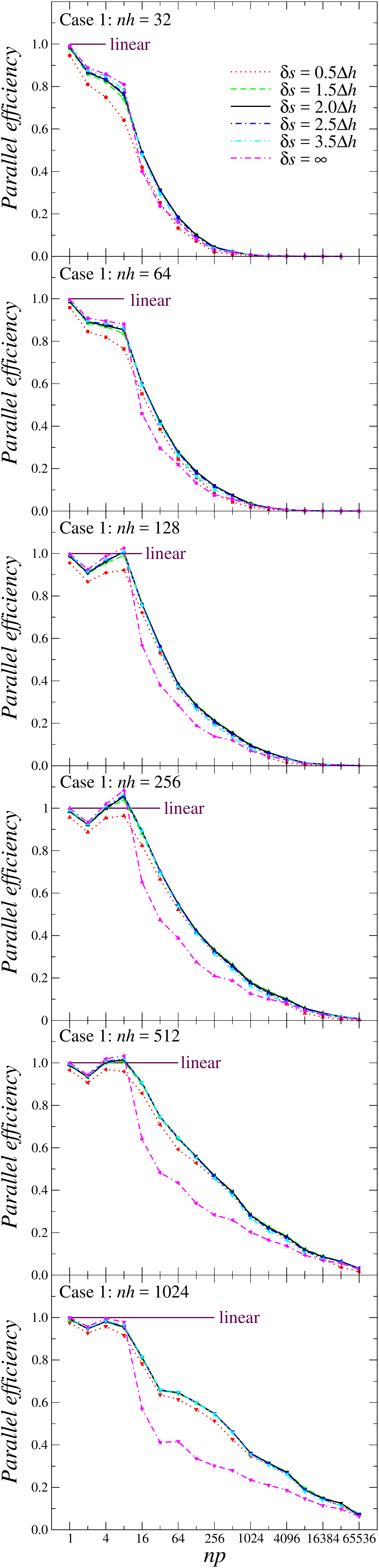}
 \includegraphics[angle=0,width=0.3\textwidth]{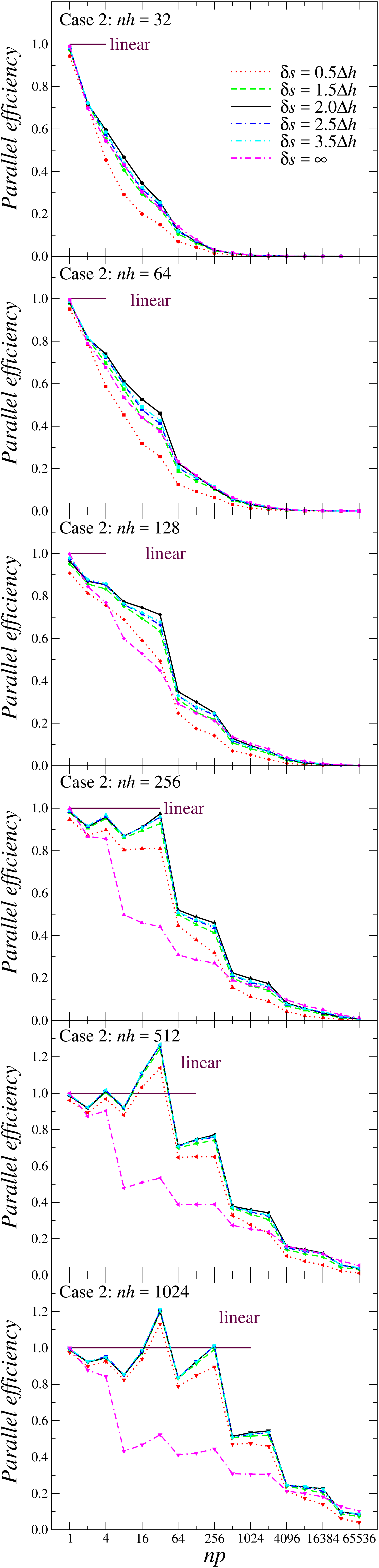}
 \includegraphics[angle=0,width=0.3\textwidth]{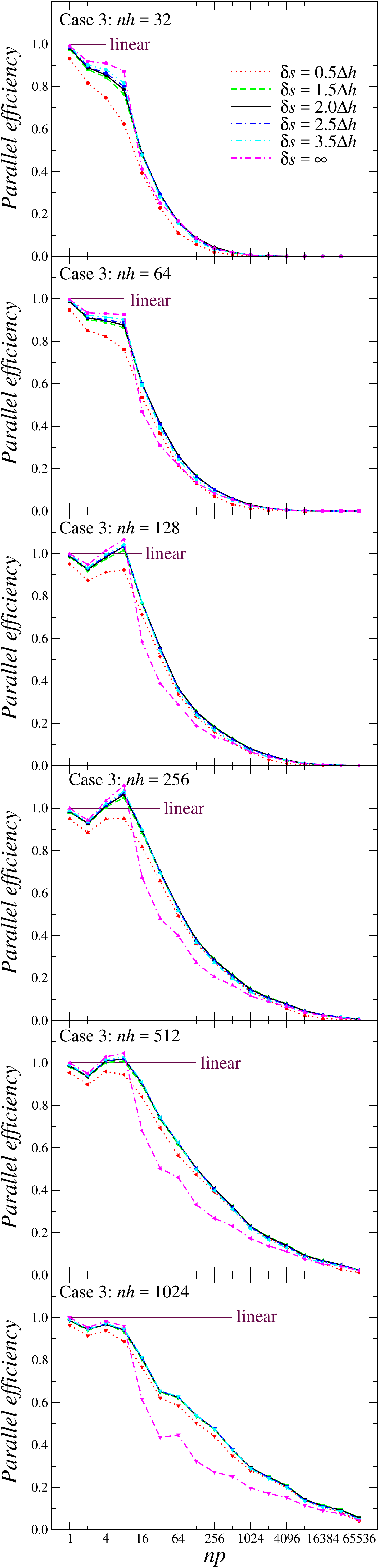}
\end{center}
 \caption{Parallel fast marching algorithm: the parallel efficiency as a function of the stride size $\delta s$, the grid size $nh$, and the number of processes $np$.}
\label{fig:efficiencies}
\end{figure}

\begin{figure}[htbp!]
\ContinuedFloat
\begin{center}
 \includegraphics[angle=0,width=0.3\textwidth]{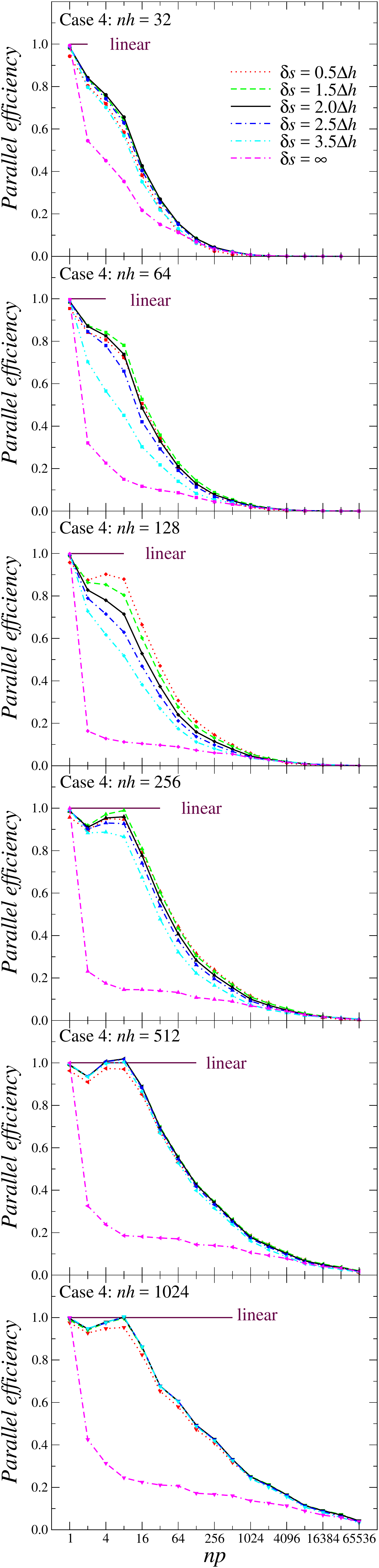}
 \includegraphics[angle=0,width=0.3\textwidth]{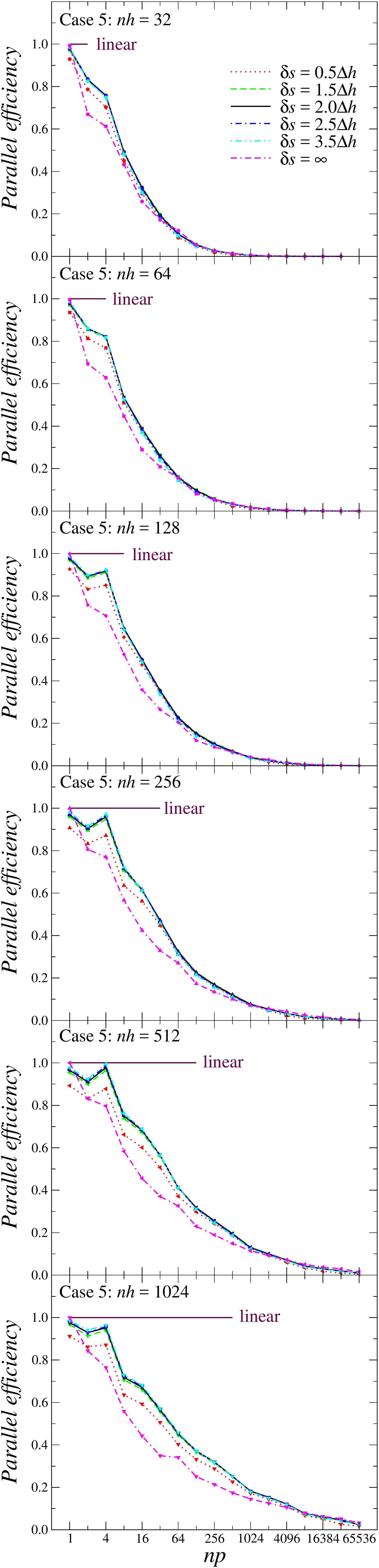}
 \includegraphics[angle=0,width=0.3\textwidth]{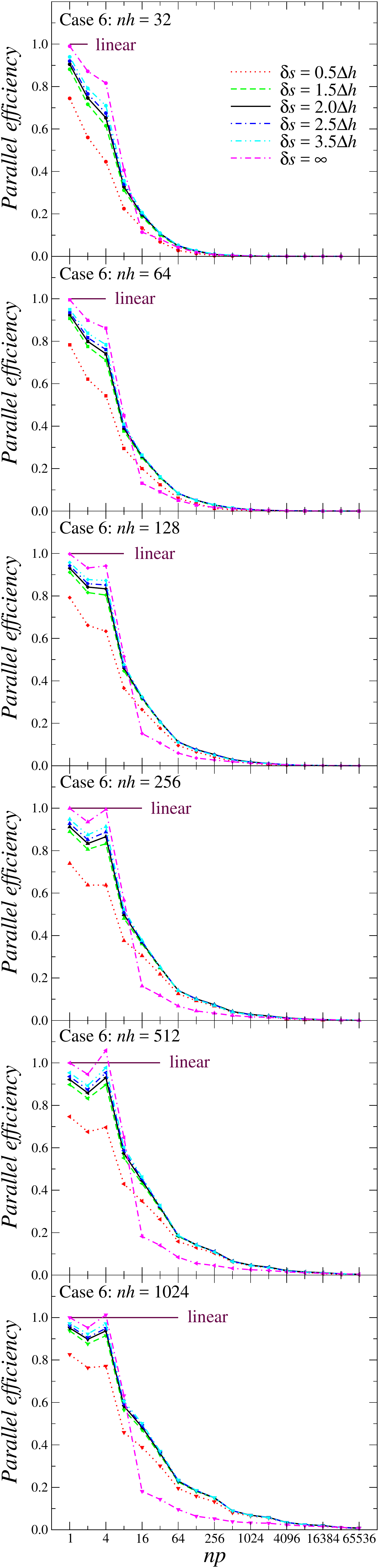}
\end{center}
 \caption{Continued. Parallel fast marching algorithm: the parallel efficiency as a function of the stride size $\delta s$, \\the grid size $nh$, and the number of processes $np$.}
\end{figure}

The parallel efficiency, which is defined as $E = \frac{T _S}{np \; T_{np}}$, is shown in Fig. \ref{fig:efficiencies}. Again, the CPU time $T_S$ from the sequential algorithm is used here instead of $T_1$. Even so, the super-linear speedup behavior is evident for several different runs, which can be identified as data above the $E = 1$ lines. Interestingly, more points above $E = 1$ can be seen for grid $nh = 512$ than grid $nh = 1024$. The efficiency of $np = 4$ or $np = 8$ is also higher than that of $np = 2$. Apparently, this is mostly the effects of improved cache performance. 

\subsection{Discussions}

There are a couple of factors that contribute to the remarkable parallel performance of the present parallel fast marching method. This first one is the improved cache performance, which is common to most parallel algorithms. Each process has its own cache and the total cache capacity available to a multi-process parallel run will be generally $np$ times of the amount available to a single process.
In addition, each process only handles a portion of the data, thus the fraction of data references readily available in its cache also becomes larger. The second one is the reduced heap size in the fast marching method. Because of its $O(N \log N)$ algorithm complexity and the fact, for example, $\frac{N}{2} \log \frac{N}{2} = \frac{1}{2} N \log N - N \log \sqrt{2} $, in theory the parallel fast marching method can achieve a super-linear speedup. This is also clearly demonstrated in some parallel computations of the present study. It should be noted that here the speedups were calculated based on the CPU times $T_S$ from the sequential algorithm. The super-linear behavior would be more prominent in more runs if the CPU times of the single-process runs $T_1$ from the parallel algorithm were used.

Several types of overheads hinder the present parallel fast marching method from achieving even better speedups. Again, some are common to most parallel algorithms. The first type is the sequential overhead in a parallel program. According to Amdahl's law, the sequential portion of code determines the maximum speedup that a parallel algorithm could possibly achieve. Here, for instance, the parameter $\texttt{bound}_{\text{band}}$ in Algorithm \ref{alg:pfmm} has to be obtained in the beginning of each restart. This portion of code is sequential: no matter how many processes are used, each process will perform the same amount of work. Actually, the MPI $\textsc{AllReduce}$ operation within this portion even involves more work with increased number of processes. Fortunately, with a domain decomposition parallelization, this is major sequential portion of code in the parallel fast marching algorithm and it only involves a few operations. On the other hand, the ghost cell overheads are also common in many domain decomposition parallel algorithms. Since each subdomain will have its own ghost cells, a multi-process run will allocate more memory than a single-process run. As the number of processes increases, the total number of ghost cells increases rapidly. For example, there are $26$ ghost cells enclosing the only single grid point in each subdomain for the parallel runs with $nh = 32$ and $np = 32$. The memory overhead and the redundant operations on ghost cells become overwhelming in this case. In addition, communication overheads are inevitable in parallel algorithms implemented with the MPI library. In the present parallel algorithm, the global reduction operation mentioned above is called once in each restart, but each call only involves three elements. The local data exchanges among neighboring subdomains are the other type of communication overheads here. The amount of updated data in the overlapping regions may vary from two layers of grid points to zero. Of course, unequal amount of data communication involves some synchronization overheads that the processes with less data to handle wait for other processes to finish. But the overall performance is better than exchanging the full overlapping regions. Note that the set of augmented tags is used to minimize data sizes involved in local data exchanges in this study. 

Apparently the MPI communications introduce two synchronization points in each restart. The load imbalance between synchronization points is the major performance penalty in the present parallel algorithm. Because the whole computational domain is equally divided into subdomains, each process handles one specified subdomain with the same amount of grid points as others. In some sense, however, each grid point is different in the fast marching method. For example, depending on the configurations of the wave fronts, the quadratic solve may involve different upwind source points. Moreover, the computational cost of a heap operation on one of the heap elements may vary a lot depending on the heap size, the type of operation, and the function value of this element. Especially, the numerical operations are concentrated on the moving wave fronts and grid points with lower function values have to be solved before those with larger ones. This makes the computational loads both spatially and temporally localized, i.e., poorly balanced, in the parallel fast marching method. Except for optimal domain decompositions, it is expected that each process performs different amounts of work in the present parallel fast marching method. Hence processes with lower loads have to idle until the heavily loaded processes reach the synchronization points. 

In this work, the global reduction operation is followed by the first marching step and then the collection of updated data in the overlapping regions before the module for synchronized local data exchanges. The present parallel algorithm does not alter the essential characteristics of the sequential fast marching algorithm. And the first marching step accounts for the major portion of the total CPU time. On the other hand, the data collection module largely represents a fixed cost of checking for updated points among all grid points in the overlapping regions. The load imbalance in the part is mostly due to the different amounts of updated data in the overlapping regions of different subdomains. A process that spends more time in the marching step is very likely to have more updated data in the overlapping regions to be collected and then sent to neighbors. Therefore, it is very unlikely that the computation imbalance can be partially offset by the communication overhead even with a non-blocking communication mode in the local data exchanges.

After the synchronized data exchanges, each process integrates the received data into its own heaps. For small or moderate numbers of processes, the cost of this part is relatively minor, compared with that of the first marching step. But for a very large $np$, the fraction of grid points within a subdomain residing in the overlapping regions will be substantial, and the integration procedure may become a significant portion in the total CPU time. In particular, if one point within an overlapping region is updated, it may be sent to and integrated by up to seven neighboring subdomains. The additional cost required for dealing with these points may greatly offset the cost savings from the shortened heaps in a smaller grid block. 
Just like the data integration module, the second marching step only accounts for a small portion of the total CPU time for small and moderate $np$, and its share may become significant for a very large $np$. On the other hand, the wave fronts only propagate a limited distance within one restart in the present restarted narrow band approach. This feature greatly confines the load imbalance in the second marching step and guarantees the overall parallel performance. 

\section{Conclusions}\label{sec:conclusions}

A highly scalable massively parallel fast marching method has been developed in this paper. A domain decomposition technique is adopted for achieving an efficient parallel algorithm capable of tackling large-scale practical applications using billions of grid points and hundreds of thousands of processes. A novel restarted narrow band approach that profoundly resembles the sequential narrow band fast marching method has been established. The fronts are advanced using essentially the sequential algorithm by a specified stride in each restart until the global narrow band width is reached or no more points have to be computed. Within each restart, simple synchronous local exchanges and global reductions are adopted for communicating updated data in the overlapping regions between neighboring subdomains and getting the latest front status, respectively. The restarted narrow band approach balances the cost associated with the number of restarts, i.e., the local data collection, communication, and integration as well as the global data reduction, and the cost of the fast marching computations extra to a sequential run. It greatly mitigates the adverse effects of the spatial-temporal load imbalance on the parallel performance. On the other hand, the independence of front characteristics is exploited to extract the masked parallelism within the fast marching method. First, special data structures are designed to advance both the positive and negative fronts concurrently in two-sided interface problems. In addition, for a newly accepted point received from a neighboring subdomain, grid points with larger function values will be refreshed only if they are influenced by its characteristics. This represents a great saving of computational cost compared with the rollback mechanism and is enabled by the augmented status tags introduced in this study. These tags are incorporated into the sequential fast marching algorithm with surprisingly few modifications. Detailed pseudo-codes for both the sequential and parallel algorithms have been provided to illustrate the simplicity of implementation and the similarity to the sequential narrow band fast marching algorithm.

Six test cases with different source configurations have been conducted to demonstrate the efficiency, flexibility, and applicability of the present parallel algorithm. These problems are extensively tested on six uniform grids ranging from $32^3$ to $1024^3$ points using different numbers of processes ranging from $1$ to $65536$. 
The accuracy of the present parallel fast marching method has been verified through comparisons of results with those from the sequential algorithm. 
Single-process runs have been performed on all grids with both the sequential and parallel algorithms. The parallel algorithm is slightly more expensive than the sequential version due to a parallelization overhead of a few percent. This overhead results from the restarted narrow band approach as extra operations on the data in the overlapping regions are required within each restart in the parallel algorithm. 

A systematical performance study has been carried out on different grids using different 3D domain decompositions. It has been verified that the number of restarts has generally a linear relationship with the grid size in one direction and is totally different from the number of iterations in iterative algorithms.  
For computations on finer grids, sustained parallel speedups have been obtained using thousands of processes and remarkable parallel efficiencies are achieved using tens of thousands of processes. 
The effects of stride sizes have been carefully studied. A stride size of $2 \Delta h$ is suggested as a rule of thumb according to the width of the overlapping regions. But other stride sizes in its neighborhood can also perform similarly well or slightly better for some cases.
Different overheads involved in the present algorithm have been discussed.

In terms of future work, apparently, our parallel algorithm can be implemented for higher-order fast marching methods with the present restarted narrow band approach in a straightforward manner. Also, the extensions on unstructured meshes are possible as evident by the few modifications in the sequential algorithm on Cartesian grids. In the present work, one layer of ghost points for a first-order scheme results in overlapping regions of a two-cell width. It is of interest to study the effects of increased widths of overlapping regions on the parallel performance. Moreover, performance data on different parallel computers are very valuable for a deeper understanding and further improvements. Although the present parallel algorithm was developed aiming at large-scale computing environments relying on message passing, it can be applied to other computing environments and expected to perform rather well. Of course, it would be interesting to compare its performance and scalability with various parallel Eikonal solvers in the literature on different parallel architectures. In order to perform a side-by-side comparison, thread-based parallel fast marching methods for shared-memory multi-core CPU or GPU systems can be derived from the present work. Along this line, it is possible to enhance the present parallel algorithm with hybrid programming modes, e.g., using domain decompositions among compute nodes (distributed memory) and multiple threads within a compute node (shared memory), for solving problems of even larger scales and achieving even better performance with millions of threads. 

\section*{Appendix}\label{sec:app}

\begin{figure}[htbp!]
\begin{center}
 \includegraphics[angle=0,width=0.5\textwidth]{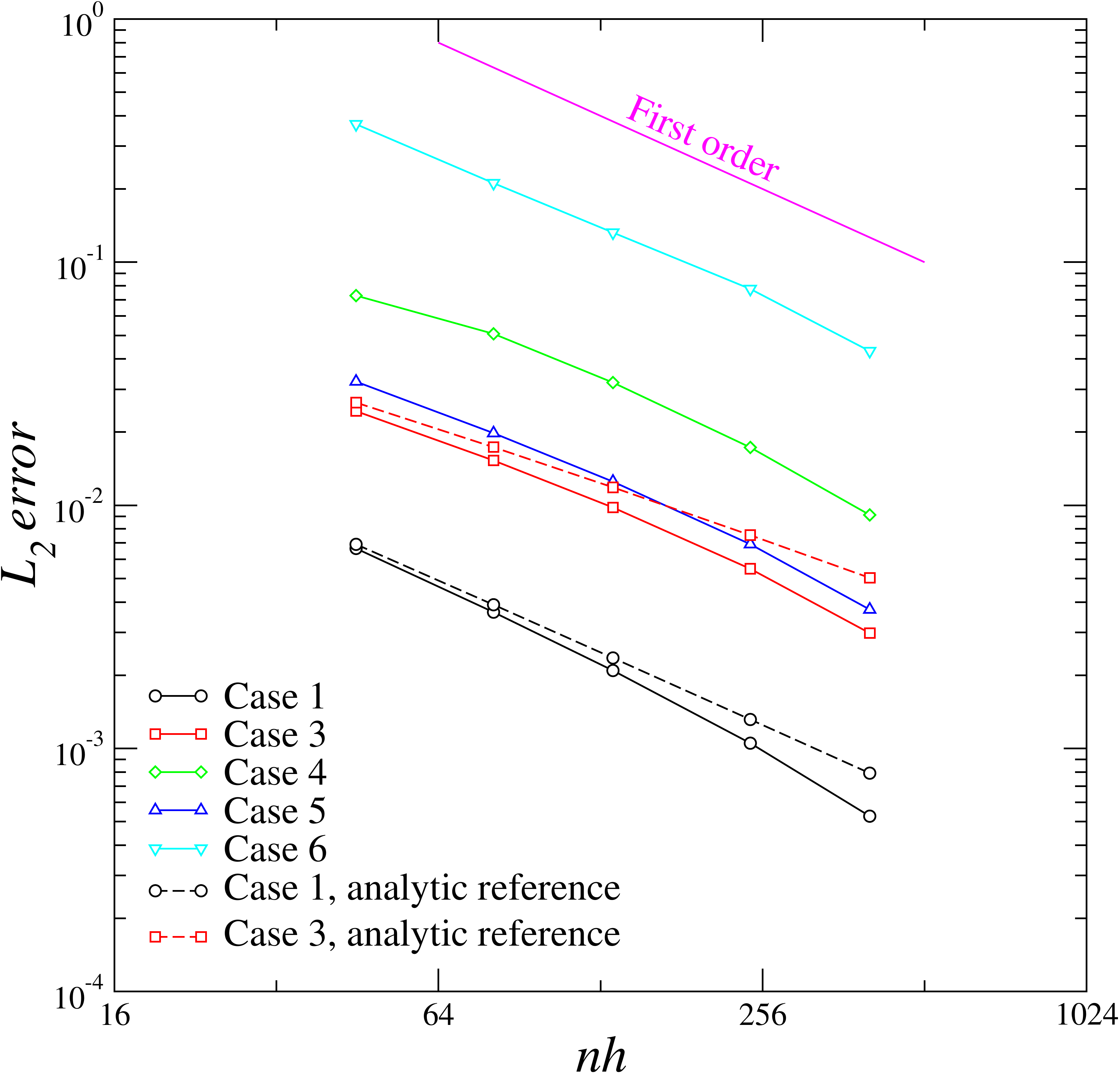}
\end{center}
 \caption{The sequential fast marching algorithm: the $L_2$ error norm as a function of the grid size $nh$.}
\label{fig:sequential_error}
\end{figure}

A grid convergence study was carried out for the sequential algorithm to show that the present implementations achieve the expected accuracy of a first-order fast marching method. Six uniform Cartesian grids with $nh = 45, 81, 135, 243, 405$, and $1215$, respectively, were employed in this study. The grid sizes were designed to collocate the points of all coarse grids on the finest grid, such that the comparisons of solutions can be directly performed without resorting to interpolations. The $L_2$ error norm is defined for the whole computational domain as
\begin{equation}
L _2 = \sqrt{\dfrac{1}{N - N _0} \sum_{i,j,k} (\psi _{i,j,k} ^R - \psi _{i,j,k} ^S) ^2}, 
\label{eq:l2norm}
\end{equation}
where $\psi _{i,j,k} ^R$ represents a reference solution computed on the finest grid or the analytical solution, if available. $N _0$ is the total number of $\texttt{FAR}$, $\texttt{BAND}$, and  $\texttt{KNOWN\_FIX}$ points, which were not included in the error norm calculation (in the present whole field computations, $\texttt{FAR}$ or $\texttt{BAND}$ point does not exist at the end). Fig. \ref{fig:sequential_error} shows the $L_2$ error norm as a function of the grid size $nh$. Case 2 was not included because boundary conditions, which were not implemented in the present algorithm, are required for the computed solution to match the analytical solution. In addition, the ghost points were defined differently on different grids (the distance of a ghost point to the domain boundary is $\frac{1}{2}\Delta h$). Thus the function values of the grid points that depend on the ghost points could not be compared directly between different grids. Nonetheless, the first-order accuracy of the present algorithms is evident for all other cases. 

\begin{figure}[htbp!]
\begin{center}
 \includegraphics[angle=0,width=0.5\textwidth]{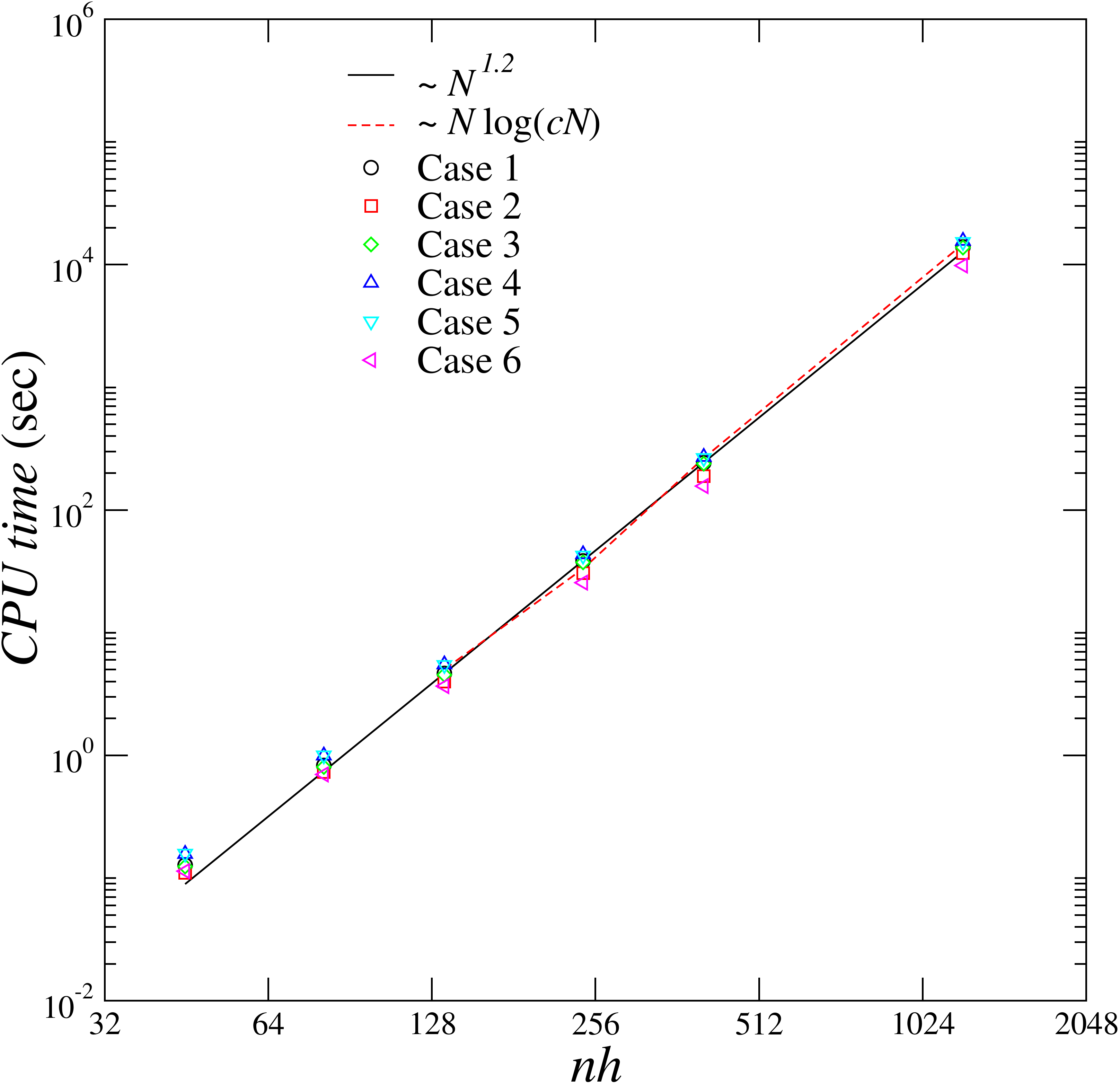}
\end{center}
 \caption{The sequential fast marching algorithm: the CPU time as a function of the grid size $nh$.}
\label{fig:sequential_cputime}
\end{figure}

Fig. \ref{fig:sequential_cputime} shows the CPU time as a function of $nh$. Although the sources and speed functions are different in these six cases, the total CPU times are in the same range. This is a desirable property of the fast marching method that distinguishes itself from other iterative methods. The $O(\log N)$ in $O(N \log N)$, the theoretical algorithm complexity of the fast marching algorithm, comes from the worst-case scenario of reordering of a heap of length $N$. In actual applications, the heap lengths are usually much smaller than $N$. In the figure, a nonlinear curve fitting given by $a + b\;N\log _{10}(c\;N)$ with $a = 6$, $b = 3 \times 10^{-6}$, and $c = 3 \times 10 ^{-7}$. On one hand, this fitting verifies the correctness of the present implementation of the fast marching algorithm. On the other hand, with such a small constant $c$ it also demonstrates the efficiency of the fast marching algorithm as a single-pass approach. Interestingly, a different fitting of $ 1 \times 10 ^{-7} \;N^{1.2}$ matches the computational results slightly better than the $N \log (N)$ one. This shows for the current test cases that the algorithm complexity of the fast marching method is only slightly higher than linear.

\section*{Acknowledgment}

This work was sponsored by the Office of Naval Research (ONR) under grant N000141-01-00-1-7, with Drs. L. Patrick Purtell, Ki-Han Kim, and Thomas C. Fu as the program managers. The simulations presented in this paper were performed at the U.S. Army Engineer Research and Development Center (ERDC) Department of Defense (DoD) Supercomputing Resource Center (DSRC) through the High Performance Computing Modernization Program (HPCMP).


\begin{thebibliography}{10}

\bibitem{Bhushan2011}
S.~Bhushan, P.~Carrica, J.~Yang, and F.~Stern.
\newblock Scalability studies and large grid computations for surface combatant
  using cfdship-iowa.
\newblock {\em International Journal of High Performance Computing
  Applications}, 25(4):466--487, 2011.

\bibitem{BreussCGV2011}
M.~Breuss, E.~Cristiani, P.~Gwosdek, and O.~Vogel.
\newblock An adaptive domain-decomposition technique for parallelization of the
  fast marching method.
\newblock {\em Applied Mathematics and Computation}, 218(1):32--44, 2011.

\bibitem{ChaconV2012}
A.~Chacon and A.~Vladimirsky.
\newblock Fast two-scale methods for eikonal equations.
\newblock {\em SIAM Journal on Scientific Computing}, 34(2):A547--A578, 2012.

\bibitem{ChaconV2015}
A.~Chacon and A.~Vladimirsky.
\newblock A parallel two-scale method for eikonal equations.
\newblock {\em SIAM Journal on Scientific Computing}, 37(1):A156--A180, 2015.

\bibitem{DetrixheGM2013}
M.~Detrixhe, F.~Gibou, and C.~Min.
\newblock A parallel fast sweeping method for the eikonal equation.
\newblock {\em Journal of Computational Physics}, 237:46--55, 2013.

\bibitem{Dijkstra59}
E.~W. Dijkstra.
\newblock A note on two problems in connexion with graphs.
\newblock {\em Numerische Mathematik}, 1:269--271, 1959.

\bibitem{Gillberg2014}
T.~Gillberg, A.~M. Bruaset, {\O}.~Hjelle, and M.~Sourouri.
\newblock Parallel solutions of static hamilton-jacobi equations for
  simulations of geological folds.
\newblock {\em Journal of Mathematics in Industry}, 4(1):1--31, 2014.

\bibitem{HelmsenPCD96}
J.~J. Helmsen, E.~G. Puckett, P.~Colella, and M.~Dorr.
\newblock Two new methods for simulating photolithography development in 3d.
\newblock {\em Proceedings of SPIE}, 2726(1):253--261, 1996.

\bibitem{Herrmann03}
M.~Herrmann.
\newblock A domain decomposition parallelization of the fast marching method.
\newblock In {\em {Annual Research Briefs}}, pages 213--225. Center for
  Turbulence Research, Stanford University, Stanford, {CA}, 2003.

\bibitem{JeongW08}
W.-K. Jeong and R.~T. Whitaker.
\newblock A fast iterative method for {E}ikonal equations.
\newblock {\em SIAM Journal on Scientific Computing}, 30(5):2512--2534, 2008.

\bibitem{RouyT92}
E.~Rouy and A.~Tourin.
\newblock A viscosity solutions approach to shape-from-shading.
\newblock {\em SIAM Journal on Numerical Analysis}, 29(3):867--884, 1992.

\bibitem{Sedgewick11}
R.~Sedgewick and K.~Wayne.
\newblock {\em Algorithms}.
\newblock Addison-Wesley, Upper Saddle River, NJ, $4^{th}$ edition, 2011.

\bibitem{Sethian99b}
J.~Sethian.
\newblock {\em Level Set Methods and Fast Marching Methods: Evolving Interfaces
  in Computational Geometry, fluid Mechanics, Computer Vision, and Materials
  Science}.
\newblock Cambridge University Press, Cambridge, $2^{nd}$ edition, 1999.

\bibitem{Sethian96}
J.~A. Sethian.
\newblock A fast marching level set method for monotonically advancing fronts.
\newblock {\em Proceedings of the National Academy of Sciences},
  93(4):1591--1595, 1996.

\bibitem{Tsitsiklis95}
J.~Tsitsiklis.
\newblock Efficient algorithms for globally optimal trajectories.
\newblock {\em IEEE Transactions on Automatic Control}, 40(9):1528--1538, 1995.

\bibitem{Tugurlan2008}
M.~C. Tugurlan.
\newblock {\em Fast marching methods-parallel implementation and analysis}.
\newblock PhD thesis, Louisiana State University, Baton Rouge, LA, 2008.

\bibitem{WeberDBBK08}
O.~Weber, Y.~S. Devir, A.~M. Bronstein, M.~M. Bronstein, and R.~Kimmel.
\newblock Parallel algorithms for approximation of distance maps on parametric
  surfaces.
\newblock {\em ACM Transactions on Graphics}, 27(104):1--16, 2008.

\bibitem{YangMBHWS10}
J.~Yang, T.~Michael, S.~Bhushan, H.~Akira, Z.~Wang, and F.~Stern.
\newblock Motion prediction using wall-resolved and wall-modeled approaches on
  a {C}artesian grid.
\newblock In {\em Proceedings of the 28th Symposium on Naval Hydrodynamics},
  September 2010.
\newblock Pasadena, CA.

\bibitem{Zhao05}
H.~Zhao.
\newblock A fast sweeping method for {E}ikonal equations.
\newblock {\em Mathematics of Computation}, 74(250):603--627, 2005.

\bibitem{Zhao07}
H.~Zhao.
\newblock Parallel implementations of the fast sweeping method.
\newblock {\em Journal of Computational Mathematics}, 25(4):421--429, 2007.

\end{thebibliography}
\end{document}